\documentclass[aps,prd,preprint,nofootinbib]{revtex4}

\usepackage{graphicx}
\usepackage{epsfig}

\newcommand{\slashp}[1]{\not{\!#1}}

\begin{document}

\title{Production of $B_c$ or $\bar{B}_c$ meson and its excited states via
$\bar{t}$-quark or $t$-quark decays}

\author{Chao-Hsi Chang$^{1,2,4}$\footnote{email:
zhangzx@itp.ac.cn}, Jian-Xiong Wang$^{3}$\footnote{email:
jxwang@mail.ihep.ac.cn} and Xing-Gang Wu$^{4}$\footnote{email:
wuxg@itp.ac.cn}}
\address{$^1$CCAST
(World Laboratory), P.O.Box 8730, Beijing 100080,
China.\footnote{Not correspondence address.}\\
$^2$Institute of Theoretical Physics, Chinese Academy of Sciences,
P.O.Box 2735, Beijing 100080, P.R. China\\
$^3$Institute of High Energy Physics, P.O.Box 918(4), Beijing 100049, China\\
$^4$Department of Physics, Chongqing University, Chongqing 400044,
P.R. China}

\date{\today}

\begin{abstract}
The production of $(b\bar{c})$-quarkonium ($\bar{B}_c$ meson and its
excited states) or $(c\bar{b})$-quarkonium ($B_c$ meson and its
excited states) via top quark $t$ or top anti-quark $\bar{t}$
decays, $t\to (b\bar{c})+c+W^+$ or $\bar{t}\to
(c\bar{b})+\bar{c}+W^-$, respectively is studied within the
framework of NRQCD. In addition to the production of the two
color-singlet $S$-wave states $|(b\bar{c})(^1S_0)_{\bf 1}\rangle$ or
$|(c\bar{b})(^1S_0)_{\bf 1}\rangle$ and $|(b\bar{c})(^3S_1)_{\bf
1}\rangle$ or $|(c\bar{b})(^3S_1)_{\bf 1}\rangle$, the production of
the $P$-wave excited $(b\bar{c})$ or $(c\bar{b})$ states, i.e. the
four color-singlet $P$-wave states $|(b\bar{c})(^1P_1)_{\bf
1}\rangle$ or $|(c\bar{b})(^1P_1)_{\bf 1}\rangle$, and
$(b\bar{c})(^3P_J)_{\bf 1}\rangle$ or $(c\bar{b})(^3P_J)_{\bf
1}\rangle$ (with $J=(1,2,3)$), is also studied. According to the
velocity scaling rule of NRQCD, for the production of $P$-wave
excited states the contributions from the two color-octet components
$|(b\bar{c})(^1S_0)_{\bf 8}\rangle$ or $|(c\bar{b})(^1S_0)_{\bf
8}\rangle$ and $|(b\bar{c})(^3S_1)_{\bf 8}\rangle$ or
$|(c\bar{b})(^3S_1)_{\bf 8}\rangle$ are also taken into account. We
quantitatively discuss the possibility and the advantages in
experimental studies of $B_c$ or $\bar{B}_c$ meson and its excited
states via the indirect production at LHC in high luminosity runs
and at LHC possible upgraded versions such as SLHC,
DLHC, TLHC etc. in future.\\


\end{abstract}

\maketitle

\section{Introduction}

Since $B_c$ meson was observed by CDF in 1998 \cite{CDF1}, the
interests in studying of the meson $B_c$ are increasing and a new
stage of the studies has been started. As pointed by the authors of
Refs.\cite{bc1,bc2}, the product cross-section at Large Hadron
Collider (LHC), CERN, is much greater than that at Tevatron,
Fermilab, therefore, more precise experimental studies of $B_c$
meson are expected at the forthcoming of LHC. Indeed, some progress
at LHC in the experimental study of $B_c$ physics can be achieved,
especially at the beginning stage of LHC (LHC just runs  at 'low'
luminosity $L=10^{32}\sim 10^{33} cm^{-2}s^{-1}$). However, when LHC
runs at higher luminosity, such as the design luminosity $L=10^{34}
cm^{-2}s^{-1}$ later on, due to the requests coming from the main
purposes of LHC (such as searching for Higgs particle and SUSY
partners etc) and the restricts from the abilities to record the
events for the detector (e.g. the detector cannot record too
frequently events etc), the condition for triggering the events
occurring in collisions has to be set such that too many $B_c$
events via the direct hadronic production according to the
theoretical estimate of `direct'
$B_c$-production\cite{bc1,bc2,bc3,bcvegpy1,bcvegpy2} are cut off. As
a result, the direct hadronic production of $B_c$ cannot be expected
to make much progress in $B_c$-meson study in the high luminosity
stage of LHC runs. Baring the situation pointed out here and the
possible upgrade for LHC (SLHC, DLHC and TLHC etc\cite{ellis1}) in
mind, the possibility to study $B_c$ meson experimentally via
indirect production of $B_c$ meson, namely, via producing a huge
amount of top antiquarks $\bar{t}$ and their decays, is worthwhile
to think seriously about. It is because that at LHC no matter how
high the luminosity, the produced top quark(s) shall never be cut
off by the trigger condition set down for any experimental purposes,
and the frequency of the $t$-quark production can be stood up for
the detector always. Furthermore, the mechanisms for producing the
doubly heavy mesons, such as $\eta_c$, $\eta_b$, $B_c$, $\cdots$ and
$J/\psi$, $\Upsilon$, $B_c^*$, $\cdots$, are interesting too. The
indirect production of $B_c$ or $\bar{B}_c$(or $B_c^{-}$), including
its excited states, via $\bar{t}$-decays or $t$-decays may offer
some knowledge on the mechanisms, therefore, this paper is devoted
to study the indirect production of $B_c$ or $\bar{B}_c$ meson via
$\bar{t}$-decays or $t$-decays. Without confusing and for
simplifying the statements, later on we will not distinguish $B_c$
and $\bar{B}_c$ unless necessary, and all results for $B_c$ and
$\bar{B}_c$ obtained in the paper are symmetric in the interchange
from particle to anti-particle.

The doubly heavy meson production via top quark decays is special
interesting from the point view of precise testing perturbative
quantum chromodynamics (pQCD)\cite{qiao}. The meson $\bar{B}_c$ is
the ground state of the heavy-flavored binding system $(b\bar{c})$,
and it is unique `doubly heavy-flavored' anti-meson in Standard
Model and is stable for strong and electromagnetic interactions.
Although in literature the `direct' hadornic production of
$\bar{B}_c$ meson has been thoroughly studied, e.g. see Refs.
\cite{bc1,bc2,bc3,bcvegpy1,bcvegpy2} (references therein) and CDF
discovered the meson which just come from the `direct' production,
as a compensation to understand the production mechanisms, it is
quite interesting that to study the production of $\bar{B}_c$ meson
`indirectly' through $t$-quark decays, especially, considering the
fact that numerous $t$-quarks may be produced at LHC. The
theoretical studies of the direct
production\cite{bc1,bc2,bc3,bcvegpy1,bcvegpy2} is based on NRQCD
\cite{nrqcd}, so now we study the indirect production based on NRQCD
too.

In the framework of effective theory of NRQCD, a doubly heavy meson
is considered as an expansion of a series Fock states. The relative
importance among the infinite ingredients is accounted by the
velocity scaling rule. Namely the physical state of $\bar{B}_c$,
$\bar{B}_c^*$, $h_{\bar{B}_c}$ and $\chi^J_{\bar{B}_c}$ can be
decomposed into a series of Fock states as follows:
\begin{eqnarray}
|\bar{B}_c\rangle = {\cal O}(v^{0})|(b\bar{c})_{\bf
1}(^{1}S_{0})\rangle+{\cal O}(v^2)|(b\bar{c})_{\bf
8}(^{1}P_{1})g\rangle  +\cdots
\nonumber\\
|\bar{B}_c^{*}\rangle = {\cal O}(v^{0})|(b\bar{c})_{\bf
1}(^{3}S_{1})\rangle+{\cal O}(v^2)|(b\bar{c})_{\bf
8}(^{3}P_{J})g\rangle  +\cdots \label{eq:1}
\end{eqnarray}
and
\begin{eqnarray}
|h_{\bar{B}_c}\rangle = {\cal O}(v^{0})|(b\bar{c})_{\bf
1}(^{1}P_{1})\rangle +{\cal O}(v^1)|(b\bar{c})_{\bf
8}(^{1}S_{0})g\rangle +\cdots
\nonumber\\
|\chi^J_{\bar{B}_c}\rangle = {\cal O}(v^{0})|(b\bar{c})_{\bf
1}(^{3}P_{J})\rangle +{\cal O}(v^1)|(b\bar{c})_{\bf
8}(^{3}S_{1})g\rangle  +\cdots, \label{eq:2}
\end{eqnarray}
here $v$ is the relative velocity between the components. The
thickened subscripts of the $(b\bar{c})$ denote for color indices,
${\bf 1}$ for color singlet and ${\bf 8}$ for color-octet; the
relevant angular momentum quantum numbers are shown in the
parentheses accordingly. According to the velocity scaling rule of
NRQCD, the probability of each Fock state in the expansion is
proportional to a definite power in $v$ as indicated as that in
Eqs.(\ref{eq:1},\ref{eq:2}). Since the value of $v^2$ is around $0.1
\sim 0.3$ \cite{nrqcd}, which is not too small, and the
contributions from the two color-octet S-wave components to the
$P$-wave production might be comparable with those from the
color-singlet components. So we shall consider the two color-octet
components: $|(b\bar{c})_{\bf 8}(^{1}S_{0})g\rangle$ and
$|(b\bar{c})_{\bf 8}(^{3}S_{1})g\rangle$, in addition to those
color-singlet components in the mesons $\bar{B}_c$, $\bar{B}_c^*$,
$h_{\bar{B}_c}$ and $\chi^J_{\bar{B}_c}$.

The calculations of the process are very complicated and lengthy by
using the conventional trace techniques to calculate the amplitude
square due to the two massive particles and bound state effects. To
shorten the calculations and to make the results more compact, we
adopt the method used in Ref.\cite{changchen} to do the
calculations. For convenience, we will call it as the `new trace
amplitude approach'. Under this approach, we first arrange the whole
amplitude into several orthogonal sub-amplitudes $M_{ss'}$ according
to the spins of the $t$-quark ($s'$) and $c$-quark ($s$), and then
do the trace of the Dirac $\gamma$ matrix strings at the amplitude
level, which result in explicit series over some independent
Lorentz-structures, and finally, we obtain the square of the
amplitude. During the calculating, some useful tricks have also been
introduced to make the expressions more compact. More detail of the
techniques and all the necessary expressions for the amplitudes of
$t(p_0)\to (b\bar{c})(p_1)+c(p_2)+W^+(p_3)$ with
$(b\bar{c})$-quarkonium in $|(b\bar{c})(^1S_0)_{\bf 1}\rangle$,
$|(b\bar{c})(^3S_1)_{\bf 1}\rangle$, $|(b\bar{c})(^1P_1)_{\bf
1}\rangle$ and $(b\bar{c})(^3P_J)_{\bf 1}\rangle$ (with $J=(1,2,3)$)
respectively are put in the APPENDIX B. While the expressions for
the contributions from the two color-octet components
$|(b\bar{c})(^1S_0)_{\bf 8}\rangle$ and $|(b\bar{c})(^3S_1)_{\bf
8}\rangle$ can be obtained from those from the two color singlet
$S$-wave components by changing the overall color-factor and the
corresponding matrix elements. This approach is different from that
of the spinor techniques or the so called `helicity amplitude
approach', which has been proposed in Ref.\cite{kleiss} and improved
in Ref.\cite{helicity}. Under the `helicity amplitude approach', one
can also derive compact results that can be further expressed by
spinor-products at the amplitude level and then do the numerical
calculations \footnote{Here, we refer to Ref.\cite{bcvegpy1} for an
example on the `helicity amplitude approach', where full processes
of the approach from the formulae deduction to the numerical
calculation can be seen explicitly. More over some tricks to
simplify the massive amplitudes can be found there.}. The above two
methods are complement to each other and both can derive simple and
compact expressions at the amplitude level. The `helicity amplitude
approach' is more suitable for the numerical calculations and is
more quicker for calculations, since full components of the helicity
amplitude can be evaluated at the amplitude level. While for the
`new trace amplitude approach', at the amplitude level only the
coefficients of the basic Lorentz structures are numerical. However
from the amplitude derived from the 'new trace amplitude approach',
one can sequentially result in the squared amplitude, which is more
easier to be compared with the results derived by the traditional
trace techniques.

According to Refs.\cite{exp}, one may expect at LHC to produce $\sim
10^8$ $t\,\bar{t}$-pairs per year under the luminosity $L=10^{34}
cm^{-2}s^{-1}$. Considering the possible upgrade for LHC and the
estimate by Refs.\cite{exp}, we will assume that one may obtain
$\sim 10^8-10^{10}$ $t\,\bar{t}$-pairs per year to examine the
possibility to observe the meson $B_c$ via decay of the produced
$t\,\bar{t}$-pairs precisely and to examine the advantages in the
indirect way to observe the meson $B_c$ and its excited states.

The paper is organized as follows. In Sec.II, we show our
calculation techniques for the process $t(p_0)\to
(b\bar{c})(p_1)+c(p_2)+W^+(p_3)$. Then we present numerical results
and make some discussions on the properties of the
$(b\bar{c})$-production through $t$-quark decays in Sec.III. The
fourth section is reserved for a summary. All necessary expressions
are put in the appendices finally.

\section{Calculation Techniques}

Under the NRQCD framework \cite{nrqcd,petrelli}, the total decay
width for the production of $(b\bar{c})$-quarkonium through the
channel $t(p_0)\to (b\bar{c})(p_1)+c(p_2)+ W^+(p_3)$ takes the form:
\begin{equation}
\Gamma =\sum_{n} H_n(t\to(b\bar{c})+c+ W^+)\times\frac{\langle{\cal
O}_n\rangle} {N_{col}} ,
\end{equation}
where $N_{col}$ refers to the number of colors, $n$ stands for the
involved state of $b\bar{c}$-quarkonium. $N_{col}=1$ for singlets
and $N_{col}=N_c^2-1$ for octets. With the help of the saturation
approximation \cite{nrqcd}, the involved decay matrix elements can
be written as \footnote{Here as suggested in Ref.\cite{petrelli}, an
overall factor $1/2N_c$ is introduced into the color-singlet matrix
elements.},
\begin{eqnarray}
\langle b\bar{c}(^1S_0)_1|{\cal O}_1(^1S_0)| b\bar{c}(^1S_0)_1
\rangle &=& |\frac{1}{\sqrt{2N_c}}\langle 0|\chi^+_c\psi_b|
b\bar{c}(^1S_0)_1 \rangle|^2[1+{\cal O}(v^4)], \\
\langle b\bar{c}(^3S_1)_1|{\cal O}_1(^3S_1)| b\bar{c}(^3S_1)_1
\rangle &=& |\frac{1}{\sqrt{2N_c}}\langle 0|\chi^+_c{\bf
\sigma}\psi_b| b\bar{c}(^3S_1)_1 \rangle|^2[1+{\cal O}(v^4)], \\
\langle b\bar{c}(^1P_1)_1|{\cal O}_1(^1P_1)| b\bar{c}(^1P_1)_1
\rangle &=& |\frac{1}{\sqrt{2N_c}}\langle
0|\chi^+_c(-\frac{i}{2}\tensor{\bf D})\psi_b|
b\bar{c}(^1P_1)_1 \rangle|^2[1+{\cal O}(v^4)], \\
\langle b\bar{c}(^3P_0)_1|{\cal O}_1(^3P_0)| b\bar{c}(^3P_0)_1
\rangle &=& |\frac{1}{\sqrt{3}\sqrt{2N_c}}\langle
0|\chi^+_c(-\frac{i}{2}\tensor{\bf D}\cdot{\bf \sigma})\psi_b|
b\bar{c}(^3P_0)_1 \rangle|^2[1+{\cal O}(v^4)], \\
\langle b\bar{c}(^3P_1)_1|{\cal O}(^3P_1)_1| b\bar{c}(^3P_1)_1
\rangle &=& |\frac{1}{\sqrt{2}\sqrt{2N_c}}\langle
0|\chi^+_c(-\frac{i}{2}\tensor{\bf D}\times{\bf \sigma})\psi_b|
b\bar{c}_1 (^3P_1)_1\rangle|^2[1+{\cal O}(v^4)], \\
\langle b\bar{c}(^3P_2)_1|{\cal O}_1(^3P_2)| b\bar{c}(^3P_2)_1
\rangle &=& |\frac{1}{\sqrt{2N_c}}\langle
0|\chi^+_c(-\frac{i}{2}\tensor{D}^{(i)}\sigma^{(j)})\psi_b|
b\bar{c}(^3P_2)_1 \rangle|^2[1+{\cal O}(v^4)],
\end{eqnarray}
where the subscript $1$ or $8$ indicates that the operator is a color
singlet or a color octet, $\psi^+$ is the Pauli-spinor field that
create a heavy quark, $\chi$ is the Pauli-spinor that create a heavy
antiquark, $D^\mu=\partial^\mu+igA^\mu$ is the gauge-covariant
derivative, $A$ is the $SU(3)$-matrix-valued gauge field. The
operator $\tensor{\bf D}$ is the difference between the covariant
derivative acting on the spinor to the right and on the spinor to
the left, which is defined by $\chi^\dag\tensor{\bf D}\psi
=\chi^\dag({\bf D}\psi)-({\bf D}\chi)^\dag\psi$. The notation
$T^{(ij)}$ is for the symmetric traceless component of a tensor:
$T^{(ij)}=(T^{ij}+T^{ji})/2-T^{kk}\delta^{ij}$. Furthermore, we have
\begin{eqnarray}
&&\frac{1}{\sqrt{2N_c}}\langle 0|\chi^+_c\psi_b| b\bar{c}(^1S_0)_1
\rangle =\frac{1}{\sqrt{4\pi}}\bar{R}_S(\Lambda)[1+{\cal
O}(v^2)] \\
&&\frac{1}{\sqrt{2N_c}}\langle 0|\chi^+_c{\bf \sigma}\psi_b|
b\bar{c}(^3S_1)_1 (\mathbf{\epsilon})\rangle =\frac{1}{\sqrt{4\pi}}
\bar{R}_S(\Lambda)\mathbf{\epsilon}[1+{\cal O}(v^2)]\\
&&\frac{1}{\sqrt{2N_c}}\langle 0|\chi^+_c{\bf
\sigma}(-\frac{i}{2}\tensor{\bf D}) \psi_b|b\bar{c}(^1P_1)_1
(\mathbf{\epsilon})\rangle =\sqrt{\frac{3}{4\pi}}
\bar{R}'_P(\Lambda) \mathbf{\epsilon}[1+{\cal O}(v^2)]\\
&&\frac{1}{\sqrt{3}}\frac{1}{\sqrt{2N_c}}\langle 0|\chi^+_c{\bf
\sigma}(\frac{1}{2}\tensor{\bf D}\cdot{\bf \sigma}) \psi_b
|b\bar{c}(^3P_0)_1 \rangle =\sqrt{\frac{3}{4\pi}}
\bar{R}'_P(\Lambda)[1+{\cal O}(v^2)] \\
&&\frac{1}{\sqrt{2}}\frac{1}{\sqrt{2N_c}}\langle 0|\chi^+_c{\bf
\sigma}(-\frac{i}{2}\tensor{\bf D}\times{\bf \sigma}) \psi_b
|b\bar{c}(^3P_1)_1 (\mathbf{\epsilon})\rangle =\sqrt{\frac{3}{4\pi}}
\bar{R}'_P(\Lambda)\mathbf{\epsilon}[1+{\cal O}(v^2)]\\
&&\frac{1}{\sqrt{2N_c}}\langle 0|\chi^+_c{\bf
\sigma}(\frac{1}{2}\tensor{D}^{(i)} \sigma^{(j)}) \psi_b
|b\bar{c}(^3P_2)_1 (\epsilon)\rangle =\sqrt{\frac{3}{4\pi}}
\bar{R}'_P(\Lambda) \epsilon^{ij}[1+{\cal O}(v^2)],
\end{eqnarray}
where $\bar{R}_S(\Lambda)$ is the average radial wavefunction for
$1S$ state averaged over a region of size $1/\Lambda$ centered at
origin, $\bar{R}'_P(\Lambda)$ is the average derivative of the
radial wavefunction at origin of size $1/\Lambda$. For convenience,
we shall take $\bar{R}_S(\Lambda)$ and $\bar{R}'_P(\Lambda)$ to be
the phenomenological values $R_S(0)$ and $R'_P(0)$, which may be
derived from the QCD potential models and relate to certain
observable such as the width for electromagnetic annihilation etc.

Although we do not know the exact values of the two decay
color-octet matrix elements, $\langle b\bar{c}(^1S_0)_8|{\cal
O}_8(^1S_0)| b\bar{c}(^1S_0)_8 \rangle$ and $\langle
b\bar{c}(^3S_1)_8|{\cal O}_8(^3S_1)| b\bar{c}(^3S_1)_8 \rangle$, we
know that they are one order in $v^2$ higher than the $S$-wave
color-singlet matrix elements according to NRQCD scale rule. More
specifically, based on the velocity scale rule, we have:
\begin{eqnarray}
\langle b\bar{c}(^1S_0)_8|{\cal O}_8(^1S_0)| b\bar{c}(^1S_0)_8
\rangle &\simeq&\Delta_S(v)^2\cdot \langle b\bar{c}(^1S_0)_1|{\cal
O}_1(^1S_0)| b\bar{c}(^1S_0)_1 \rangle \nonumber\\
\langle b\bar{c}(^3S_1)_8|{\cal O}_8(^3S_1)| b\bar{c}(^3S_1)_8
\rangle &\simeq& \Delta_S(v)^2\cdot \langle b\bar{c}(^3S_1)_1|{\cal
O}_1(^3S_1)| b\bar{c}(^3S_1)_1 \rangle\,,
\end{eqnarray}
where the second equation comes from the vacuum-saturation
approximation. $\Delta_S(v)$ is of the order $v^2$ or so, and we
take it to be within the region of 0.10 to 0.30, which is in
consistent with the identification: $\Delta_S(v)\sim\alpha_s(Mv)$
and has covered the possible variation due to the different ways to
obtain the wave functions at the origin ($S$-wave) and the first
derivative of the wave functions at the origin ($P$-wave) etc.

\begin{figure}
\centering
\includegraphics[width=0.7\textwidth]{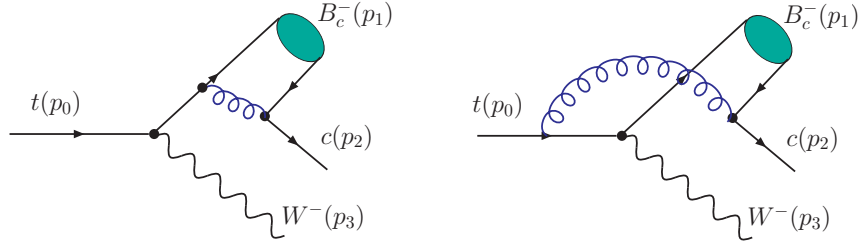}
\caption{Feynman diagrams for $t(p_0)\to (b\bar{c})(p_1)+c(p_2)+
W^+(p_3)$, where $b\bar{c}$-quarkonium is in eight states, i.e. the
six color-singlet states $(^1S_0)_1$, $(^3S_1)_1$, $(^1P_1)_1$ and
$(^3P_J)_1$ (with $J=(1,2,3)$), and two color-octet states
$(^1S_0)_8$ and $(^3S_1)_8$ respectively.. The left and the right
diagram are for the hard scattering amplitudes ${\cal A}_1$ and
${\cal A}_2$ respectively. } \label{feyn}
\end{figure}

The Feynman diagrams for the process of $t(p_0)\to
b\bar{c}(p_1)+c(p_2)+ W^+(p_3)$ are shown in Fig.(\ref{feyn}), where
$b\bar{c}$-quarkonium is in eight states: six color-singlet
components $(^1S_0)_1$, $(^3S_1)_1$, $(^1P_1)_1$ and $(^3P_J)_1$
(with $J=(1,2,3)$), and two color-octet components $(^1S_0)_8$ and
$(^3S_1)_8$ respectively. Based on the phase-space integration
simplification as shown in the Appendix A, the decay width of the
process can be written in the form:
\begin{equation}\label{width1}
d\Gamma=\frac{3|\bar{M}|^2}{256\pi^3 m_t^3}\times\frac{\langle{\cal
O}_n\rangle} {N_{col}}ds_1 ds_2 ,
\end{equation}
where the extra factor $3$ in the numerator comes from the sum of
the $c$-quark color, $|\bar{M}|^2$ is the mean square of the hard
scattering amplitude, i.e. $|\bar{M}|^2=\frac{1}{2\times3}\sum|{\cal
A}_1 +{\cal A}_2|^2$ with ${\cal A}_1$ and ${\cal A}_2$ are two
amplitudes of the process, $s_1=(p_1+p_2)^2$ and $s_2=(p_2+p_3)^2$.

We deal with either $L=0$ or $L=1$ state here. The two hard
scattering amplitudes of the process, which correspond to the left
and right diagram of Fig.(\ref{feyn}), can be written as
\begin{eqnarray}\label{swave1}
{\cal A}^{S=0,L=0}_1 &=&i{\cal C}\bar{u}_i(p_2,s)\left[\gamma_\mu
\frac{\Pi^0_{p_1}(q)}{(p_2+p_{11})^2}\gamma_\mu
\frac{\slashp{p_1}+\slashp{p_2}+m_b}
{(p_1+p_2)^2-m_b^2}\slashp{\varepsilon(p_3)}P_L \right] u_j(p_0,s')|_{q=0} \\
{\cal A}^{S=0,L=0}_2 &=&i{\cal C}\bar{u}_i(p_2,s)\left[\gamma_\mu
\frac{\Pi^0_{p_1}(q)}{(p_2+p_{11})^2}\slashp{\varepsilon(p_3)}P_L
\frac{\slashp{p_{12}}+\slashp{p_3}+m_t}{(p_{12}+p_3)^2-m_t^2}
\gamma_\mu \right] u_j(p_0,s')|_{q=0} \label{swave2}
\end{eqnarray}
and
\begin{eqnarray}\label{swave3}
{\cal A}^{S=1,L=0}_1 &=&i{\cal C}\bar{u}_i(p_2,s)\left[\gamma_\mu
\frac{\varepsilon^\alpha_s(p_1)\Pi^\alpha_{p_1}(q)}{(p_2+p_{11})^2}\gamma_\mu
\frac{\slashp{p_1}+\slashp{p_2}+m_b}
{(p_1+p_2)^2-m_b^2}\slashp{\varepsilon(p_3)}P_L \right] u_j(p_0,s')|_{q=0} \\
{\cal A}^{S=1,L=0}_2 &=&i{\cal C}\bar{u}_i(p_2,s)\left[\gamma_\mu
\frac{\varepsilon^\alpha_s(p_1)\Pi^\alpha_{p_1}(q)}
{(p_2+p_{11})^2}\slashp{\varepsilon(p_3)}P_L
\frac{\slashp{p_{12}}+\slashp{p_3}+m_t}{(p_{12}+p_3)^2-m_t^2}
\gamma_\mu \right] u_j(p_0,s')|_{q=0} \label{swave4}
\end{eqnarray}
and
\begin{eqnarray}
{\cal A}^{S=0,L=1}_1 &=&i{\cal C} \varepsilon^{\alpha}_l(p_1)
\bar{u}_i(p_2,s) \frac{d}{dq_\alpha} \left[\gamma_\mu
\frac{\Pi^0_{p_1}(q)}{(p_2+p_{11})^2}\gamma_\mu
\frac{\slashp{p_1}+\slashp{p_2}+m_b} {(p_1+p_2)^2-m_b^2}
\slashp{\varepsilon(p_3)}P_L \right] u_j(p_0,s')|_{q=0}\label{pwave1} \\
{\cal A}^{S=0,L=1}_2 &=&i{\cal C} \varepsilon^{\alpha}_l(p_1)
\bar{u}_i(p_2,s) \frac{d}{dq_\alpha} \left[\gamma_\mu
\frac{\Pi^0_{p_1}(q)}{(p_2+p_{11})^2} \slashp{\varepsilon(p_3)}P_L
\frac{\slashp{p_{12}} +\slashp{p_3}+m_t}{(p_{12}+p_3)^2-m_t^2}
\gamma_\mu \right] u_j(p_0,s')|_{q=0}\label{pwave2}
\end{eqnarray}
and
\begin{eqnarray}
{\cal A}^{S=1,L=1}_1 &=&i{\cal C} \varepsilon^{J}_{\alpha\beta}(p_1)
\bar{u}_i(p_2,s) \frac{d}{dq_\alpha} \left[\gamma_\mu
\frac{\Pi^\beta_{p_1}(q)}{(p_2+p_{11})^2}\gamma_\mu
\frac{\slashp{p_1}+\slashp{p_2}+m_b} {(p_1+p_2)^2-m_b^2}
\slashp{\varepsilon(p_3)}P_L \right] u_j(p_0,s')|_{q=0}\label{pwave3} \\
{\cal A}^{S=1,L=1}_2 &=&i{\cal C} \varepsilon^{J}_{\alpha\beta}(p_1)
\bar{u}_i(p_2,s) \frac{d}{dq_\alpha} \left[\gamma_\mu
\frac{\Pi^\beta_{p_1}(q)}{(p_2+p_{11})^2}
\slashp{\varepsilon(p_3)}P_L \frac{\slashp{p_{12}}
+\slashp{p_3}+m_t}{(p_{12}+p_3)^2-m_t^2} \gamma_\mu \right]
u_j(p_0,s')|_{q=0}\label{pwave4}
\end{eqnarray}
with the color factor ${\cal C}={\cal C}_s$ or ${\cal C}_o$ for
color-singlet and color-octet respectively, ${\cal
C}_s=\frac{4gg_s^2}{3\sqrt{6}}\delta_{ij}$ and ${\cal
C}_o=\frac{gg_s^2}{\sqrt{2}}(\sqrt{2}T^aT^bT^a)_{ij}$ ($\sqrt{2}T^b$
stands for the color of the color-octet $b\bar{c}$ state).
$\varepsilon(p_3)$ is the polarization vector of $W^+$,
$P_L=\frac{1-\gamma_5}{2}$ and $P_R=\frac{1+\gamma_5}{2}$. $q$,
$p_{11}$ and $p_{12}$ are the relative momentum between the two
constitute quarks of $(b\bar{c})$-quarkonium and the momenta of
these two constitute quarks respectively. More explicitly, we have
\begin{equation}
p_{11}=\frac{m_c}{M}p_1+q \;\;{\rm and}\;\;
p_{12}=\frac{m_b}{M}p_2-q,
\end{equation}
where $M\simeq m_b+m_c$. $\varepsilon^\alpha_{s}(p_1)$ and
$\varepsilon^\alpha_{l}(p_1)$ are the polarization vectors relating
to the spin and the orbit angular momentum of
$(b\bar{c})$-quarkonium, $\varepsilon^{J}_{\alpha\beta}(p_1)$ is the
polarization tensor for the spin triplet $P$-wave states with $J=0$,
$1$ and $2$ respectively. The covariant form of the projectors can
be conveniently written as
\begin{equation}
\Pi^0_{p_1}(q)= \frac{-\sqrt{M}}{4m_bm_c}(\slashp{p_{11}}-m_c)
\gamma_5(\slashp{p_{12}}+m_b),
\end{equation}
and
\begin{equation}
\Pi^\alpha_{p_1}(q)=
\frac{-\sqrt{M}}{4m_bm_c}(\slashp{p_{11}}-m_c)
\gamma^\alpha(\slashp{p_{12}}+m_b),
\end{equation}
To do the simplification of the projector, the following
simplification shall be useful:
\begin{equation}
\Pi^0_{p_1}(0)=\frac{1} {2\sqrt{M}}\gamma_5
(\slashp{p_1}+M)\;\;,\;\; \Pi^\alpha_{p_1}(0)=\frac{1}
{2\sqrt{M}} \gamma^\alpha(\slashp{p_1}+M),
\end{equation}
and
\begin{equation}
\frac{d}{dq_\alpha}\Pi^{0}_{p_1}(q)|_{q=0}=\frac{\sqrt{M}}{4m_b
m_c}\gamma_5 \gamma_\alpha(\slashp{p_1}+m_b-m_c) , \label{pwave5}
\end{equation}
\begin{equation}
\frac{d}{dq_\alpha}\Pi^{\beta}_{p_1}(q)|_{q=0}=-\frac{\sqrt{M}}{4m_b
m_c} \left[\gamma_\alpha\gamma_\beta(\slashp{p_1}+
m_b-m_c)-2g_{\alpha\beta}(\slashp{p^0_{11}-m_c})\right] .
\label{pwave6}
\end{equation}
Here the properties: $p_1^\alpha\varepsilon^\alpha=0$ and
$p_1^\alpha\varepsilon^{\alpha\beta}
=p_1^\beta\varepsilon^{\alpha\beta}=0$. $p^0_{11}=\frac{m_c}{M}p_1$
and $p^0_{12}=\frac{m_b}{M}p_2$, are applied. After substituting all
these relations into the amplitudes and doing the possible
simplifications, the amplitudes then be squared, summed over the
freedoms in final state and averaged over the ones in initial state.
And the selection of the appropriate total angular momentum quantum
number is done by performing the proper polarization sum. If
defining
\begin{equation}
\Pi_{\alpha\beta}=-g_{\alpha\beta}+\frac{p_{1\alpha}
p_{1\beta}}{M^2}\,,
\end{equation}
the sum over polarization for a spin triplet S-state or a spin
singlet P-state is given by
\begin{equation}
\sum_{J_z}\varepsilon_\alpha \varepsilon^*_{\alpha'}
=\Pi_{\alpha\alpha'} ,\label{3s1}
\end{equation}
where $J_z=s_z$ or $l_z$ respectively. In the case of $^3P_J$
states, as for the three multiplets
$\varepsilon^{J}_{\alpha\beta}(p_1)$ with $J=0$, 1 and 2, the sum
over the polarization is given by
\begin{eqnarray}\label{3pja}
\varepsilon^{(0)}_{\alpha\beta} \varepsilon^{(0)*}_{\alpha'\beta'}
&=& \frac{1}{3} \Pi_{\alpha\beta}\Pi_{\alpha'\beta'} \\
\sum_{J_z}\varepsilon^{(1)}_{\alpha\beta}
\varepsilon^{(1)*}_{\alpha'\beta'} &=& \frac{1}{2}
(\Pi_{\alpha\alpha'}\Pi_{\beta\beta'}-
\Pi_{\alpha\beta'}\Pi_{\alpha'\beta}) \label{3pjb}\\
\sum_{J_z}\varepsilon^{(2)}_{\alpha\beta}
\varepsilon^{(2)*}_{\alpha'\beta'} &=& \frac{1}{2}
(\Pi_{\alpha\alpha'}\Pi_{\beta\beta'}+
\Pi_{\alpha\beta'}\Pi_{\alpha'\beta})-\frac{1}{3}
\Pi_{\alpha\beta}\Pi_{\alpha'\beta'} .\label{3pjc}
\end{eqnarray}

All the terms for the squared hard scattering amplitudes
$|\bar{M}|^2$ for S-wave and P-wave states are put in Appendix B
accordingly, where the detail of the calculation techniques are also
attached for convenience. To effectuate the calculations and to make
the results more compact, we adopt the method used in
Ref.\cite{changchen}. As stated in the Introduction, we call it the
`new trace amplitude approach' for convenience. Under the approach,
we arrange the whole amplitude into several orthogonal
sub-amplitudes $M_{ss'}$ according to the spins of the $t$-quark
($s'$) and $c$-quark ($s$) first, and then do the trace of the Dirac
$\gamma$ matrix strings at the amplitude level by properly dealing
with the massive spinors, which result in explicit series over some
independent Lorentz-structures. The expressions for these
coefficients of all the considered channels are put in Appendix B.
With the help, one can do the square of the amplitude easily. As a
cross-check of our results, we adopt the traditional trace
techniques and also the FDC\cite{fdc} package to derive the
numerical results of the mentioned processes. Indeed we find a well
agreement among these methods.

As a comparison and for later usages, let us present the width for
the two body decay $t(p_1)\to b(p_2)+ W^+(p_3)$, which is dominant
for the $t$-quark decay:
\begin{equation}
\Gamma=\frac{G_F m_t^2|\mathbf{\vec{p}_2}|}
{4\sqrt{2\pi}}\left[(1-y^2)^2+x^2(1+y^2-2x^2)\right] ,
\end{equation}
where $|\mathbf{\vec{p}_2}|=\frac{m_t}{2}
\sqrt{(1-(x-y)^2)(1-(x+y)^2)}$, $m_w=m_t x$ and $m_b=m_t y$.

\section{Numerical Results of Differential Cross-Sections}

In numerical calculations, we take the parameters as follows:
\begin{eqnarray}
m_b=4.9\ {\rm GeV},\; m_c=1.5\ {\rm GeV},\; m_t=176\ {\rm GeV},\;
m_w=80.22\ {\rm GeV},\; \alpha_s(2m_c)=0.26 ,
\end{eqnarray}
and $g=2\sqrt{2}m_w\sqrt{G_F/\sqrt{2}}$. Then, the decay width
of $t\to W^{+} + b$ is
\begin{equation}
\Gamma(t\to W^{+} + b)=1.59{\rm GeV}.
\end{equation}
And the decay widths of $t\to (b\bar{c}) + W^{+} + c$ are:
\begin{eqnarray}
\Gamma_{t\to(b\bar{c})[(^3S_1)_1]}&=&0.79 \;{\rm MeV} \\
\Gamma_{t\to(b\bar{c})[(^1S_0)_1]}&=&0.57 \;{\rm MeV} \\
\Gamma_{t\to(b\bar{c})[(^1P_1)_1]}&=&0.057 \;{\rm MeV} \\
\Gamma_{t\to(b\bar{c})[(^3P_0)_1]}&=&0.034 \;{\rm MeV} \\
\Gamma_{t\to(b\bar{c})[(^3P_1)_1]}&=&0.070 \;{\rm MeV} \\
\Gamma_{t\to(b\bar{c})[(^3P_2)_1]}&=&0.075 \;{\rm MeV} \\
\Gamma_{t\to(b\bar{c})[(^3S_1)_8]}&=&0.091\times{v^4} \;{\rm MeV} \\
\Gamma_{t\to(b\bar{c})[(^1S_0)_8]}&=&0.070\times{v^4} \;{\rm MeV}
\end{eqnarray}
where $v^2\simeq (0.1\sim 0.3)$. It may be more useful to show the
ratio between the decay width of $t\to (b\bar{c}) + W^{+} + c$ and
$t\to W^{+} + b$, since uncertanties from the electro-weak coupling
can be cancelled out:
\begin{eqnarray}
\frac{\Gamma_{t\to(b\bar{c})[(^3S_1)_1]}}{\Gamma(t\to W^{+} + b)}&=& 4.97\times10^{-4}\\
\frac{\Gamma_{t\to(b\bar{c})[(^1S_0)_1]}}{\Gamma(t\to W^{+} + b)}&=& 3.58\times10^{-4}\\
\frac{\Gamma_{t\to(b\bar{c})[(^1P_1)_1]}}{\Gamma(t\to W^{+} + b)}&=& 0.36\times10^{-4}\\
\frac{\Gamma_{t\to(b\bar{c})[(^3P_0)_1]}}{\Gamma(t\to W^{+} + b)}&=& 0.21\times10^{-4}\\
\frac{\Gamma_{t\to(b\bar{c})[(^3P_1)_1]}}{\Gamma(t\to W^{+} + b)}&=& 0.44\times10^{-4}\\
\frac{\Gamma_{t\to(b\bar{c})[(^3P_2)_1]}}{\Gamma(t\to W^{+} + b)}&=& 0.47\times10^{-4}\\
\frac{\Gamma_{t\to(b\bar{c})[(^3S_1)_8]}}{\Gamma(t\to W^{+} + b)}&=& 0.57\times10^{-4} v^4\\
\frac{\Gamma_{t\to(b\bar{c})[(^1S_0)_8]}}{\Gamma(t\to W^{+} + b)}&=& 0.44\times10^{-4} v^4.
\end{eqnarray}

\begin{figure}
\centering
\includegraphics[width=0.45\textwidth]{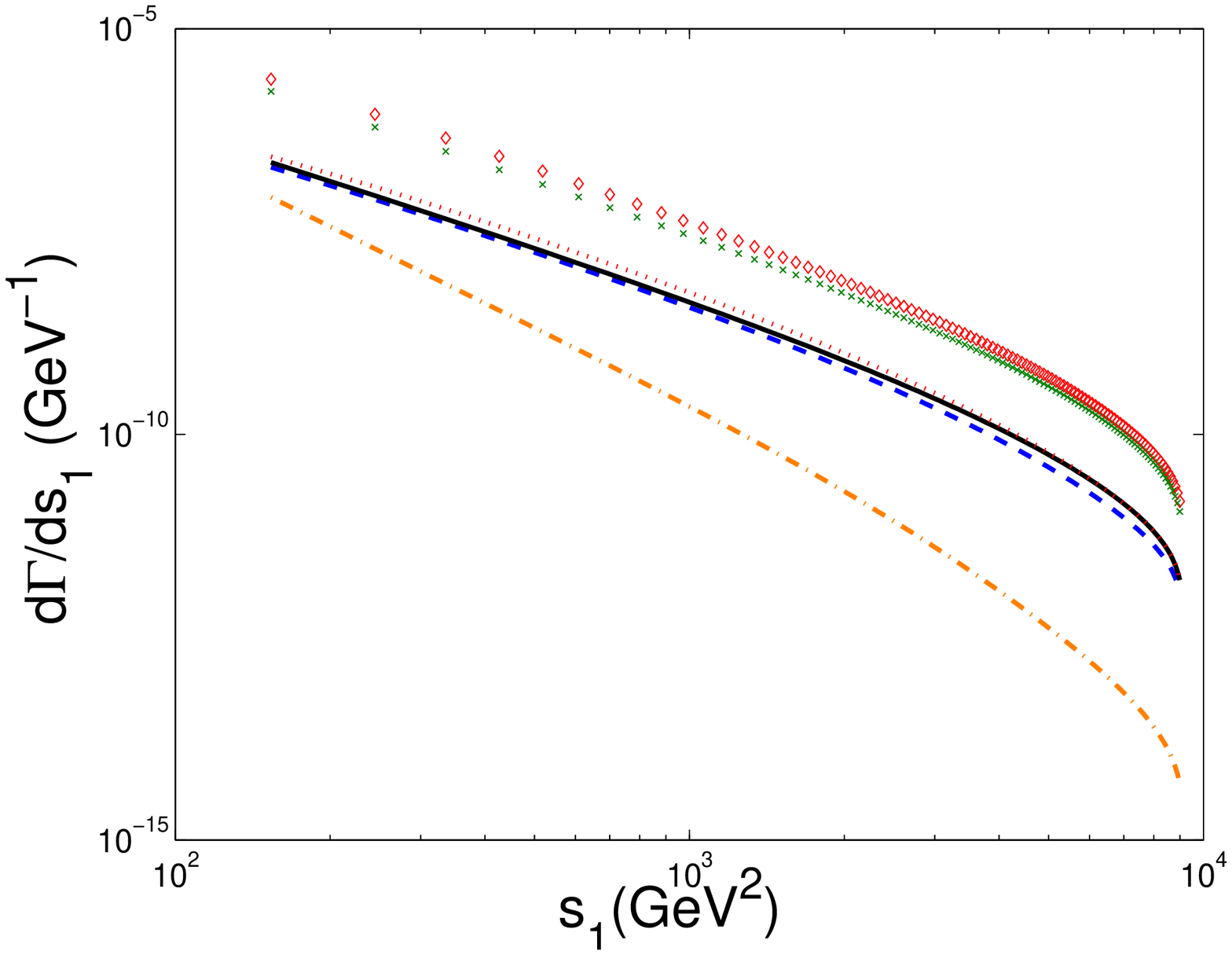}%
\hspace{0.2cm}
\includegraphics[width=0.45\textwidth]{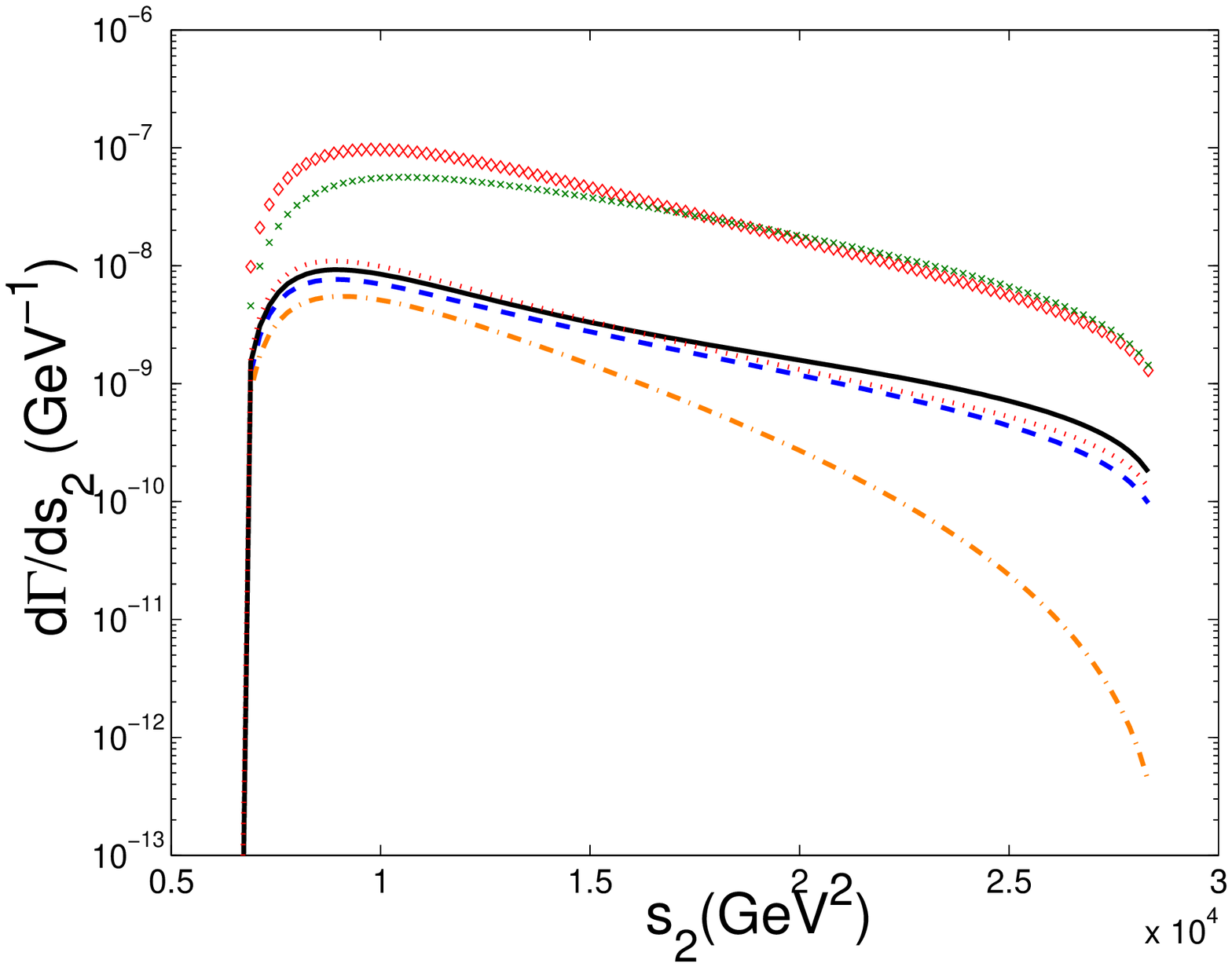}\hfill
\caption{The invariant mass distributions for $t(p_0)\to
(b\bar{c})(p_1)+c(p_2)+ W^+(p_3)$. The left is for $d\Gamma/ds_1$
and the right is for $d\Gamma/ds_2$. The diamond line, the cross
line, the dotted line, the solid line, the dashed line and the
dash-dot line are for $(b\bar{c})$-quarkonium in Fock states:
$(^3S_1)_1$, $(^1S_0)_1$, $(^3P_2)_1$, $(^3P_1)_1$, $(^1P_1)_1$ and
$(^3P_0)_1$ respectively.} \label{diss1s2}
\end{figure}

\begin{figure}
\centering
\includegraphics[width=0.45\textwidth]{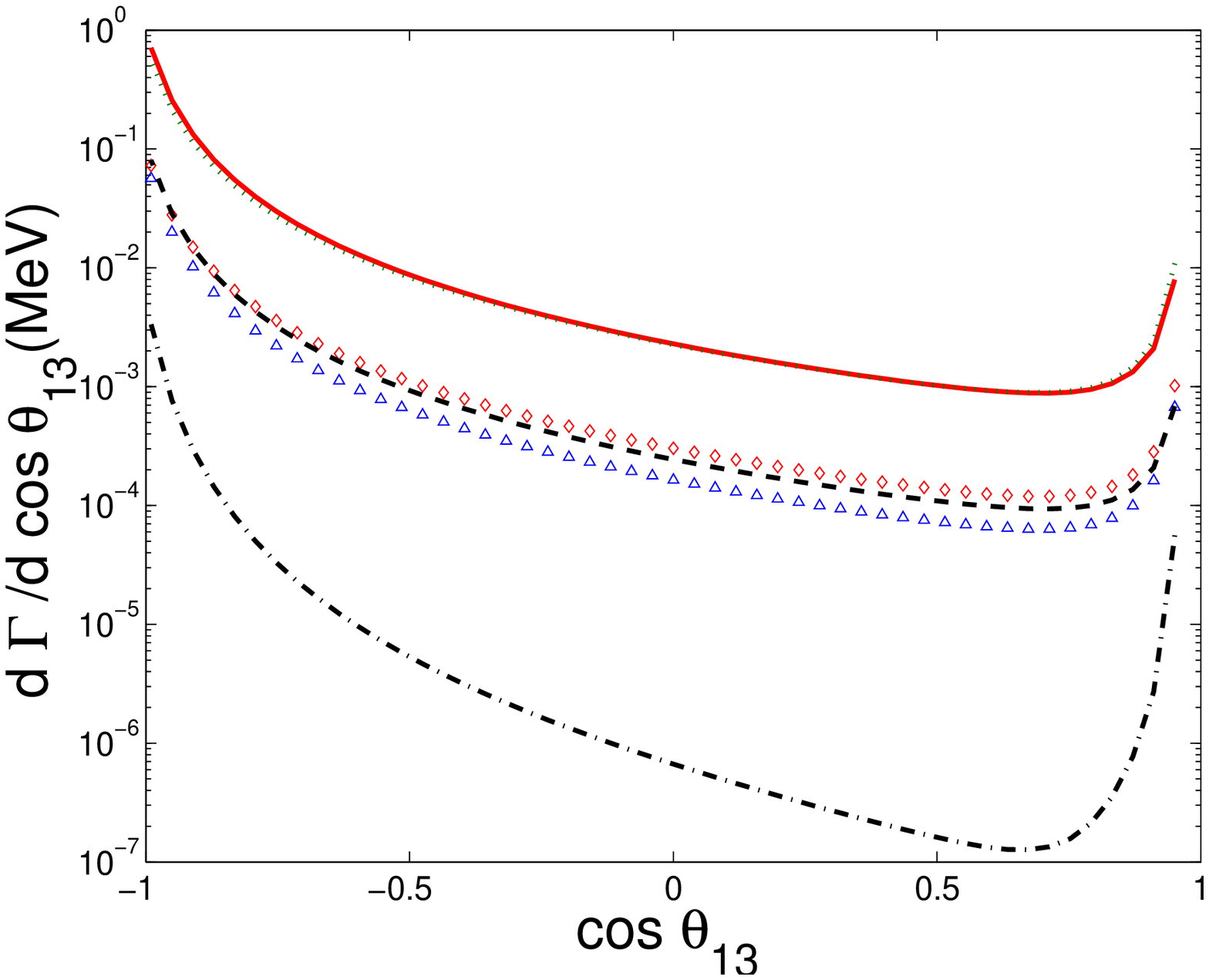}%
\hspace{0.3cm}
\includegraphics[width=0.45\textwidth]{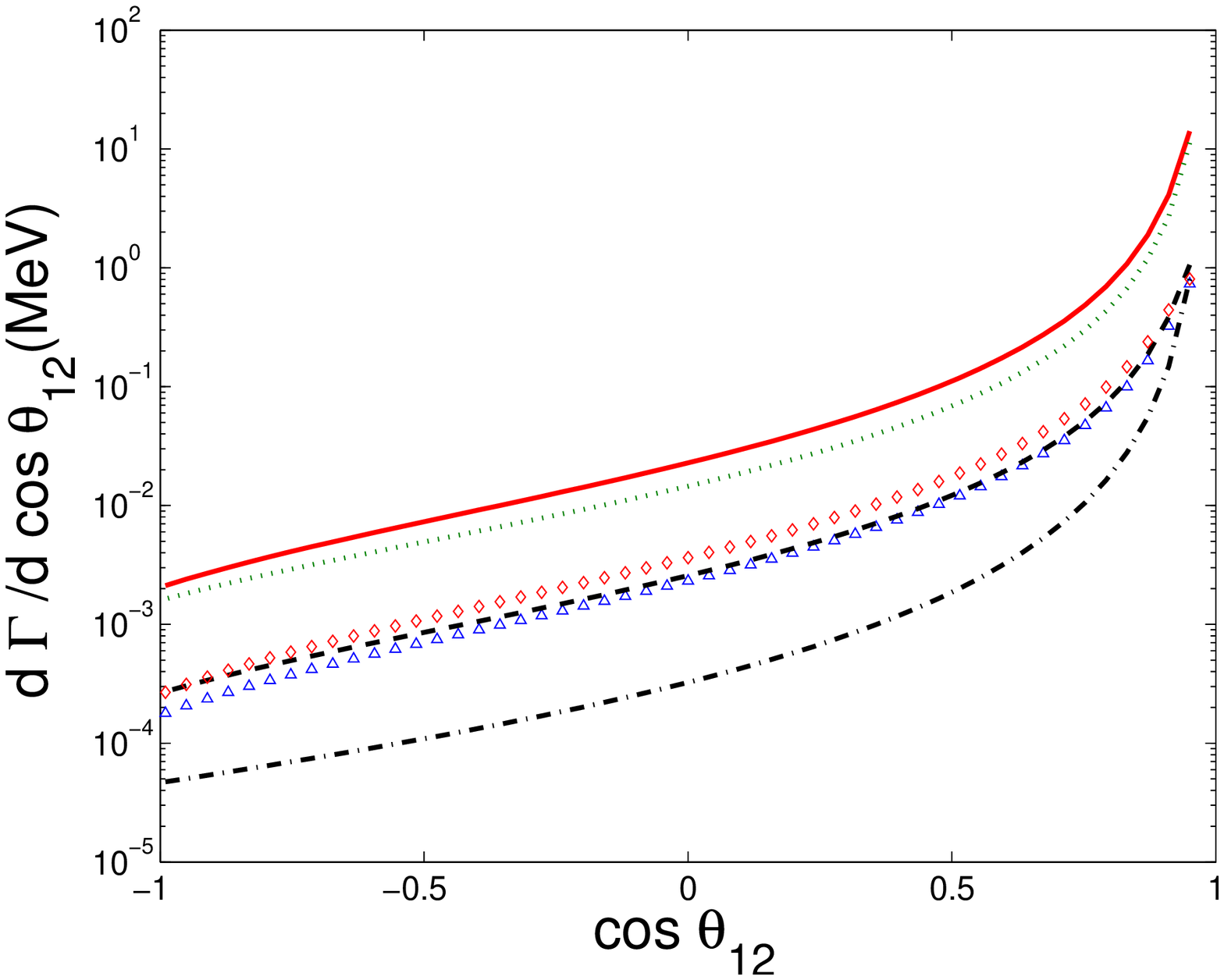}\hfill
\caption{The differential distributions $d\Gamma/d\cos\theta_{13}$
(Left) and $d\Gamma/d\cos\theta_{12}$ (Right) for $t(p_0)\to
(b\bar{c})(p_1)+c(p_2)+ W^+(p_3)$. The solid line, the dotted line,
the diamond line, the dashed line, the triangle line and the
dash-dot line are for $(b\bar{c})$-quarkonium in Fock states:
$(^3S_1)_1$, $(^1S_0)_1$, $(^3P_2)_1$, $(^3P_1)_1$, $(^1P_1)_1$ and
$(^3P_0)_1$ respectively.} \label{discos}
\end{figure}

Let us show some more characteristics of the decay $t\to (b\bar{c})
+ W^{+} + c$. The differential distributions of the invariant masses
$s_1$ and $s_2$, i.e. $d\Gamma/ds_1$ and $d\Gamma/ds_2$ are shown in
Fig.\ref{diss1s2}. While the differential distributions of
$\cos\theta_{13}$ and $\cos\theta_{12}$, i.e.
$d\Gamma/d\cos\theta_{13}$ and $d\Gamma/d\cos\theta_{12}$ is shown
in Fig.\ref{discos}, where $\theta_{13}$ is the angle between
$\vec{p}_1$ and $\vec{p}_3$, and $\theta_{12}$ is the angle between
$\vec{p}_1$ and $\vec{p}_2$ respectively in the $t$-quark rest frame
($\vec{p}_0=0$). It can be found that the largest differential
cross-section of $d\Gamma/d\cos\theta_{13}$ is achieved when
$\theta_{13}=180^{\circ}$, i.e. the $(b\bar{c})$-quarkonium and
$W^+$ moving back to back in the rest frame of $t$-quark. And the
largest differential cross-section of $d\Gamma/d\cos\theta_{12}$ is
achieved when $\theta_{12}=0^{\circ}$, i.e. the
$(b\bar{c})$-quarkonium and $c$-quark moving in the same direction.

\section{Discussions and Summary}

In the present paper, we have studied the decay channel $t(p_0)\to
b\bar{c}(p_1)+c(p_2)+ W^+(p_3)$ in the leading $\alpha_s$
calculation but with the $v^2$-expansion up to $v^4$, where
$b\bar{c}$-quarkonium is in one of the eight states: the six
color-singlet states $(^1S_0)_1$, $(^3S_1)_1$, $(^1P_1)_1$ and
$(^3P_J)_1$ (with $J=(1,2,3)$), and two color-octet states
$(^1S_0)_8$ and $(^3S_1)_8$ respectively. In literature, only
$(^3S_1)_1$ state has been studied \cite{qiao}, however it can be
found that all the other considered states can be sizable in
addition to the $(^3S_1)_1$ state.

As mentioned in the Introduction, about $10^8$ $t\,\bar{t}$ per year
will be produced in the stage of high luminosity run at LHC and
$t$-quark events always trigger the detector, then according to the
present estimate it is possible to accumulate about $10^5$ $B_c$
events a year via $\bar{t}$-quark decay at LHC. Moreover, the
indirect production of the $(b\bar{c})$-quarkonium may be traced
back to $t$-quark decay and has the characteristics in
$\theta_{13}\,, \theta_{12}$ shown in Figs.(2,3) etc, that may be
used to identify the $(b\bar{c})$-quarkonium events. Thus there may
be some advantages in $(b\bar{c})$-quarkonium studies via the
indirect production in comparison with the direct production.
Especially when LHC is really upgraded to SLHC, DLHC and TLHC, so
$10^9\sim 10^{10}$ $t\,\bar{t}$ per year may be produced, one may
expect much progress in $(b\bar{c})$-quarkonium studies is achieved
then.

\vspace{10mm}

\noindent {\bf\Large Acknowledgments:} This work was supported in
part by the Natural Science Foundation of China (NSFC) and by the
Grant from Chongqing University. \\

\appendix

\section{Formulae for the phase space integration}

In this section, we derive the phase space of $t(m_t, p_0)\to
(b\bar{c})(m_1, p_1) +c(m_2, p_2)+ W^+(m_3, p_3)$. The decay width
is proportional to the phase space:
\begin{equation}
d\Gamma\propto\frac{1}{2p_0^0}\prod_{i=1}^3 \frac{d^3p_i}{(2\pi)^3
(2p^0_i)}(2\pi)^4\delta^4(p_0- p_1- p_2- p_3),
\end{equation}
where $p_i=(p_i^0,\vec{p}_i)=(p_i^0,p_i^1,p_i^2,p_i^3)$.
Furthermore, in the rest frame of top quark ($p_0^0=m_t$), we have
\begin{eqnarray}
\frac{d\Gamma}{ds_1 ds_2}&\propto& \frac{1}{(2m_t)(2\pi)^5} d^4p_1
d^4p_2 d^4p_3 \delta(p_1^2-m_1^2) \delta(p_2^2-m_2^2)
\delta(p_3^2-m_3^2)\theta(p_1^0)\theta(p_2^0)\cdot\nonumber\\
&&\theta(p_3^0)\delta(s_1-(p_1+p_2)^2) \delta(s_2-(p_2+p_3)^2)
\delta^4(p_0- p_1-p_2- p_3) \nonumber\\
&\propto& \frac{1}{(2m_t)(2\pi)^5} d^4p_1 d^4p_3 \delta(p_1^2-m_1^2)
\delta((p_0-p_1-p_3)^2-m_2^2)\delta(p_3^2-m_3^2)\theta(p_1^0)\cdot\nonumber\\
&&\theta(p_3^0)\theta(m_t-p_1^0-p_3^0)
\delta(s_1-(p_0-p_3)^2)\delta(s_2-(p_0-p_1)^2) \nonumber\\
&\propto& \frac{1}{2^8 m_t^3\pi^5} d^3\vec{p}_1 d^3\vec{p}_3
\delta({p_1^0}^{2}-\vec{p}_1^2-m_1^2)
\delta({p_3^0}^{2}-\vec{p}_3^2-m_3^2)
\theta(p_1^0)\theta(p_3^0)\cdot\nonumber\\
&&\theta(m_t-p_1^0-p_3^0) \delta(s_2+m_3^2-m_2^2-2m_tp_3^0
+2p_1^0p_3^0-2\vec{p}_1\cdot\vec{p}_3)\nonumber \\
&\propto& \frac{|\vec{p}_1||\vec{p}_3|}{2^{10} m_t^3\pi^5} d\Omega_1
\sin\theta_{13}d\theta_{13}d\phi_{3} \theta(p_1^0)
\theta(p_3^0)\theta(m_t-p_1^0-p_3^0)\cdot\nonumber\\
&& \delta(s_1+s_2-m_t^2-m_2^2
+2p_1^0p_3^0-2|\vec{p}_1|\cdot|\vec{p}_3|\cos\theta_{13})\label{costhe}\\
&\propto& \frac{1}{2^8 m_t^3\pi^3}\theta(p_1^0)\theta(p_3^0)
\theta(m_t-p_1^0-p_3^0)\theta(X) ,
\end{eqnarray}
where $p_1^0=\frac{m_t^2+m_1^2-s_2}{2m_t}$ and
$p_3^0=\frac{m_t^2+m_3^2-s_1}{2m_t}$,
$|\vec{p}_3|=\sqrt{{p_3^0}^2-m_3^2}$ and
$|\vec{p}_1|=\sqrt{{p_1^0}^2-m_1^2}$. The step function $\theta(X)$
is determined by ensuring $|\cos\theta_{13}|\leq1$, where
\begin{displaymath}
\cos\theta_{13}=\frac{s_1+s_2-m_t^2-m_2^2
+2p_1^0p_3^0}{2|\vec{p}_1||\vec{p}_3|} .
\end{displaymath}
All these step functions leads to the integration ranges:
\begin{eqnarray}
s_1^{\rm min} &=& m_1^2+m_2^2-\frac{\left[ (s_2-m_t^2+m^2_{1})
(s_2-m_3^2+m^2_2)+ \sqrt{\lambda(s_2,m_t^2,m^2_1)
\lambda(s_2,m_3^2,m^2_2)} \right]}{2s_2} \\
s_1^{\rm max} &=& m^2_{1}+m_2^2-\frac{\left[ (s_2-m_t^2+m^2_{1})
(s_2-m_3^2+m^2_2)- \sqrt{\lambda(s_2,m_t^2,m^2_{1})
\lambda(s_2,m_w^2,m^2_2)} \right]}{2s_2} \\
s_2^{\rm min} &=& (m_2+m_3)^2 \\
s_2^{\rm max} &=& (m_t-m_{1})^2
\end{eqnarray}
where $\lambda(x,y,z)=(x-y-z)^2-4yz$.

Furthermore, we can obtain the $\cos\theta_{13}$ distribution from
Eq.(\ref{costhe}):
\begin{eqnarray}
\frac{d\Gamma}{dcos\theta_{13} ds_1}&\propto& \frac{J} {2^7
m_t^3\pi^3}\theta(p_1^0)\theta(p_3^0)
\theta(m_t-p_1^0-p_3^0)\theta(Y),
\end{eqnarray}
where the extra Jacoiban
\begin{equation}
J=\frac{-|\vec{p}_1||\vec{p}_3|}{\left|1 - \frac{p_3^0} {2 m_t} +
\frac{|\vec{p}_3| ( {m_1}^2 + {m_t}^2 - s_2 ) \cos\theta_{13}}{
{m_t}{\sqrt{{m_1}^4 + {( {m_t}^2 - s_2 ) }^2 - 2{m_1}^2( {m_t}^2 +
s_2 )}}}\right|}
\end{equation}
and
\begin{eqnarray}
s_2 &=&\frac{1} {{|\vec{p}_3|}^2\cos\theta_{13}^2-( m_t - p_3^0
)^2}\Bigg\{( {m_1}^2 + {m_t}^2 ) {|\vec{p}_3|}^2 \cos\theta_{13}^2-(
m_t - p_3^0 ) (m_t( {m_2}^2 - {m_3}^2
+\nonumber\\
&& m_t p_3^0 )-{m_1}^2p_3^0) - {m_t}|\vec{p}_3|
\cos\theta_{13}\Big[- 2{m_1}^2({m_2}^2
+2(m_t-p_3^0)^2-s_1-2|\vec{p}_3|^2\cos\theta_{13}^2)\nonumber\\
&&  +{m_1}^4 + (m_2^2-s_1)^2 \Big]^{1/2}\Bigg\}
\end{eqnarray}
The $\theta(X)$ function determines the boundary of $s_1$:
\begin{equation}
s_{1min}\leq s_1\leq (m_t-m_3)^2
\end{equation}
where
\begin{eqnarray}
s_{1min} &=& \frac{{m_2}^2{m_t}^2 + {m_1}^2(
{m_t}^2\cos\theta_{13}^2 + {m_3}^2(\cos\theta_{13}^2 -1 ) )+m_1 m_t
\sqrt{Y}}{{m_t}^2 -{m_1}^2( 1-\cos\theta_{13}^2 ) }
\end{eqnarray}
with
\begin{eqnarray}
Y&=& {( {m_1}^2 - {m_2}^2 + {m_3}^2 -{m_t}^2 ) }^2 -\Big( {m_1}^4
-2(m_2^2-3m_3^2+m_t^2)m_1^2
+\nonumber\\
&& ((m_2-m_3)^2-m_t^2)((m_2+m_3)^2-m_t^2)\Big) \cos\theta_{13}^2+
4{m_1}^2{m_3}^2\cos\theta_{13}^4 .
\end{eqnarray}
It should be noted that the integration over $\cos\theta_{13}$
should be from $1$ to $-1$. The distribution for $\theta_{12}$ can
be obtained in a similar way.

\section{Amplitude of the process $t(p_0)\to (b\bar{c})(p_1)+c(p_2)+W^+(p_3)$}

Now let us illustrate the method to calculate the amplitude squared.
In general, the amplitude for the process in which two massive
fermions with spin projection $s$ and $s'$ are involved can be
written as
\begin{equation}
M_{ss'}=\bar{u}_s(p_2) A u_{s'}(p_0) ,
\end{equation}
where $p_2$ and $p_0$ denote the momenta of the top quark and the
charm quark with mass $m_t$ and $m_c$, and $A$ is an explicit string
of Dirac $\gamma$ matrices for the process, which can be read from
Eqs.(\ref{swave1},\ref{swave2},\ref{swave3},\ref{swave4},\ref{pwave1},
\ref{pwave2},\ref{pwave3},\ref{pwave4}).

Let us introduce a massless spinor $u_{-}(k_0)$ with a light-like
momentum $k_0$ and negative helicity first. Thus $u_{-}(k_0)$ is
satisfied with the following projection relation:
\begin{equation}
u_{-}(k_0)\bar{u}_{-}(k_0)=\omega_{-} \slashp{k_0} ,
\end{equation}
where $\omega_{-}=(1-\gamma_5)/2$. By introducing another spacelike
vector $k_1$ which satisfies the relations:
\begin{equation}
k_1\cdot k_1=-1 , k_0\cdot k_1=0 ,
\end{equation}
then the other, massless, independent and positive helicity spinor
$u_{+}(k_0)$ may be constructed:
\begin{equation}
u_{+}(k_0)=\slashp{k_1} u_{-}(k_0) .
\end{equation}
It is easy to check that $u_{+}(k_0)$ is satisfied with the
projection
\begin{equation}
u_{+}(k_0)\bar{u}_{+}(k_0)=\omega_{+} \slashp{k_0} ,
\end{equation}
where $\omega_{+}=(1+\gamma_5)/2$. Using these massless spinors, one
can construct the massive spinors for the fermion and antifermion as
follows:
\begin{eqnarray}
u_s(q)&=&(\slashp{q}+m)u_{-}(k_0)/\sqrt{2k_0\cdot q} , \\
u_{-s}(q)&=&(\slashp{q}+m)u_{+}(k_0)/\sqrt{2k_0\cdot q} ,
\end{eqnarray}
with the spin vector $s_\mu$:
\begin{displaymath}
s_\mu=\frac{q_\mu}{m}-\frac{m}{q\cdot k_0}k_{0\mu} .
\end{displaymath}
Using the above identities, we can write down the amplitude $M_{\pm
s\pm s'}$ with four possible spin projections in the trace form for
the $\gamma$-matrices:
\begin{eqnarray}
M_{ss'} &=& N Tr[(\slashp{p_0}+m_t)\cdot
\omega_{-}\slashp{k_0}\cdot(\slashp{p_2}+m_c)\cdot A] \\
M_{-s-s'} &=& N Tr[(\slashp{p_0}+m_t)\cdot
\omega_{+}\slashp{k_0}\cdot(\slashp{p_2}+m_c)\cdot A] \\
M_{-ss'} &=& N Tr[(\slashp{p_0}+m_t)\cdot
\omega_{-}\slashp{k_0}\slashp{k_1}\cdot(\slashp{p_2}+m_c)\cdot A]\\
M_{s-s'} &=& N Tr[(\slashp{p_0}+m_t)\cdot\omega_{+}\slashp{k_1}
\slashp{k_0}\cdot(\slashp{p_2}+m_c)\cdot A]
\end{eqnarray}
with the normalization constant $N=1/\sqrt{4(k_0\cdot p_0) (k_0\cdot
p_2)}$. Thus the squared unpolarized matrix elements can be written
as
\begin{displaymath}
|M|^2=|M_{ss'}|^2+ |M_{-s-s'}|^2+ |M_{-ss'}|^2+ |M_{s-s'}|^2 .
\end{displaymath}

Next, we are to simplify the calculation. For such purpose, we
recombine these $M_{\pm s \pm s'}$ into $M_n$ ($n=1,\cdots,4$) as
follows:
\begin{eqnarray}
M_1 &=& \frac{1}{\sqrt{2}}(M_{ss'}+M_{-s-s'})=\frac{N}{\sqrt{2}}
Tr[(\slashp{p_0}+m_t)\cdot \slashp{k_0}\cdot(\slashp{p_2}+m_c)\cdot A] \\
M_2 &=& \frac{1}{\sqrt{2}}(M_{ss'}-M_{-s-s'})=\frac{N}{\sqrt{2}}
Tr[(\slashp{p_0}+m_t)\cdot \slashp{k_0}\gamma_5\cdot(\slashp{p_2}+m_c)\cdot A] \\
M_3 &=& \frac{1}{\sqrt{2}}(M_{s-s'}-M_{-ss'})=\frac{N}{\sqrt{2}}
Tr[(\slashp{p_0}+m_t)\cdot \slashp{k_1}\slashp{k_0}\cdot(\slashp{p_2}+m_c)\cdot A] \\
M_4 &=& \frac{1}{\sqrt{2}}(M_{s-s'}+M_{-ss'})=\frac{N}{\sqrt{2}}
Tr[(\slashp{p_0}+m_t)\cdot \gamma_5
\slashp{k_1}\slashp{k_0}\cdot(\slashp{p_2}+m_c)\cdot A]
\end{eqnarray}
and then we have $|M|^2=|M_1|^2+ |M_2|^2+ |M_3|^2+ |M_4|^2$. In
order to write down $A$ as explicitly and simply as possible, we set
the vector $k_0$:
\begin{equation}
k_0=p_2-\alpha p_0 ,
\end{equation}
where the coefficient $\alpha$ is determined by the requirement that
$k_0$ be a light-like vector:
\begin{displaymath}
\alpha=\frac{p_0\cdot p_2}{m_t^2}\pm\frac{\Delta}{m_t^2}
\end{displaymath}
with $\Delta=\sqrt{(p_0\cdot p_2)^2-m_t^2 m_c^2}$ . Furthermore, if
choosing $k_1$: $k_1\cdot p_0=0$ and $k_1\cdot p_2=0$. And then the
resultant $M_n$ can be simplified as
\begin{eqnarray}\label{am1}
M_1 &=& L_1 Tr[(\slashp{p_0}+m_t)\cdot(\slashp{p_2}+m_c)\cdot A] \\
\label{am2}
M_2 &=& -L_2 Tr[(\slashp{p_0}+m_t)\cdot\gamma_5\cdot(\slashp{p_2}+m_c)\cdot A] \\
\label{am3}
M_3 &=& L_2 Tr[(\slashp{p_0}+m_t)\cdot\slashp{k_1}\cdot(\slashp{p_2}+m_c)\cdot A] \\
M_4 &=& L_1 Tr[(\slashp{p_0}+m_t)\cdot\gamma_5
\cdot\slashp{k_1}\cdot(\slashp{p_2}+m_c)\cdot A] \label{am4}
\end{eqnarray}
with
\begin{displaymath}
2L_{1}=\frac{1}{\sqrt{p_0\cdot p_2+m_c m_t}}\;\;{\rm and}\;\;
2L_{2}=\frac{1}{\sqrt{p_0\cdot p_2-m_c m_t}} .
\end{displaymath}
The value of $k_1$ is arbitrary, and we take its explicit form as
\begin{equation}
k^\mu_1=i \kappa \epsilon^{\mu\nu\rho\sigma}
p_{0\nu}p_{1\rho}p_{2\sigma} ,
\end{equation}
where $\kappa$ is a suitable normalization constant and
$\slashp{k_1}$ can be expressed as
\begin{displaymath}
\slashp{k_1}=\kappa \gamma_5 \left[(p_0\cdot
p_1)\slashp{p_2}+(p_2\cdot p_1)\slashp{p_0}-(p_2\cdot
p_0)\slashp{p_1}-\slashp{p_2}\cdot
\slashp{p_1}\cdot\slashp{p_0}\right] .
\end{displaymath}
Substituting $k_1$ into Eqs.(\ref{am3},\ref{am4}), we obtain
\begin{eqnarray}
M_3 &=& M'_3+\kappa [(p_0\cdot p_1) m_c -(p_1\cdot p_2)m_t]M_2 \\
M_4 &=& M'_4-\kappa [(p_0\cdot p_1) m_c +(p_1\cdot p_2)m_t]M_1
\end{eqnarray}
where
\begin{eqnarray}
M'_3 &=& \frac{\kappa}{4L_2}Tr[(\slashp{p_0}+m_t)
\cdot\gamma_5\cdot\slashp{p_1}\cdot(\slashp{p_2}+m_c)\cdot A]\\
M'_4&=& \frac{\kappa}{4L_1}Tr[(\slashp{p_0}+m_t)
\cdot\slashp{p_1}\cdot(\slashp{p_2}+m_c)\cdot A]
\end{eqnarray}

The amplitudes for the process $t(p_0)\to
(b\bar{c})(p_1)+c(p_2)+W^+(p_3)$ can be expanded over some basic
Lorentz structures:
\begin{equation}\label{amat}
M_i(n)=\sum^m_{j=1} A^i_j(n) B_j(n) (i=1-4)\;\;,\;\;
M'_i(n)=\sum^m_{j=1} A^{i'}_j(n) B_j(n) \;\; (i=3,4)
\end{equation}
where $m$ is the number of basic Lorentz structure $B_j(n)$, whose
value dependents on the $(b\bar{c})$-quarkonium state $n$: e.g.
$m=3$ for $n=(b\bar{c})[^1S_0]_1$, $m=11$ for
$n=(b\bar{c})[^3S_1]_1$ and $(b\bar{c})[^1P_1]_1$ and $m=30$ for
$n=(b\bar{c})[^3P_J]_1$. As for $A^3_j(n)$ and $A^4_j(n)$, they can
be expressed by
\begin{eqnarray}\label{relat1}
A^3_j(n) &=& A^{3'}_j(n)+\kappa [(p_0\cdot p_1) m_c -(p_1\cdot
p_2)m_t] A^2_j(n) \\
A^4_j(n) &=& A^{4'}_j(n)-\kappa [(p_0\cdot p_1) m_c +(p_1\cdot
p_2)m_t] A^1_j(n) \label{relat2}
\end{eqnarray}
The explicit expression for $A^{1,2}_j(n)$ and $A^{3',4'}_j(n)$ of
each state shall be listed in the following subsections.

To short the notation, we define some dimensionless parameters
\begin{displaymath}
r_1=\frac{m_b}{m_t},\;\; r_2=\frac{m_c}{m_t},\;\;
r_3=\frac{m_w}{m_t},\;\; r_4=\frac{M}{m_t}
\end{displaymath}
and
\begin{eqnarray}
&&u=p_1\cdot p_3/m_t^2=\frac{1}{2m_t^2}(s_3-m_{B_c}^2-m_w^2),
v=p_2\cdot p_3/m_t^2=\frac{1}{2m_t^2} (s_2-m_c^2-m_w^2), \nonumber\\
&&w=p_0\cdot p_3/m_t^2=\frac{1}{2m_t^2}(m_t^2+m_w^2-s_1), x=p_1\cdot
p_2/m_t^2 =\frac{1}{2m_t^2}(s_1-m_{B_c}^2-m_c^2), \nonumber\\
&&y=p_0\cdot p_1/m_t^2=\frac{1}{2m_t^2}(m_t^2+m_{B_c}^2-s_2),
z=p_0\cdot p_2/m_t^2= \frac{1}{2m_t^2}(m_t^2+m_c^2-s_3),\nonumber
\end{eqnarray}
where $s_1=(p_1+p_2)^2$, $s_2=(p_2+p_3)^2$ and $s_3=(p_1+p_3)^2$,
which satisfy the relation:
$s_1+s_2+s_3=m_t^2+m_c^2+m_w^2+m_{B_c}^2$. And the short notations
for the denominators are
\begin{eqnarray}
d_1 &=& \frac{1}{(p_2+p^0_{11})^2}\frac{1}{(p_1+p_2)^2-m_b^2} \;,\;
d_2 = \frac{1}{(p_2+p^0_{11})^2}\frac{1}{(p_3+p^0_{12})^2-m_t^2},\nonumber\\
d_3 &=& \frac{m_t^2}{(p_2+p^0_{11})^4}\frac{1}{(p_1+p_2)^2-m_b^2}
\;,\;
d_4 = \frac{m_t^2}{(p_2+p^0_{11})^4}\frac{1}{(p_3+p^0_{12})^2-m_t^2},\nonumber\\
d_5 &=& \frac{1}{(p_2+p^0_{11})^2}\frac{m_t^2}
{((p_3+p^0_{12})^2-m_t^2)^2}. \nonumber
\end{eqnarray}
Furthermore, the following relations are useful to short the
expressions:
\begin{displaymath}
y+z+w=1 ,\;\; x+u+r^2_4=y, \;\; x+v+r_2^2=z ,\;\; u+v+r_3^2=w .
\end{displaymath}

\subsection{Coefficients for spin-singlet S-wave state: $(b\bar{c})[^1S_0]_1$}

There are three basic Lorentz structures $B_j$ for the case of
$(b\bar{c})[^1S_0]_1$, which are
\begin{equation}
B_1=\frac{p_1\cdot\epsilon(p_3)}{m_t}\;,\; B_2=
\frac{p_2\cdot\epsilon(p_3)}{m_t} \;,\;
B_3=\frac{i}{m_t^3}\varepsilon(p_1,p_2,p_3,\epsilon(p_3)),
\end{equation}
where the short notation
$\varepsilon(p_1,p_2,p_3,\epsilon(p_3))=\varepsilon^{\mu\nu\rho\sigma}
p_{1\mu}p_{2\nu} p_{3\rho} \epsilon_\sigma(p_3)$. The values of the
coefficients $A^{1}_j$ and $A^{3'}_j$ are
\begin{eqnarray}
A^1_1 &=& \frac{2{L_1}{{m_t}}^{\frac{7}{2}}
}{{{r_4}}^{\frac{3}{2}}}\Big( {d_2}( {{r_1}}^3{r_2} + {{r_1}}^2( 2(
v + x )+ {r_2} + 4{{r_2}}^2 )-{r_2}( 3 z-2 v + r2 (u - w + 2) - x)\nonumber\\
&&+ {r_1}( -v + {r_2}( -2 + 2u + 3v + 4x + {r_2}( -2 + 3{r_2} )+
{{r_3}}^2 ))) + {d_1}{r_4}( 2x( {r_1} + 2{r_2} ) +\nonumber\\
&& ( v +{r_2} - {r_1}{r_2} + 4{{r_2}}^2 ) {r_4} ) \Big) \\
A^1_2 &=& \frac{-2{L_1}{{m_t}}^{\frac{7}{2}}}{{\sqrt{{r_4}}}}\Big(
-2x{d_1}( 1 + {r_2} )+{d_1}( u + {r_1} + {{r_1}}^2 + 3{r_1}{r_2} -
{r_2}( 3 + 2{r_2} )) {r_4}+ \nonumber\\
&&{d_2}( 1 + {r_2} ) ( u + ( 1 + {r_1} ) {r_4} )\Big)\\
A^1_3 &=& \frac{-2{L_1}{{m_t}}^{\frac{7}{2}}}{{\sqrt{{r_4}}}}\Big(
{d_2}( 1 + {r_2} )+ {d_1}{r_4}\Big) \\
A^{3'}_1 &=& \frac{\kappa {{m_t}}^{\frac{9}{2}}}{2{L_2}
{{r_4}}^{\frac{3}{2}}}\Big( {d_1}{r_4}( -2x ( v + {r_1} + {r_2}( 2 +
{r_2} ))+ ( v( {r_1} - 3{r_2} ) + {r_2}(
{r_1} + {{r_1}}^2 + 3{r_1}{r_2} -\nonumber\\
&& {r_2}( 3 + 2{r_2} ))) {r_4} ) + {d_2}( 4x^2{r_1} + 2x( {{r_1}}^3
+ {r_2}( -1 + v + {r_2} )+ {{r_1}}^2( -1 + 3{r_2} )+\nonumber\\
&& {r_1}( -1 + 2u + v + 2{{r_2}}^2 ))+ {r_4}( ( 1 + {r_1} ) {r_2} (
2u + 2{{r_1}}^2 + ( -1 + {r_2} ) {r_2} + {r_1}( -1 + 3{r_2}
))+\nonumber\\
&& v( 2u + ( -1 + {r_2} ) {r_4} ))) \Big)\\
A^{3'}_2 &=& \frac{-\kappa {{m_t}}^{\frac{9}{2}}}{2
{L_2}{\sqrt{{r_4}}}}\Big( ( {d_2}( 2u( u + x )+ {{r_1}}^4 - ( u + 2x
) {r_2} - {{r_2}}^3 + {{r_1}}^3( -1 + 3{r_2} )+{{r_2}}^2( u -
{{r_3}}^2 )+ \nonumber\\
&& {{r_1}}^2( 2( u + x )+ 3( -1 + {r_2}) {r_2} - {{r_3}}^2 )+
{r_1}( -u - 2x + {r_2}( 3u + 2x +( -3 + {r_2} ) {r_2} -\nonumber\\
&&  2{{r_3}}^2 ) ))+ {d_1}( -4x^2 + u( {r_1} - 3{r_2} ) {r_4} + ( -1
+{r_1} - 4{r_2} ) {{r_4}}^3 - 2x( u + 4{r_2}{r_4} ))) \Big)\\
A^{3'}_3 &=& \frac{-\kappa
{{m_t}}^{\frac{9}{2}}}{2{L_2}{\sqrt{{r_4}}}} \Big( ( {d_2}( 2( u + x
)+ 2{{r_1}}^2 + 3{r_1}{r_2} + {{r_2}}^2 + {r_4} )+{d_1}( -2x + (
{r_1} - 3{r_2} ) {r_4} ))\Big)
\end{eqnarray}
And the values of the coefficients $A^2_j$ and $A^{4'}_j$ are
\begin{eqnarray}
A^2_1&=&-\frac{A^1_1}{L_1}L_2
-\frac{4L_2{{m_t}}^{\frac{7}{2}}}{{{r_4}}^{\frac{3}{2}}} \Big(
2x{d_2}{r_2}-{d_2}( -v + ( {r_1} - 3{r_2} ) {r_2} ) {r_4} -
{d_1}{r_2}{{r_4}}^2 \Big)\\
A^2_2 &=& -\frac{A^1_2}{L_1}L_2-\frac{4L_2{{m_t}}^{\frac{7}{2}}}
{{\sqrt{{r_4}}}}\Big( {d_1} ( -2x + {r_1}{r_4} -3{r_2}{r_4} )+{d_2}(
u + {{r_4}}^2 )\Big) \\
A^2_3 &=&-\frac{A^1_3}{L_1}L_2- \frac{4L_2 {d_2}
{{m_t}}^{\frac{7}{2}}}{{\sqrt{{r_4}}}} \\
A^{4'}_1 &=& \frac{A^{3'}_1}{L_1}L_2-\frac{\kappa
{{m_t}}^{\frac{9}{2}}}{L_1\sqrt{{r_4}}}\Big({d_2}( {{r_1}}^2{r_2} +
{r_2}( 2 u - 2 v + x + z)+{r_1}( z-2 v - 3 x ))
+\nonumber\\
&& {d_1}( {r_1} - 3{r_2} ) {r_2}{r_4} - 2x{d_1}( {r_2} +
{r_4})\Big)\\
A^{4'}_2 &=& \frac{A^{3'}_2}{L_1}L_2-\frac{\kappa
{{m_t}}^{\frac{9}{2}}}{L_1{\sqrt{{r_4}}}}\Big (
{d_1}{{r_4}}^3 + {d_2}{r_4}( x+y )  \Big)\\
A^{4'}_3 &=& \frac{A^{3'}_3}{L_1}L_2+\frac{\kappa
{{m_t}}^{\frac{9}{2}}{d_2}{\sqrt{{r_4}}}}{L_1}
\end{eqnarray}
Here for convenience an overall factor ${\cal C}_s$ has been
contracted out from these coefficients. Then the square of the
amplitude $|M_i|^2$ can be conveniently obtained with the help of
Eqs.(\ref{amat},\ref{relat1},\ref{relat2}).

\subsection{Amplitude for spin-triplet S-wave state: $[^3S_1]_1$}

There are eleven basic Lorentz structures $B_j$ for the case of
$(b\bar{c})[^3S_1]$, which are
\begin{eqnarray}
B_1 &=& \epsilon(s_z)\cdot\epsilon(p_3),\; B_2=
\frac{i}{m_t^2}\varepsilon(p_1,p_2,\epsilon(s_z),\epsilon(p_3)),\;
B_3=\frac{i}{m_t^2}
\varepsilon(p_1,p_3,\epsilon(s_z),\epsilon(p_3)),\nonumber\\
B_4 &=& \frac{i}{m_t^2}
\varepsilon(p_2,p_3,\epsilon(s_z),\epsilon(p_3)),\;
B_5=\frac{p_1\cdot\epsilon(p_3) p_2\cdot\epsilon(s_z)}{m_t^2},\;
B_6=\frac{p_1\cdot\epsilon(p_3) p_3\cdot\epsilon(s_z)}{m_t^2},\nonumber\\
B_7 &=&\frac{p_2\cdot\epsilon(p_3) p_2\cdot\epsilon(s_z)}{m_t^2},
B_8 =\frac{p_2\cdot\epsilon(p_3) p_3\cdot\epsilon(s_z)}{m_t^2},
B_9=\frac{ip_1\cdot
\epsilon(p_3)}{m_t^4} \varepsilon(p_1,p_2,p_3,\epsilon(s_z)), \nonumber\\
B_{10}&=&\frac{i}{m_t^4} \varepsilon(p_1,p_2,p_3,\epsilon(p_3))p_3
\cdot \epsilon(s_z),\; B_{11}=\frac{i}{m_t^4}
\varepsilon(p_1,p_2,p_3,\epsilon(p_3))p_2 \cdot \epsilon(s_z),
\end{eqnarray}
where the polarization vector $\epsilon(s_z)$ related to the spin
angular momentum of the spin-triplet state. The values of the
coefficients $A^{1}_j$ and $A^{3'}_j$ are
\begin{eqnarray}
A^1_1 &=& \frac{-2{L_1}{{m_t}}^{\frac{7}{2}}}{{\sqrt {{r_4}}}} \Big(
-( {d_1}{r_4}( x - {r_1}( z + {r_2})+ {r_2}( y+z+{r_2})) )+
{d_2}( 2u( v + x ) + x( {r_1}-1 ) \nonumber\\
&&+ 2x( {r_1}-1 ) {r_4} + v( 2{r_1}-1 ) {r_4} + 2{{r_2}}^2( u + (
{r_1}-1  ) {r_4})+ {r_2}( u - x + ( u + x ) {r_1} +\nonumber\\
&& ( 2 r_1 + w - 2) {r_4} +( -1 + {r_1} ) {{r_4}}^2 ) )\Big) \\
A^1_2 &=& \frac{-2{L_1}{{m_t}}^{\frac{7}{2}}}{{\sqrt{{r_4}}}} \Big (
{d_2}( 1 + 2u + {r_1} + {r_2} + {r_1}{r_2} + 2{{r_3}}^2 ) + {d_1}(
1 + {r_1} ) {r_4} \Big) \\
A^1_3 &=& \frac{2{L_1}{{m_t}}^{\frac{7}{2}}}{{\sqrt{{r_4}}}} \Big(
-{d_1}{r_2}{r_4}+{d_2}( 4( v + x )+ {r_2}( 1 -
{r_1} + 2{r_2} + {r_4} ))\Big)\\
A^1_4 &=& \frac{-2{L_1}{{m_t}}^{\frac{7}{2}}}{{\sqrt{{r_4}}}} \Big(
2u{d_2} + {d_1}( -{r_1} + {r_2} ) {r_4} +{d_2}{r_4}( -1 -
{r_2} + 2{r_4} )\Big)\\
A^1_5 &=& \frac{-2{L_1}{{m_t}}^{\frac{7}{2}}}{{\sqrt{{r_4}}}}\Big(
{d_2}( 1 + {r_1} ) ( 1 + {r_2} )+{d_1}( 1 + {r_1} + 2{r_2} )
{r_4} \Big) \\
A^1_6 &=& \frac{2{L_1}{{m_t}}^{\frac{7}{2}}}{{\sqrt{{r_4}}}} \Big(
-{d_1}{r_2}{r_4}+{d_2}( 2( v + x )+ {r_2}( 1 -
{r_1} + 2{r_2} + {r_4} ))\Big)\\
A^1_7 &=& -4{d_1}{L_1}{{m_t}}^{\frac{7}{2}} {\sqrt{{r_4}}}( 1 +
{r_2} )\\
A^1_8 &=& 2{L_1}{{m_t}}^{\frac{7}{2}}{\sqrt{{r_4}}}( {d_2} +
{d_1}{r_1} - {d_1}{r_2} +{d_2}{r_2} )\\
A^1_9 &=& \frac{-4{d_2}{L_1}{{m_t}}^{\frac{7}{2}}}{{\sqrt{{r_4}}}}\\
A^1_{10} &=&
\frac{4{d_2}{L_1}{{m_t}}^{\frac{7}{2}}}{{\sqrt{{r_4}}}}\\
A^{3'}_1 &=& \frac{-\kappa{{m_t}}^{\frac{9}{2}}}
{2{L_2}{\sqrt{{r_4}}}} \Big(( {d_1}{r_4}( -( x ( 2( u + x )+ {r_1}
))- ( u + x ) {{r_2}}^2 + ( v - x ) {{r_4}}^2 + {r_2}( x + ( u + x )
{r_1}+  \nonumber\\
&& ( 1 + {r_1} ) {{r_4}}^2))+ {d_2}( 2x( x + ( u + x ) {r_1} )+ x(
2(v-1) - 2{r_1} + {{r_3}}^2 ) {r_4} - ( v - x ) ( 1 + {r_1} )
{{r_4}}^2 \nonumber\\
&& + v{{r_4}}^3 - {{r_2}}^2{r_4}( u + {r_4} + {r_1}{r_4} )+
{r_2}( u + {r_4} + {r_1}{r_4} ) ( 2( u + x )+ {r_4}( -1 + 2{r_4} ))))\Big)\\
A^{3'}_2 &=& \frac{\kappa {{m_t}}^{\frac{9}{2}}}{2
{L_2}{\sqrt{{r_4}}}} \Big( -( {d_1}{r_4}( 2x+ {r_2} + {{r_2}}^2 -
{r_1}( 1 + {r_2} )+ {{r_4}}^2 ))+ {d_2}(-2x{r_1}
+ 2( u + x + u{r_2} )+ \nonumber\\
&& {r_4}( {{r_3}}^2 + {r_4} - {r_1}{r_4} ))\Big)  \\
A^{3'}_3 &=& \frac{-\kappa
{{m_t}}^{\frac{9}{2}}}{2{L_2}{\sqrt{{r_4}}}}\Big(  ( {d_1}( 2x
-{r_1}{r_2} + {{r_2}}^2 ) {r_4} + {d_2}( 2x( 1 + {r_1} )- 2u{r_2}
+2v{r_4} + \nonumber\\
&& {r_2}( 1 + {r_2} - 2{r_4} ) {r_4} ))\Big) \\
A^{3'}_4 &=& \frac{\kappa {{m_t}}^{\frac{9}{2}}
{\sqrt{{r_4}}}}{2{L_2}}\Big({d_1}{{r_4}}^2 + {d_2}
( 2( u + x )+ {r_4}( 1 + {r_1} + 2{r_2} + {r_4} ))\Big) \\
A^{3'}_5 &=& \frac{\kappa {{m_t}}^{\frac{9}{2}}}{2
{L_2}{\sqrt{{r_4}}}} \Big( {d_1}{r_4}( 2v -2x + {r_1} + {r_2} +
{r_1}{r_2} + {{r_2}}^2 - {{r_4}}^2 )+ {d_2} ( 2( u
+ x )- 2x{r_1} + \nonumber\\
&& {r_4}( {{r_3}}^2 + {r_4} - {r_1}{r_4} ))\Big)  \\
A^{3'}_6 &=& \frac{-\kappa
{{m_t}}^{\frac{9}{2}}}{2{L_2}{\sqrt{{r_4}}}} \Big((  {d_1}( 2x
-{r_1}{r_2} + {{r_2}}^2 ) {r_4} + {d_2}( {{r_2}}^2{r_4} + 2( x
+x{r_1} + v{r_4} )+ \nonumber\\
&&{r_2}( -2( u + x )+ {r_4} - 2{{r_4}}^2 ) ))\Big) \\
A^{3'}_7 &=& - \frac{\kappa{d_1}{{m_t}}^{\frac{9}{2}}
{\sqrt{{r_4}}}}{{L_2}}( u + 2x + {{r_4}}^2 )\\
A^{3'}_8 &=& \frac{\kappa {{m_t}}^{\frac{9}{2}}{\sqrt{{r_4}}}
}{2{L_2}} \Big({d_1}{{r_4}}^2 + {d_2} ( 2( u + x )+ {r_4}( 1 + {r_1}
+{r_4} ))\Big)\\
A^{3'}_9 &=&\frac{\kappa {d_2}{{m_t}}^{\frac{9}{2}}{r_2}}{{L_2}{\sqrt{{r_4}}}} \\
A^{3'}_{11} &=&- \frac{\kappa{d_1}{{m_t}}^{\frac{9}{2}}
{\sqrt{{r_4}}}}{{L_2}}
\end{eqnarray}
and $A^{1}_{11}=A^{3'}_{10}=0$. And the values of the coefficients
$A^2_j$ and $A^{4'}_j$ are
\begin{eqnarray}
A^2_1&=&-\frac{A^1_1}{L_1}L_2+\frac{4{L_2}
{{m_t}}^{\frac{7}{2}}}{{\sqrt{{r_4}}}} \Big( {d_1}( x - {r_1}{r_2}
+{{r_2}}^2 ){r_4} + {d_2}( ( v + 2x ) {r_4} + 2{{r_2}}^2{r_4}
-\nonumber\\
&& {r_1}( x + 2{r_2}{r_4} )+{r_2}(y-2u)  ) \Big)\\
A^2_2 &=& -\frac{A^1_2}{L_1}L_2-\frac{4{L_2}
{{m_t}}^{\frac{7}{2}}}{{\sqrt{{r_4}}}}(
{d_1}{r_4} +{d_2}{r_4} ) \\
A^2_3 &=&-\frac{A^1_3}{L_1}L_2 +\frac{4{d_2}{L_2}
{{m_t}}^{\frac{7}{2}}{r_2}}{{\sqrt{{r_4}}}}\\
A^2_4 &=&-\frac{A^1_4}{L_1}L_2 +4{d_2}{L_2}{{m_t}}^{\frac{7}{2}}{\sqrt{{r_4}}}\\
A^2_5 &=&-\frac{A^1_5}{L_1}L_2 -\frac{4{L_2}{{m_t}}^{\frac{7}{2}}
}{{\sqrt{{r_4}}}}( {d_1}{r_4} +{d_2}{r_4} )\\
A^2_6 &=&-\frac{A^1_6}{L_1}L_2 +\frac{4{d_2}{L_2}{{m_t}}^{\frac{7}{2}}{r_2}}{{\sqrt{{r_4}}}}\\
A^2_7 &=&-\frac{A^1_7}{L_1}L_2 -8{d_1}{L_2}{{m_t}}^{\frac{7}{2}}{\sqrt{{r_4}}}\\
A^2_8 &=&-\frac{A^1_8}{L_1}L_2 +4{d_2}{L_2}{{m_t}}^{\frac{7}{2}}{\sqrt{{r_4}}}\\
A^{4'}_1 &=& \frac{A^{3'}_1}{L_1}L_2-\frac{\kappa
{{m_t}}^{\frac{9}{2}}}{{L_1}{\sqrt{{r_4}}}}\Big( -2x^2{d_2} - x{r_4}
( -( {d_1} + 2{d_2} )( {r_1} - {r_2})+ {d_2}{r_4} ) +\nonumber\\
&& {r_4}( -u{d_2}{r_2} + v{d_2}{r_4} + {r_2}{r_4}(
-{d_1}{r_4} - {d_2}{r_4} ) )\Big)\\
A^{4'}_2 &=& \frac{A^{3'}_2}{L_1}L_2-\frac{\kappa
{{m_t}}^{\frac{9}{2}}}{{L_1} {\sqrt{{r_4}}}}( {d_1}(
{r_1} - {r_2} ) {r_4} + {d_2}( 2( u + x )+ {{r_4}}^2 ))\\
A^{4'}_3 &=& \frac{A^{3'}_3}{L_1}L_2+\frac{\kappa
{d_2}{{m_t}}^{\frac{9}{2}}}{{L_1}{\sqrt{{r_4}}}}( 2x +
{r_2}{r_4} )\\
A^{4'}_4 &=& \frac{A^{3'}_4}{L_1}L_2-\frac{\kappa {d_2}{{m_t}}^{\frac{9}{2}}
{{r_4}}^{\frac{3}{2}}}{{L_1}}\\
A^{4'}_5 &=& \frac{A^{3'}_5}{L_1}L_2-\frac{\kappa
{{m_t}}^{\frac{9}{2}}}{{L_1}{\sqrt{{r_4}}}}(
{d_1}{{r_4}}^2 + {d_2} ( u + x +y))\\
A^{4'}_6 &=& \frac{A^{3'}_6}{L_1}L_2+\frac{\kappa
{d_2}{{m_t}}^{\frac{9}{2}} }{{L_1}{\sqrt{{r_4}}}}( 2x
+ {r_2}{r_4} )\\
A^{4'}_8 &=& \frac{A^{3'}_8}{L_1}L_2-\frac{\kappa
{d_2}{{m_t}}^{\frac{9}{2}}{{r_4}}^{\frac{3}{2}}}{{L_1}}
\end{eqnarray}
and for $j=9,10,11$ and $m=7,9,10,11$, we have
\begin{displaymath}
A^2_j =-\frac{A^1_j}{L_1}L_2 \;\;{\rm and}\;\; A^{4'}_m =
\frac{A^{3'}_m}{L_1}L_2 .
\end{displaymath}
Here for convenience an overall factor ${\cal C}_s$ has been
contracted out from these coefficients. Then the square of the
amplitude $|M_i|^2$ can be conveniently obtained with the help of
Eqs.(\ref{amat},\ref{relat1},\ref{relat2}) and Eq.(\ref{3s1}).

\subsection{Amplitude for spin-singlet P-wave state: $[^1P_1]_1$}

The basic structures are similar to the case of $^3S_1$, only one
needs to replace the polarization vector $\epsilon(s_z)$ related to
the spin angular momentum of the spin-triplet S-state to the present
$\epsilon(l_z)$ that is related to the radial angular momentum of
the spin-singlet P-wave state. Secondly, we need to change
coefficients there to the present case. The values of the
coefficients $A^{1}_j$ and $A^{3'}_j$ are
\begin{eqnarray}
A^1_1 &=& \frac{{L_1}{{m_t}}^{\frac{5}{2}}}
{{r_1}{r_2}{{r_4}}^{\frac{3}{2}}} \Big( -( {d_1}( {r_1} - {r_2} )
{{r_4}}^2 ( x + {r_1}( {r_2}+z ) + {r_2}( r_2 + y + z)))
- 2{d_5}{r_1}{r_2} ( 2u{r_1}+ \nonumber\\
&& (r_1^2 + r_3^2-1 ) {r_4} ) ( x + 2( v + x ) {r_4} +
2{{r_2}}^2{r_4} + {r_2}( 2r_4+y) )+ {d_2}{r_4}( {{r_1}}^2 ( x + y {r_2})\nonumber\\
&& + {r_4}( x - 2u( v + x ) + 2{{r_2}}^3 - {{r_2}}^2( -2 + 3u + v +
{{r_3}}^2 )+{r_2}( v-u + 3x + {{r_4}}^2 )) \nonumber\\
&&- {r_1}( {r_2}( x - (w - 2) {r_4} + 2{{r_4}}^2 )+  {r_4}( v+
2v{r_4} + 2x( 1 + {r_4} )) \nonumber\\
&& +{{r_2}}^2( u +x + {r_4}( 2 + 3{r_4} ))))\Big) \\
A^1_2 &=& \frac{{L_1}{{m_t}}^{\frac{5}{2}}}{{r_1}
{r_2}{{r_4}}^{\frac{3}{2}}}( -{d_1}( {r_1}-1) ( {r_1} - {r_2} )
{{r_4}}^2 + 2{d_5}{r_1}{r_2}( 1 + {r_2} ) ( 2u{r_1} + (
{{r_1}}^2 + {{r_3}}^2 -1 ) {r_4} )-\nonumber\\
&& {d_2}{r_4}( -{{r_1}}^2( 1 + {r_2})+ {r_1}{r_2}( 1 + {r_2} )+
{r_4} + ( {r_2} + 2( u + {{r_3}}^2 )) {r_4} ))\\
A^1_3 &=& \frac{{L_1}{{m_t}}^{\frac{5}{2}}}
{{r_1}{r_2}{{r_4}}^{\frac{3}{2}}}( {d_2}{r_4}( {{r_1}}^2{r_2} + (
r_2 + v + x + 3 z ) {r_4} - {r_1}{r_2}( {r_2} + {r_4} ) )
+{r_2}( {d_1}( {r_1} - {r_2} ) {{r_4}}^2 +\nonumber\\
&& 2{d_5}{r_1}{r_2}( 2u{r_1} +(r_1^2 + r_3^2-1 ) {r_4} ))) \\
A^1_4 &=& -\frac{{L_1}{{m_t}}^{\frac{5}{2}}
{\sqrt{{r_4}}}}{{r_1}{r_2}} ( {d_1}( r_2^2-r_1^2)+{d_2}( {r_1}( 1 +
{r_2} )- {r_2}( 1 + {r_2} )+ 2( u + {{r_4}}^2 ))) \\
A^1_5 &=& -\frac{{L_1}{{m_t}}^{\frac{5}{2}}}
{{r_1}{r_2}{{r_4}}^{\frac{3}{2}}} ( {d_2}( 1 + {r_2} ) ( {r_2} +
{r_1}( 1 - {r_1} + {r_2} )) {r_4} + {d_1}( -1 + {r_1} - 2{r_2} ) (
{r_1} - {r_2} ) {{r_4}}^2 + \nonumber\\
&& 2{r_1}{r_2} ( {d_5}( 1 + {r_2} ) ( 2u{r_1} + (r_1^2 + r_3^2-1 )
{r_4} )+ 2( {d_4} ( {{r_1}}^3{r_2} + r_1^2(r_2 - 2 (v + x - 2 z))\nonumber\\
&& - {r_2} (r_2 (u - w + 2)-2 v - x + 3 z)+
{r_1}( -v + {r_2}( -2 + 2u + 3v + 4x + \nonumber\\
&& {r_2}( -2 + 3{r_2} )+ r_3^2 ))) + {d_3}{r_4} ( 2x( {r_1} + 2{r_2}
)+ ( v + {r_2} - {r_1}{r_2} +
4{{r_2}}^2 ) {r_4} )))) \\
A^1_6 &=& \frac{{L_1}{{m_t}}^{\frac{5}{2}}}
{{r_1}{r_2}{{r_4}}^{\frac{3}{2}}}( {d_2}{r_4}( {{r_1}}^2{r_2} + (
r_2 - v - x + 3 z) {r_4} - {r_1}{r_2}( {r_2} + {r_4} ))\nonumber\\
&& +{r_2}( {d_1}( {r_1} - {r_2} ) {{r_4}}^2 + 2{d_5}{r_1}( 2v(
2{r_1}-1) {r_4} +4(r_1-1) r_2^2 {r_4} + 4x( {r_1} + (r_1-1) r_4)\nonumber\\
&& + {r_2}( 2( u + 2x ) {r_1} + ( -3 + 2v - ( -4 + {r_1} ) {r_1} +
{{r_3}}^2 ){r_4} + 2( -1 + {r_1} ) {{r_4}}^2)) )) \\
A^1_7 &=& \frac{2{L_1}{{m_t}}^{\frac{5}{2}}}
{{r_1}{r_2}{\sqrt{{r_4}}}}( {d_1}( {r_1} - {r_2} ) ( 1 + {r_2} )
{r_4} +2{r_1}{r_2}( {d_4}( 1 + {r_2} )( u + {r_4} + {r_1}{r_4} ) -
{d_3}( 2x - 2u{r_4} +\nonumber\\
&& 2{{r_2}}^2{r_4} + {{r_4}}^2 - 2{{r_4}}^3 + {r_1}( u - 2( 1 +
{r_2} ) {r_4} + {{r_4}}^2 )+ {r_2}( 2 r_4 - u + 2 y)) )) \\
A^1_8 &=& -\frac{{L_1}{{m_t}}^{\frac{5}{2}}}
{{r_1}{r_2}{\sqrt{{r_4}}}}( {d_2}( {r_1} - {r_2} ) ( 1 + {r_2}-r_4 )
{r_4} +4{d_5}{r_1}{r_2}( 1 + {r_2} ) ( u + {r_4} + {r_1}{r_4} )) \\
A^1_9 &=& \frac{-2{d_2}{L_1}{{m_t}}^{\frac{5}{2}}{\sqrt{{r_4}}}}{{r_1}{r_2}}\\
A^1_{10} &=& \frac{2{L_1}{{m_t}}^{\frac{5}{2}}}
{{r_1}{r_2}{\sqrt{{r_4}}}}( -2{d_5}{r_1}{r_2}( 1 + {r_2} )+ {d_2}{r_4} )\\
A^1_{11} &=& \frac{4{L_1}{{m_t}}^{\frac{5}{2}}}{{\sqrt{{r_4}}}} (
{d_4}( 1 + {r_2} )+ {d_3}{r_4} )\\
A^{3'}_1 &=& \frac{-\kappa {{m_t}}^{\frac{7}{2}}}
{4{L_2}{r_1}{r_2}{{r_4}}^{\frac{3}{2}}} \Big( {d_1}( {r_1} - {r_2} )
{{r_4}}^2 ( 2x( u + x )- x{r_1} + ( u + x ) {{r_2}}^2 + ( x-v )
{{r_4}}^2 + {r_2}( x( r_1 -1)\nonumber\\
&& +u{r_1} + ({r_1}  -1) {{r_4}}^2 ))- 2{d_5}{r_1}{r_2}( 2u{r_1} + (
-1 + {{r_1}}^2 + {{r_3}}^2 ) {r_4} ) ( -2x^2 + {r_4}( -2u{r_2} +
v{r_4} \nonumber\\
&&+ {r_2}( 1 + {r_2} - 2{r_4} ) {r_4} )- x( 2u + {r_4}( -2 + 2{r_2}
+ {r_4} ) ) ) + {d_2}{r_4}( 2x^2( {r_1}( -{r_1} + {r_2} ) + {r_4}
)+\nonumber\\
&& x( -{{r_1}}^2( 2u + {{r_4}}^2 ) + {r_4}( {r_2}( 2( -1 + u + v ) +
2{r_2} + {{r_3}}^2 )+ {{r_4}}^2 ) + {r_1}( 2u{r_2} - ( -2 + 2v
+\nonumber\\
&& 2{r_2} + {{r_3}}^2 ) {r_4} + ( -2 + 3{r_2} ) {{r_4}}^2 ) ) +
{r_4}( -v{r_4} ( -( {r_1} - {r_2} ) ( {r_1} - {r_4} )+ {r_4} )+ \nonumber\\
&& {r_2}( u( 2u + {r_2} - {{r_2}}^2 )+ {{r_1}}^2( 1 + {r_2} ) {r_4}
+ ( -1 + 2u + {r_2} ) {{r_4}}^2 + u{r_1}( -1 + {r_2} + 2{r_4}
)-\nonumber\\
&& {r_1}{r_4}( {r_2} + {{r_2}}^2 - 2(-1 + {r_4} ) {r_4} ))) ) \Big) \\
A^{3'}_2 &=& \frac{\kappa {{m_t}}^{\frac{7}{2}}}
{4{L_2}{r_1}{r_2}{{r_4}}^{\frac{3}{2}}}\Big( 2{d_5}{r_1}{r_2}(
2u{r_1} + ( -1 + {{r_1}}^2 + {{r_3}}^2 ) {r_4} ) ( 2x + {{r_4}}^2 )
+{d_1}( {r_1} - {r_2} ) {{r_4}}^2 ( 2x + {r_1} + \nonumber\\
&& {r_2} + {r_1}{r_2} + {{r_2}}^2 + {{r_4}}^2 )+ {d_2}{r_4}( 2x(
{{r_1}}^2 - {r_1}{r_2} + {r_4} )+ {r_4}( 2u( 1 + {r_2} ) +
{r_2}{{r_3}}^2 + \nonumber\\
&& {{r_1}}^2{r_4} +{{r_4}}^2 - {r_1}({{r_3}}^2 + {r_2}{r_4} )))\Big)\\
A^{3'}_3 &=& \frac{\kappa {{m_t}}^{\frac{7}{2}}}{4{L_2}
{r_1}{r_2}{{r_4}}^{\frac{3}{2}}} \Big( {d_1}( {r_1} - {r_2} ) ( 2x +
{r_2}( {r_1} + {r_2} )) {{r_4}}^2 +4x{d_5}{r_1}{r_2}( 2u{r_1} + (
r_1^2 + r_3^2-1 ) {r_4} )\nonumber\\
&& +{d_2}{r_4}( 2v( {r_1} - {r_2} ) {r_4} - 2x( {r_1}( {r_2}-{r_1})+
{r_4} )+ {r_2}{r_4}( {r_1}( 1 +{r_2} )-{r_2}( 1 + {r_2} )+ \nonumber\\
&& 2( u + {{r_4}}^2 )) )\Big)\\
A^{3'}_4 &=& \frac{-\kappa
{{m_t}}^{\frac{7}{2}}{\sqrt{{r_4}}}}{4{L_2}{r_1}{r_2}}\Big( {d_1}(
{r_1} - {r_2} ) {{r_4}}^2 +2{d_5}{r_1}{r_2}( 2u{r_1} + ( -1 +
{{r_1}}^2 + {{r_3}}^2 ) {r_4} )+\nonumber\\
&& {d_2}( -2( u + x ) {r_2} +{{r_1}}^2{r_4} - ( 1 + 3{r_2} )
{{r_4}}^2 + {r_1}( 2( u + x ) - {r_2}{r_4} + {{r_4}}^2 ))\Big)\\
A^{3'}_5 &=& \frac{\kappa {{m_t}}^{\frac{7}{2}}}{4{L_2}{r_1}{r_2}
{{r_4}}^{\frac{3}{2}}}\Big( {d_1}( {r_1} - {r_2} ) ( 2x -2v + {r_1}
+ {{r_1}}^2 - {r_2} + 3{r_1}{r_2} ) {{r_4}}^2 + {d_2}{r_4}( (
2u + {{r_1}}^2 + {{r_1}}^3 + \nonumber\\
&& {r_2}( {r_2} + {{r_3}}^2 )- {r_1}( ( -2 + {r_2} ) {r_2} +
{{r_3}}^2 )) {r_4} + 2x( {{r_1}}^2 - {r_1}{r_2} + {r_4} ))-
2{r_1}{r_2}( {d_5} ( 2u{r_1} + \nonumber\\
&& ( -1 + {{r_1}}^2 + {{r_3}}^2 ) {r_4} ) ( u+x+y )+ 2( {d_3}{r_4}(
-2x ( v + {r_1} + {r_2}( 2 + {r_2} )) + ( v( {r_1} -
3{r_2} )+ \nonumber\\
&& {r_2}( {r_1} + {{r_1}}^2 + 3{r_1}{r_2} - {r_2}( 3 + 2{r_2} )))
{r_4} ) + {d_4}( 4x^2{r_1} + 2x( {{r_1}}^3 + {r_2}( -1 + v + {r_2} )
+ \nonumber\\
&& {{r_1}}^2( 3{r_2}-1)+ {r_1}(2u-1 + v + 2{{r_2}}^2 )) + {r_4}( ( 1
+ {r_1} ) {r_2} ( 2u + 2{{r_1}}^2 + ( {r_2}-1 ) {r_2}
+ \nonumber\\
&& {r_1}( -1 + 3{r_2} )) +v( 2u + ( -1 + {r_2} ) {r_4} )) )))\Big)\\
A^{3'}_6 &=& \frac{\kappa {{m_t}}^{\frac{7}{2}}}{4{L_2}
{r_1}{r_2}{{r_4}}^{\frac{3}{2}}}\Big( {d_1}( {r_1} - {r_2} ) ( 2x +
{r_2}( {r_1} + {r_2} )) {{r_4}}^2 +4{d_5}{r_1}{r_2}( 4x^2{r_1} - ( 1
+ {r_1} ) {r_2}{r_4}( -2u +\nonumber\\
&& ( 1 + {r_2} - 2{r_4} ) {r_4} ) + v{r_4}( 2u - ( 1 + {r_1} ) {r_4}
+ {{r_4}}^2 )+ 2x( 2u{r_1} + ( v + ( 1 + {r_1} ) ( {r_2} -1))
{r_4} \nonumber\\
&& + {r_1}{{r_4}}^2 ))+ {d_2}{r_4}( 2v( {r_1} - {r_2} ) {r_4} + 2x(
{{r_1}}^2 - {r_1}{r_2} + ( -1 + {r_2} ) {r_4} )+ {r_2}{r_4} ( {r_1}(
1 + {r_2} )- \nonumber\\
&& {r_2}( 1 + {r_2} )+ 2( u +{{r_4}}^2 )))\Big)\\
A^{3'}_7 &=& \frac{\kappa {{m_t}}^{\frac{7}{2}}}{2{L_2}
{r_1}{r_2}{\sqrt{{r_4}}}}\Big( {d_1}( {r_1} - {r_2} ) {r_4} ( x+y
)+2{d_4}{r_1}{r_2}( 2u( u + x )+ ( u + 2x ) ({r_1} -1){r_4} + \nonumber\\
&& u{{r_4}}^2 - {{r_3}}^2{{r_4}}^2 + ({r_1} -1) {{r_4}}^3 )+
2{d_3}{r_1}{r_2}( -4x^2 + 2x( -u + 2{r_1}{r_4} - 2{r_4}( {r_2} +
{r_4} ))+ \nonumber\\
&&  {r_4}({r_4}( {r_2} + {{r_2}}^2 - 4{r_2}{r_4} - {r_4}( 2 + {r_4}
) + {r_1}( 1 + {r_2} +2{r_4} ))- u (2{r_2} + {r_4}-2{r_1} )  ))\Big)\\
A^{3'}_8 &=& \frac{-\kappa
{{m_t}}^{\frac{7}{2}}}{4{L_2}{r_1}{r_2}{\sqrt{{r_4}}}}\Big( {d_1}(
{r_1} - {r_2} ) {{r_4}}^3 +2{d_5}{r_1}{r_2}( 4u( u + x )+ ( 4x(
r_1-1)+u( 4{r_1}-2 )) {r_4}\nonumber\\
&&  + (2u -1 + {{r_1}}^2 - {{r_3}}^2 ) {{r_4}}^2 + 2( -1 + {r_1} )
{{r_4}}^3 )+ {d_2}{r_4}( -2( u + x ) {r_2} + {{r_1}}^2{r_4} -  \nonumber\\
&& ( 1 +{r_2} ) {{r_4}}^2 +{r_1}( 2( u + x )- {r_2}{r_4} + {{r_4}}^2 ) )\Big)\\
A^{3'}_9 &=& \frac{-\kappa {m_t}^{\frac{7}{2}}}
{2{L_2}{r_1}{{r_4}}^{\frac{3}{2}}} \Big ( -{d_2}{{r_4}}^2+
2{d_5}{r_1}( 2u{r_1} + ( -1 + {{r_1}}^2 + {{r_3}}^2 )
{r_4} )\Big)\\
A^{3'}_{10} &=& -\frac{\kappa {d_5}{{m_t}}^{\frac{7}{2}}}
{{L_2}{\sqrt{{r_4}}}}( 2( u + x )  + {r_4}( 1 + {r_1} + {r_4} )  )\\
A^{3'}_{11} &=& \frac{\kappa {{m_t}}^{\frac{7}{2}}}{2
{L_2}{r_1}{r_2}{\sqrt{{r_4}}}}\Big( {d_1}( {r_1} - {r_2} ) {r_4} +
2{r_1}{r_2}( -2x{d_3} + {d_3}( 2{r_1} - 2{r_2} - {r_4} ) {r_4} +
{d_4}( 2( u + x )\nonumber\\
&& +{r_4}( 1 + {r_1} + {r_4} ) )) \Big)
\end{eqnarray}
And the values of the coefficients $A^2_j$ and $A^{4'}_j$ are
\begin{eqnarray}
A^2_1&=&-\frac{A^1_1}{L_1}L_2+\frac{2{L_2}{{m_t}}^{\frac{5}{2}}}{{r_1}{r_2}{{r_4}}^{\frac{3}{2}}}
( -{d_1}( {r_1} - {r_2} ) {{r_4}}^2 ( x + {r_2}{r_4} )-
2{d_5}{r_1}{r_2}( x + 2{r_2}{r_4} ) ( 2u{r_1} + \nonumber\\
&&( -1 + {{r_1}}^2 + {{r_3}}^2 ) {r_4} )+{d_2}{r_4}( x{{r_1}}^2 +
{r_2}{r_4}(y-2 u - v + 2 z )-{r_1}( x( {r_2} + 2{r_4} )\nonumber\\
&&  + {r_4}( v + 2{r_2}( {r_2} + {r_4} ) )) ) )\\
A^2_2 &=& -\frac{A^1_2}{L_1}L_2+\frac{2{L_2}
{{m_t}}^{\frac{5}{2}}}{{r_1}{r_2}{{r_4}}^{\frac{3}{2}}} (
4u{d_5}{{r_1}}^2{r_2} + {r_4}( {r_1}( {d_2}( {r_1} - {r_2} )+
2{d_5}{r_2}( -1 + {{r_1}}^2 + {{r_3}}^2 ))+\nonumber\\
&& ( {d_1}{r_1} - ( {d_1} +{d_2} ) {r_2} ) {r_4} )) \\
A^2_3 &=&-\frac{A^1_3}{L_1}L_2 +\frac{2{d_2}
{L_2}{{m_t}}^{\frac{5}{2}}{\sqrt{{r_4}}}}{{r_1}}\\
A^2_4 &=&-\frac{A^1_4}{L_1}L_2
+\frac{2{d_2}{L_2}{{m_t}}^{\frac{5}{2}}{\sqrt{{r_4}}}}{{r_1}{r_2}}
({r_2} -{r_1})\\
A^2_5 &=&-\frac{A^1_5}{L_1}L_2
-\frac{{L_2}{{m_t}}^{\frac{5}{2}}}{{r_1} {r_2}{{r_4}}^{\frac{3}{2}}}
( 8u{d_5}{{r_1}}^2{r_2} +8{d_4}{r_1}{r_2}( -2x{r_2} + ( -v + ( {r_1}
- 3{r_2} ) {r_2} ) {r_4} )+\nonumber\\
&& {r_4}( 4{d_5}{r_1}{r_2} ( -1 + {{r_1}}^2 + {{r_3}}^2 )+ 2( -(
{d_1}{r_1} )+ {d_1}{r_2} + 4{d_3}{r_1}{{r_2}}^2 ) {r_4} + {d_2}(
-2{{r_1}}^2 +  \nonumber\\
&& 3{r_1}{r_2}+ {{r_2}}^2 + {r_4} + (-1 + {r_2} ) {r_4} )))\\
A^2_6 &=&-\frac{A^1_6}{L_1}L_2
+\frac{2{L_2}{{m_t}}^{\frac{5}{2}}}{{r_1}{{r_4}}^{\frac{3}{2}}}(
{d_2}{{r_4}}^2 + 4{d_5}{r_1}(2r_1 x + r_2 r_4)-r_4(r_2 r_4-v+2z)))\\
A^2_7 &=&-\frac{A^1_7}{L_1}L_2
+\frac{4{L_2}{{m_t}}^{\frac{5}{2}}}{{r_1}{r_2}{\sqrt{{r_4}}}} (
{d_1}( {r_1} - {r_2} ) {r_4} +2{r_1}{r_2}( -2x{d_3} + {d_3}( 2{r_1}
- 2{r_2} - {r_4} ) {r_4} +\nonumber\\
&& {d_4}( u + {{r_4}}^2 )) )\\
A^2_8 &=&-\frac{A^1_8}{L_1}L_2
-\frac{2{L_2}{{m_t}}^{\frac{5}{2}}}{{r_1}{r_2}{\sqrt{{r_4}}}} (
{d_2}( {r_1} - {r_2} ) {r_4} +4{d_5}{r_1}{r_2}( u + {{r_4}}^2 ) )\\
A^2_{10} &=&-\frac{A^1_8}{L_1}L_2 -\frac{8{d_5}{L_2}{{m_t}}^{\frac{5}{2}}}{{\sqrt{{r_4}}}}\\
A^2_{11} &=&-\frac{A^1_8}{L_1}L_2 +\frac{8{d_4}{L_2}{{m_t}}^{\frac{5}{2}}}{{\sqrt{{r_4}}}}\\
A^{4'}_1 &=& \frac{A^{3'}_1}{L_1}L_2-\frac{\kappa
{{m_t}}^{\frac{7}{2}}}{2{L_1}{r_1}{r_2}{\sqrt{{r_4}}}} (
2{d_5}{r_1}{r_2}( 2x + {r_2}{r_4} ) ( 2u{r_1} + ( -1 + {{r_1}}^2 +
{{r_3}}^2 ) {r_4} )+\nonumber\\
&& {d_1}( {r_1} - {r_2} ) {r_4} ( x{r_1} + {r_2}(y-u))+{d_2}{r_4}(
-2x^2 - {{r_1}}^2{r_2}{r_4} + v{{r_4}}^2 - x{{r_4}}^2+ \nonumber\\
&& 2x{r_1}( {r_2} + {r_4} )- {{r_2}}^2( y + x)+
{r_1}{r_2}( u + {r_4}( {r_2} + 2{r_4} )) ))\\
A^{4'}_2 &=& \frac{A^{3'}_2}{L_1}L_2-\frac{\kappa
{{m_t}}^{\frac{7}{2}}{\sqrt{{r_4}}}}{2{L_1} {r_1}{r_2}}
( {d_1}( {r_1} - {r_2} ) {r_4} +{d_2}( y+u + x))\\
A^{4'}_3 &=& \frac{A^{3'}_3}{L_1}L_2-\frac{\kappa
{d_2}{{m_t}}^{\frac{7}{2}}{\sqrt{{r_4}}}}{2
{L_1}{r_1}{r_2}}( -2x + ( {r_1} - {r_2} ) {r_2} )\\
A^{4'}_4 &=& \frac{A^{3'}_4}{L_1}L_2-\frac{\kappa
{d_2}{{m_t}}^{\frac{7}{2}}{{r_4}}^{\frac{5}{2}}}
{2{L_1}{r_1}{r_2}}\\
A^{4'}_5 &=& \frac{A^{3'}_5}{L_1}L_2+\frac{\kappa
{{m_t}}^{\frac{7}{2}}}{2{L_1} {r_1}{r_2}{{r_4}}^{\frac{3}{2}}}(
2{d_4}{r_2}( -4x{r_1}( {r_1} - {r_2} ){r_4} + {r_1}{r_4}( -2v{r_4} +
{r_2}( 4u + 3{{r_1}}^2 + \nonumber\\
&& {r_2}( 2{r_2}-1)+ {r_1}( 5{r_2}-1 - {r_4} ) + {r_4} ) ) ) +
{r_4}(-x(8{d_3}{r_1}{r_2}({r_1}+2{r_2})+2{d_2}{r_4})-\nonumber\\
&& {r_4}( {d_1}{( {r_1} - {r_2} ) }^2 - 4{d_3}{r_1}( {r_1} - 3{r_2}
){{r_2}}^2 + {d_2}( 2u + {{r_4}}^2 )) ))\\
A^{4'}_6 &=& \frac{A^{3'}_6}{L_1}L_2-\frac{\kappa
{{m_t}}^{\frac{7}{2}}}{2{L_1}{r_1} {r_2}{\sqrt{{r_4}}}} ( {d_2}( -2x
+ ( {r_1} - {r_2} ) {r_2} ) {r_4} - 4{d_5}{r_1}{r_2}( v{r_4} +
{{r_2}}^2{r_4} + \nonumber\\
&& {r_1}( 2x + {r_2}{r_4} )- 2 y {r_2}))\\
A^{4'}_7 &=& \frac{A^{3'}_6}{L_1}L_2+\frac{2\kappa
{{m_t}}^{\frac{7}{2}}{\sqrt{{r_4}}}}{{L_1}}(
{d_3}{{r_4}}^2 + {d_4}( x+y ))\\
A^{4'}_8 &=& \frac{A^{3'}_8}{L_1}L_2-\frac{\kappa
{{m_t}}^{\frac{7}{2}}{\sqrt{{r_4}}}}{2{L_1}{r_1}{r_2}}
( {d_2}{{r_4}}^2 + 4{d_5}{r_1}{r_2}( x+y ))\\
A^{4'}_{10} &=& \frac{A^{3'}_8}{L_1}L_2+\frac{2\kappa {d_5}
{{m_t}}^{\frac{7}{2}}{\sqrt{{r_4}}}}{{L_1}}\\
A^{4'}_{11} &=& \frac{A^{3'}_8}{L_1}L_2-\frac{2\kappa
{d_4}{{m_t}}^{\frac{7}{2}}{\sqrt{{r_4}}}}{{L_1}}
\end{eqnarray}
and
\begin{equation}
A^2_{10} =-\frac{A^1_{10}}{L_1}L_2 \;\;{\rm and}\;\; A^{4'}_{10} =
\frac{A^{3'}_{10}}{L_1}L_2 .
\end{equation}
Here for convenience an overall factor ${\cal C}_s$ has been
contracted out from these coefficients. Then the square of the
amplitude $|M_i|^2$ can be conveniently obtained with the help of
Eqs.(\ref{amat},\ref{relat1},\ref{relat2}) and Eq.(\ref{3s1}).

\subsection{Amplitude for spin-triplet P-wave state: $[^3P_J]_1$}

There are totally thirty independent basic Lorentz structures $B_j$
for the case of $(b\bar{c})[(^3P_J)_{\bf 1}]$, which are
\begin{eqnarray}
B_1 &=& \frac{1}{m_t}p_{2\alpha}\epsilon_\beta(p_3)
\varepsilon^J_{\alpha\beta},\;
B_2=\frac{1}{m_t}p_{2}\cdot\epsilon(p_3)\varepsilon^J_{\alpha\alpha},\;
B_3=\frac{1}{m_t}p_{1}\cdot\epsilon(p_3)\varepsilon^J_{\alpha\alpha},\nonumber\\
B_4 &=& \frac{1}{m_t}p_{3\alpha}\epsilon_\beta(p_3)
\varepsilon^J_{\alpha\beta},\;
B_5=\frac{i\varepsilon^J_{\alpha\beta}}{m_t^3}\varepsilon(p_1,p_2,p_3,\alpha)
\epsilon_\beta(p_3),\;
B_6=\frac{i\varepsilon^J_{\alpha\alpha}}{m_t^3}\varepsilon(p_1,p_2,p_3,\epsilon(p_3)),
\nonumber\\
B_7 &=& \frac{i\varepsilon^J_{\alpha\beta}}{m_t^3}
\varepsilon(p_1,p_2,\alpha,\epsilon(p_3))p_{3\beta} ,\; B_8 =
\frac{i\varepsilon^J_{\alpha\beta}}{m_t^3}
\varepsilon(p_1,p_2,\alpha,\epsilon(p_3))p_{2\beta}
,\nonumber\\
B_9 &=& \frac{i\varepsilon^J_{\alpha\beta}}{m_t^3}
\varepsilon(p_1,p_2,\alpha,\beta)p_{1}\cdot\epsilon(p_3),\; B_{10}=
\frac{i\varepsilon^J_{\alpha\beta}}{m_t^3}\varepsilon(p_1,p_3,\alpha,\epsilon(p_3))
p_{2\beta} ,\nonumber\\
B_{11} &=& \frac{i\varepsilon^J_{\alpha\beta}}{m_t^3}
\varepsilon(p_1,p_3,\alpha,\epsilon(p_3))p_{3\beta},\;
B_{12}=\frac{i\varepsilon^J_{\alpha\beta}}{m_t^3}
\varepsilon(p_1,p_3,\alpha,\beta)p_{1}\cdot\epsilon(p_3),\nonumber\\
B_{13} &=& \frac{i\varepsilon^J_{\alpha\beta}}{m_t^3}
\varepsilon(p_1,p_3,\alpha,\beta)p_{2}\cdot\epsilon(p_3),\;
B_{14}=\frac{i\varepsilon^J_{\alpha\beta}}{m_t^3}
\varepsilon(p_2,p_3,\alpha,\epsilon(p_3))p_{2\beta},\nonumber\\
B_{15} &=& \frac{i\varepsilon^J_{\alpha\beta}}{m_t^3}
\varepsilon(p_2,p_3,\alpha,\epsilon(p_3))p_{3\beta},\;
B_{16}=\frac{i\varepsilon^J_{\alpha\beta}}{m_t^3}
\varepsilon(p_2,p_3,\alpha,\beta)p_{1}\cdot\epsilon(p_3),\nonumber\\
B_{17} &=& \frac{i\varepsilon^J_{\alpha\beta}}{m_t^3}
\varepsilon(p_1,\alpha,\epsilon(p_3),\beta),\;
B_{18}=\frac{i\varepsilon^J_{\alpha\beta}}{m_t^3}
\varepsilon(p_2,\alpha,\epsilon(p_3),\beta),\;
B_{19}=\frac{i\varepsilon^J_{\alpha\beta}}{m_t^3}
\varepsilon(p_3,\alpha,\epsilon(p_3),\beta)\nonumber\\
B_{20} &=& \frac{i\varepsilon^J_{\alpha\beta}}{m_t^5}
\varepsilon(p_1,p2,p3,\alpha)p_1\cdot\epsilon(p_3) p_{2\beta},\;
B_{21}=\frac{i\varepsilon^J_{\alpha\beta}}{m_t^5}
\varepsilon(p_1,p2,p3,\alpha)p_1\cdot\epsilon(p_3)
p_{3\beta},\nonumber\\
B_{22} &=& \frac{i\varepsilon^J_{\alpha\beta}}{m_t^5}
\varepsilon(p_1,p2,p3,\cdot\epsilon(p_3))p_{3\alpha} p_{2\beta},\;
B_{23}=\frac{i\varepsilon^J_{\alpha\beta}}{m_t^5}
\varepsilon(p_1,p2,p3,\cdot\epsilon(p_3))p_{3\alpha}
p_{3\beta},\nonumber\\
B_{24} &=& \frac{i\varepsilon^J_{\alpha\beta}}{m_t^5}
\varepsilon(p_1,p2,p3,\cdot\epsilon(p_3))p_{2\alpha} p_{2\beta},\;
B_{25}=\frac{\varepsilon^J_{\alpha\beta}}{m_t^3}
p_{1}\cdot\epsilon(p_3)p_{2\alpha}p_{3\beta},\nonumber\\
B_{26} &=& \frac{\varepsilon^J_{\alpha\beta}}{m_t^3}
p_{1}\cdot\epsilon(p_3)p_{3\alpha}p_{3\beta},\; B_{27}=
\frac{\varepsilon^J_{\alpha\beta}}{m_t^3}
p_{1}\cdot\epsilon(p_3)p_{2\alpha}p_{2\beta},\;
B_{28}=\frac{\varepsilon^J_{\alpha\beta}}{m_t^3}
p_{2}\cdot\epsilon(p_3)p_{2\alpha}p_{2\beta},\nonumber\\
B_{29} &=& \frac{\varepsilon^J_{\alpha\beta}}{m_t^3}
p_{2}\cdot\epsilon(p_3)p_{3\alpha}p_{2\beta},\;
B_{30}=\frac{\varepsilon^J_{\alpha\beta}}{m_t^3}
p_{2}\cdot\epsilon(p_3)p_{3\alpha}p_{3\beta} .
\end{eqnarray}

Noting the fact that $\varepsilon^{0,2}_{\alpha\beta}$ is the
symmetric tensor and $\varepsilon^{1}_{\alpha\beta}$ is the
anti-symmetric tensor, and the fact that
$\varepsilon^{1}_{\alpha\alpha}=\varepsilon^{2}_{\alpha\alpha}=0$,
one may observe that the terms involving the following coefficients
do not have any contributions to the square of the amplitude, so one
can safely set the following coefficients to zero:
\begin{eqnarray}
&& A^i_j(^3P_0)=0 \;\;\;\;\;\;\;\;{\rm for}\;\;
i=(1-4),j=(9,12,13,16,17,18,19)\\
&& A^i_j(^3P_1)=0 \;\;\;\;\;\;\;\;{\rm for}\;\;
i=(1-4),j=(2,3,6,24,26,27,28,30)\\
&& A^i_j(^3P_2)=0 \;\;\;\;\;\;\;\;{\rm for}\;\;
i=(1-4),j=(2,3,6,9,12,13,16,17,18,19) .
\end{eqnarray}

The coefficients $A^{1}_j$, $A^{2}_J$, $A^{3'}_j$ and $A^{4'}_j$
that are the same for all the three $[^3P_J]_1$ states (with
$J=1,2,3$):
\begin{eqnarray}
A^1_{20} &=&
\frac{8{d_4}{L_1}{{m_t}}^{\frac{5}{2}}}{{\sqrt{{r_4}}}},\;\;
A^1_{21} =
\frac{-8{d_5}{L_1}{{m_t}}^{\frac{5}{2}}}{{\sqrt{{r_4}}}},\;\;
A^1_{22} =
\frac{-8{d_4}{L_1}{{m_t}}^{\frac{5}{2}}}{{\sqrt{{r_4}}}}\\
A^1_{23} &=&
\frac{8{d_5}{L_1}{{m_t}}^{\frac{5}{2}}}{{\sqrt{{r_4}}}},\;\;
A^{3'}_{20} = \frac{-2\kappa
{d_4}{{m_t}}^{\frac{7}{2}}{r_2}}{{L_2}{\sqrt{{r_4}}}},\;\;
A^{3'}_{21} = \frac{2\kappa
{d_5}{{m_t}}^{\frac{7}{2}}{r_2}}{{L_2}{\sqrt{{r_4}}}} \\
A^{3'}_{22} &=& A^{3'}_{23}=0
\end{eqnarray}
and for $j=20,21,22,23$: $ A^2_j =-\frac{A^1_j}{L_1}L_2 \;\;{\rm
and}\;\; A^{4'}_j = \frac{A^{3'}_j}{L_1}L_2$.

The coefficients $A^{1}_j$, $A^{2}_J$, $A^{3'}_j$ and $A^{4'}_j$
that are the same for both $[^3P_0]_1$ and $[^3P_0]_2$:
\begin{eqnarray}
A^1_1 &=& \frac{4{L_1}{{m_t}}^{\frac{5}{2}}}{{\sqrt{{r_4}}}}( {d_5}(
1 + {r_2} ) ( 2u{r_1} + (r_1^2 +r_3^2 -1 ) {r_4} )-
{d_3}{r_4}( x - {r_1}( z + {r_2})+\nonumber\\
&&{r_2}( r_2 + y + z ))+{d_4}( 2u( v+ x ) +
 x( {r_1}-1 )+ 2x({r_1}-1 ) {r_4} +v( 2{r_1}-1) {r_4}+  \nonumber\\
&& 2{{r_2}}^2( u +  ( {r_1}-1 ) {r_4} )+
{r_2}( u - x + ( u + x ) {r_1} +( w-2 +2{r_1} ) {r_4}+({r_1}-1) {{r_4}}^2 )))\\
A^1_4 &=& \frac{-4{d_5}{L_1}{{m_t}}^{\frac{5}{2}}}{{\sqrt{{r_4}}}} (
-x + 2u( v + x )+ ( u - v - 3x ) {r_2} +( -1 + 3u + v ) {{r_2}}^2 -
3{{r_2}}^3 +\nonumber\\
&& {{r_1}}^2( r_2 - v - x + 3 z )+ {r_1}( -v - x +
{r_2}( -1 + 3v + 3x + {r_2}( -2+ 3{r_2} ))))\\
A^1_5 &=& 0\\
A^1_7 &=& \frac{-4{d_5}{L_1}{{m_t}}^{\frac{5}{2}}}{{\sqrt{{r_4}}}} (
1 + 2u + {r_1}{r_2} + 2{{r_3}}^2 + {r_4} )\\
A^1_8 &=& \frac{4{L_1}{{m_t}}^{\frac{5}{2}}}{{\sqrt{{r_4}}}}( {d_3}(
1 + {r_1} ) {r_4} +{d_4}( 1 + 2u + {r_1}{r_2} + 2{{r_3}}^2 +
{r_4} ))\\
A^1_{10} &=& \frac{-4{L_1}{{m_t}}^{\frac{5}{2}}} {{\sqrt{{r_4}}}}(
{d_4} ( r_2 + v + x + 3 z ) - {d_3}{r_2}{r_4}
)\\
A^1_{11} &=& \frac{4{d_5}{L_1}{{m_t}}^{\frac{5}{2}}}{{\sqrt{{r_4}}}}
( r_2 + v + x + 3 z )\\
A^1_{14} &=&
\frac{4{L_1}{{m_t}}^{\frac{5}{2}}}{{\sqrt{{r_4}}}}(2u{d_4} + (
{d_3}( -{r_1} + {r_2} )+ {d_4}( -1 + 2{r_1} + {r_2} )) {r_4} )\\
A^1_{15} &=&
\frac{-4{d_5}{L_1}{{m_t}}^{\frac{5}{2}}}{{\sqrt{{r_4}}}}( 2u +
2{{r_1}}^2 + ( -1 + {r_2} ) {r_2} +{r_1}( -1 + 3{r_2} ))\\
A^1_{25} &=& \frac{-4{L_1}{{m_t}}^{\frac{5}{2}}} {{\sqrt{{r_4}}}}(
{d_5}( 1 + {r_1} ) ( 1 + {r_2} )+ {d_4} ( r_2 - v - x + 3 z )- {d_3}{r_2}{r_4} )\\
A^1_{29} &=& -4{L_1}{{m_t}}^{\frac{5}{2}}{\sqrt{{r_4}}}( {d_4} +
{d_3}{r_1} - {d_3}{r_2} + {d_4}{r_2} )\\
A^{3'}_1 &=& \frac{\kappa {{m_t}}^{\frac{7}{2}}}{{L_2}
{\sqrt{{r_4}}}}( {d_4} ( 2x^2 + 2u^2{r_2} + 2{{r_1}}^4{r_2} + ( u -
v + 3x ) {{r_2}}^2 + ( -1 + u + v ) {{r_2}}^3 + {{r_2}}^4
+\nonumber\\
&& {{r_1}}^3( x + {r_2} + 5{{r_2}}^2 )+ {{r_1}}^2( -v - x + {r_2} (
3 r_2 + 4 u - 3 v + 4 z - 1))+\nonumber\\
&& x{r_2}( u + v + w - 2 )+ {r_1}( ( u - 2v + 2x )
{r_2} + ( -2 + 5u + 2v + 3x ) {{r_2}}^2 + \nonumber\\
&& 3{{r_2}}^3 + {{r_2}}^4 +x( u + v + w + 2 x - 2 )))
+{d_3}( -2x( u + x )+x{r_2} + {{r_1}}^3{r_2} + \nonumber\\
&&  ( v-u - 2x ) {{r_2}}^2 + {{r_2}}^3 - x{r_1}( 1 + {r_2} )+
{{r_1}}^2( v - x + {r_2} + 2{{r_2}}^2 )+ {r_1}{r_2}( u + 2v
+\nonumber\\
&& {r_2}( 2 + {r_2} ) ) ) {r_4} +{d_5}( 2u{r_1} + ( -1 + {{r_1}}^2 +
{{r_3}}^2 ) {r_4} ) ( x+y ))\\
A^{3'}_4 &=& -\frac{\kappa
{d_5}{{m_t}}^{\frac{7}{2}}}{{L_2}{\sqrt{{r_4}}}}( 2x^2 + 2u^2{r_2} +
( -1 + 2u + 2v ) x{r_2} +2{{r_1}}^4{r_2} + ( u - v + 3x ) {{r_2}}^2
+\nonumber\\
&& ( -1 + u + v ) {{r_2}}^3 + {{r_2}}^4 + {{r_1}}^3{r_2}( 1 + 5{r_2}
) +{{r_1}}^2( -v - x + {r_2} ( -1 + 4u + v + \nonumber\\
&& 3x + {r_2}( 3 + 4{r_2} ) ))+{r_1}( 2x^2 + x( -1 + 2v + {r_2}( 2 +
3{r_2} )) + \nonumber\\
&& {r_2}( u - 2v +{r_2}( -2 + 5u + 2v + {r_2}( 3 + {r_2} )))) )\\
A^{3'}_5 &=& \frac{\kappa {d_5}{{m_t}}^{\frac{7}{2}}}{{L_2}
{\sqrt{{r_4}}}} ( 2u{r_1} + ( -1 + {{r_1}}^2 + {{r_3}}^2 ) {r_4} )\\
A^{3'}_7 &=& -\frac{\kappa {d_5}{{m_t}}^{\frac{7}{2}}
}{{L_2}{\sqrt{{r_4}}}}( -2( u + x )+ {{r_1}}^3 + {{r_1}}^2( -1 +
2{r_2} )+{r_1}( 2x + ( -2 + {r_2} ) {r_2} - {{r_3}}^2)-\nonumber\\
&& {r_2}( 2u + {r_2} + {{r_3}}^2 ))\\
A^{3'}_8 &=& \frac{\kappa
{{m_t}}^{\frac{7}{2}}}{{L_2}{\sqrt{{r_4}}}}( {d_4} ( -2( u + x ) +
{{r_1}}^3 + {{r_1}}^2( -1 + 2{r_2} )+ {r_1}( 2x + ( -2 + {r_2} )
{r_2} - {{r_3}}^2 )-\nonumber\\
&& {r_2}( 2u + {r_2} + {{r_3}}^2 )) + {d_3}( 2x + {{r_1}}^2 + {r_1}(
-1 + {r_2} ) + {r_2} + 2{{r_2}}^2 ) {r_4} )\\
A^{3'}_{10} &=& \frac{\kappa {{m_t}}^{\frac{7}{2}}}{{L_2}
{\sqrt{{r_4}}}}( {d_3} ( 2x{r_1} + 2x{r_2} - {{r_1}}^2{r_2} +
{{r_2}}^3 )+{d_4}( 2x - 2{{r_1}}^2{r_2} + {r_1}( 2( v + x ) + {r_2}
- 3{{r_2}}^2 )\nonumber\\
&& + {r_2}( -2u + 2v + {r_2} -{{r_2}}^2 )))\\
A^{3'}_{11} &=& \frac{\kappa
{d_5}{{m_t}}^{\frac{7}{2}}}{{L_2}{\sqrt{{r_4}}}}( -2x +
2{{r_1}}^2{r_2} + {r_2}( 2u - 2v - {r_2} + {{r_2}}^2 )+ {r_1}( -2(
v + x )+ {r_2}( -1 + 3{r_2} )))\\
A^{3'}_{14} &=& -\frac{\kappa {{m_t}}^{\frac{7}{2}}{\sqrt{{r_4}}} (
{d_3}{{r_4}}^2 + {d_4}( 2( u + x )+ 2{{r_1}}^2 +
5{r_1}{r_2} + 3{{r_2}}^2 + {r_4} )) }{{L_2}}\\
A^{3'}_{15} &=& \frac{\kappa
{d_5}{{m_t}}^{\frac{7}{2}}{\sqrt{{r_4}}}}{{L_2}}( 2( u + x )+
( 1 + 2{r_1} + 3{r_2} ) {r_4} )\\
A^{3'}_{25} &=& -\frac{\kappa {{m_t}}^{\frac{7}{2}}}{{L_2}
{\sqrt{{r_4}}}} ( {d_4}( -2x + 2{{r_1}}^2{r_2} + {r_2}( 2( u - v + x
)+ ( -1 + {r_2} ) {r_2} )+ {r_1}( -2( v + x )+ \nonumber\\
&& {r_2}( -1 + 3{r_2} ) ) )+ {d_5}( -2( u + x )+ {{r_1}}^3 +
{{r_1}}^2( 2{r_2}-1 )+{r_1}(2x +({r_2}-2){r_2} -{{r_3}}^2 )\nonumber\\
&& - {r_2}( {r_2} + {{r_3}}^2 ))+ {d_3}( -2x + ( {r_1} -
{r_2} ) {r_2} ){r_4} )\\
A^{3'}_{29} &=& -\frac{\kappa
{{m_t}}^{\frac{7}{2}}{\sqrt{{r_4}}}}{{L_2}}( {d_3}{{r_4}}^2 + {d_4}(
2( u + x )+ 2{{r_1}}^2 + 3{r_1}{r_2} + {{r_2}}^2 + {r_4} ))
\end{eqnarray}
and
\begin{eqnarray}
A^2_1&=&-\frac{A^1_1}{L_1}L_2-\frac{8{L_2}{{m_t}}^{\frac{5}{2}}}
{{\sqrt{{r_4}}}}( {d_3}( x - {r_1}{r_2} + {{r_2}}^2 ) {r_4} - {d_5}(
2u{r_1} + ( {{r_1}}^2 + {{r_3}}^2-1 ) {r_4} )+ \nonumber\\
&& {d_4}( ( v + 2x ) {r_4} +
2{{r_2}}^2{r_4} -{r_1}( x + 2{r_2}{r_4} )+ {r_2}( y-2u ) ))\\
A^2_4&=&-\frac{A^1_4}{L_1}L_2+\frac{8{d_5}{L_2}{{m_t}}^{\frac{5}{2}}}{{\sqrt{{r_4}}}}(
( v + 2x ) {r_4} + 2{{r_2}}^2{r_4} -{r_1}( x + 2{r_2}{r_4} )+ {r_2}(y-2u))\\
A^2_5&=& -\frac{A^1_5}{L_1}L_2 \\
A^2_7&=&-\frac{A^1_7}{L_1}L_2-8{d_5}{L_2}{{m_t}}^{\frac{5}{2}}{\sqrt{{r_4}}}\\
A^2_8&=&-\frac{A^1_8}{L_1}L_2+\frac{8{L_2}{{m_t}}^{\frac{5}{2}}
}{{\sqrt{{r_4}}}}( {d_3}{r_4} +{d_4}{r_4} )\\
A^2_{10}&=&-\frac{A^1_{10}}{L_1}L_2-\frac{8{d_4}{L_2}{{m_t}}^{\frac{5}{2}}{r_2}}{{\sqrt{{r_4}}}}\\
A^2_{11}&=&-\frac{A^1_{11}}{L_1}L_2+\frac{8{d_5}{L_2}{{m_t}}^{\frac{5}{2}}{r_2}}{{\sqrt{{r_4}}}}\\
A^2_{14}&=&-\frac{A^1_{14}}{L_1}L_2-8{d_4}{L_2}{{m_t}}^{\frac{5}{2}}{\sqrt{{r_4}}}\\
A^2_{15}&=&-\frac{A^1_{15}}{L_1}L_2+8{d_5}{L_2}{{m_t}}^{\frac{5}{2}}{\sqrt{{r_4}}}\\
A^2_{25}&=&-\frac{A^1_{25}}{L_1}L_2-\frac{8{L_2}{{m_t}}^{\frac{5}{2}}
}{{\sqrt{{r_4}}}}( {d_4}{r_2} +{d_5}{r_4} )\\
A^2_{29}&=&-\frac{A^1_{29}}{L_1}L_2-8{d_4}{L_2}{{m_t}}^{\frac{5}{2}}{\sqrt{{r_4}}}\\
A^{4'}_1 &=& \frac{A^{3'}_1}{L_1}L_2-\frac{2\kappa
{{m_t}}^{\frac{7}{2}}}{{L_1}{\sqrt{{r_4}}}}( 2x^2{d_4} + x{r_4}( -(
( {d_3} + 2{d_4} )( {r_1} - {r_2} ))+ {d_4}{r_4} ) +
\nonumber\\
&& {r_4}( -( v{d_4}{r_4} )+ {d_3}{r_2}{{r_4}}^2 + {d_4}{r_2}( u +
{{r_4}}^2 )))\\
A^{4'}_4 &=& \frac{A^{3'}_4}{L_1}L_2-\frac{2\kappa
{d_5}{{m_t}}^{\frac{7}{2}}}{{L_1}{\sqrt{{r_4}}}} (x( 2{r_1} - 2{r_2}
- {r_4} ) {r_4}-2x^2 +{r_4}(v{r_4}-u{r_2} - {r_2}{{r_4}}^2 ))\\
A^{4'}_5 &=& \frac{A^{3'}_5}{L_1}L_2\\
A^{4'}_7 &=& \frac{A^{3'}_7}{L_1}L_2-\frac{2\kappa
{d_5}{{m_t}}^{\frac{7}{2}} }{{L_1}{\sqrt{{r_4}}}}( 2(
u + x ) + {{r_4}}^2 )\\
A^{4'}_8 &=& \frac{A^{3'}_8}{L_1}L_2+\frac{2\kappa
{{m_t}}^{\frac{7}{2}}}{{L_1}{\sqrt{{r_4}}}}( {d_3}(
{r_1} - {r_2} ) {r_4} +{d_4}( 2( u + x )+ {{r_4}}^2 ) )\\
A^{4'}_{10} &=& \frac{A^{3'}_{10}}{L_1}L_2-\frac{2\kappa
{d_4}{{m_t}}^{\frac{7}{2}} }{{L_1}{\sqrt{{r_4}}}}( 2x
+ {r_2}{r_4} )\\
A^{4'}_{11} &=& \frac{A^{3'}_{11}}{L_1}L_2+\frac{2\kappa
{d_5}{{m_t}}^{\frac{7}{2}} }{{L_1}{\sqrt{{r_4}}}}( 2x
+ {r_2}{r_4} )\\
A^{4'}_{14} &=& \frac{A^{3'}_{14}}{L_1}L_2+\frac{2\kappa
{d_4}{{m_t}}^{\frac{7}{2}}{{r_4}}^{\frac{3}{2}}}{{L_1}}\\
A^{4'}_{15} &=& \frac{A^{3'}_{15}}{L_1}L_2-\frac{2\kappa
{d_5}{{m_t}}^{\frac{7}{2}}{{r_4}}^{\frac{3}{2}}}{{L_1}}\\
A^{4'}_{25} &=& \frac{A^{3'}_{25}}{L_1}L_2-\frac{2\kappa
{{m_t}}^{\frac{7}{2}}}{{L_1}{\sqrt{{r_4}}}}( 2u{d_5} +
2x( {d_4} + {d_5} )+ {d_4}{r_2}{r_4} + {d_5}{{r_4}}^2 )\\
A^{4'}_{29} &=& \frac{A^{3'}_{29}}{L_1}L_2+\frac{2\kappa
{d_4}{{m_t}}^{\frac{7}{2}}{{r_4}}^{\frac{3}{2}}}{{L_1}}.
\end{eqnarray}

The remaining coefficients for the case of $[^3P_0]_1$:
\begin{eqnarray}
A^1_2 &=&
\frac{{L_1}{{m_t}}^{\frac{5}{2}}}{{r_1}{r_2}{\sqrt{{r_4}}}}( {d_2}(
1 + {r_2} ) ( {{r_1}}^2 - {{r_1}}^3 + {r_2}( u + {r_2} )+ {r_1}( -u
+ {r_2}( 2 + {r_2} )))+{d_1}( {{r_1}}^4 + \nonumber\\
&& 2x{r_1}( 1 + {r_2} ) + {{r_1}}^3(7{r_2}-1 ) + {r_1}{r_2}( 6u
+ ( {r_2}-7 ) {r_2} ) + {{r_1}}^2( u + {r_2}(9{r_2} -5))-\nonumber\\
&& {r_2}( 2x( 1 + {r_2} ) + {r_2}({r_2}( 3 + 2{r_2} )-u))) +
2{d_5}{r_1}{r_2}(1 + {r_2} )(-2u{r_1} - \nonumber\\
&& ({{r_1}}^2-1 + {{r_3}}^2 ) {r_4} ))\\
A^1_3 &=& -\frac{{L_1}{{m_t}}^{\frac{5}{2}}}
{{r_1}{r_2}{{r_4}}^{\frac{3}{2}}}( {d_2}( -{{r_1}}^4{r_2} +
{{r_1}}^3( r_2 + v + x + z)+ {r_1}{r_2}( 2x + {r_2} (
2( u + v + 2x )+ \nonumber\\
&& {r_2}( 3{r_2}-1)) ) + {{r_1}}^2( v + {r_2}( 3 r_2 - u - v - w - 3
x + 5 z + 2 )- {{r_2}}^2( v + 2x +\nonumber\\
&& 3{{r_2}}^2 - {r_2}( -2 + v + {{r_3}}^2 ) )) + {d_1} (
{{r_1}}^3{r_2} + {{r_2}}^2( r_2 - 3 v + 4 z )+
{{r_1}}^2( v + 2x +\nonumber\\
&& {r_2} + 6{{r_2}}^2 ) + {r_1}{r_2}( 6 r_2 - 7 v - 3 x + 13 z))
{r_4} + 2{d_5}{r_1}{r_2} ( 2( v + x ) + \nonumber\\
&& {r_2}( 2 + {r_1} + 3{r_2} ) ) ( 2u{r_1} + ( -1 +
{{r_1}}^2 + {{r_3}}^2 ) {r_4} ))\\
A^1_6 &=& \frac{{L_1}{{m_t}}^{\frac{5}{2}}}{{r_1}{r_2}
{\sqrt{{r_4}}}}( -{d_2}( {r_1} - {r_2} ) ( 1 + {r_2} )
+{d_1}( {{r_1}}^2 + 6{r_1}{r_2} + {{r_2}}^2 ))\\
A^1_{24} &=& 0\\
A^1_{26} &=& \frac{4{d_5}{L_1}{{m_t}}^{\frac{5}{2}}}{{\sqrt{{r_4}}}}
( r_2 - v - x + 3 z )\\
A^1_{27} &=& \frac{4{L_1}{{m_t}}^{\frac{5}{2}}}{{ \sqrt{{r_4}}}}(
{d_4}( 1 + {r_1} ) ( 1 + {r_2} )+ {d_3}( 1 + {r_1} + 2{r_2}
) {r_4} )\\
A^1_{28} &=& 8{d_3}{L_1}{{m_t}}^{\frac{5}{2}} {\sqrt{{r_4}}}( 1 +
{r_2} )\\
A^1_{30} &=& 4{d_5}{L_1}{{m_t}}^{\frac{5}{2}} {\sqrt{{r_4}}}( 1 +
{r_2} )\\
A^{3'}_2 &=& \frac{-\kappa
{{m_t}}^{\frac{7}{2}}}{4{L_2}{r_1}{r_2}{\sqrt{{r_4}}}}(
2{d_5}{r_1}{r_2}( 2u{r_1} + ( {{r_1}}^2-1 + {{r_3}}^2 ) {r_4} )
( x+y )+{d_1}( 4x^2( {r_2}-{r_1})\nonumber\\
&& + ( {{r_1}}^3 + {{r_1}}^2( 1 + 6{r_2} )+ {r_2}( 3u + {r_2} +
4{{r_2}}^2 )+ {r_1}( u + {r_2}( 6 + 13{r_2} ))) {{r_4}}^2
+\nonumber\\
&& 2x( u( {r_2} -{r_1}) + 4{r_2}{{r_4}}^2 ) ) +{d_2}( {r_2}( -4u(
u + x ) - ( u + 2x ) ( 2{r_1}-1 ) {r_4} -2u{{r_4}}^2 + \nonumber\\
&& 2{{r_3}}^2{{r_4}}^2 + ( 1 - 2{r_1} ) {{r_4}}^3 )+
{r_4}( -{{r_3}}^2{{r_4}}^2 + {r_1}( 1 + {r_4} ) ( x+y ) +u( u + x + y ))))\\
A^{3'}_3 &=& \frac{\kappa {{m_t}}^{\frac{7}{2}}}{4{L_2}
{r_1}{r_2}{{r_4}}^{\frac{3}{2}}}( 2x{d_1} ( {{r_1}}^2 - {r_1}( v + (
-5 + {r_2} ) {r_2} )+ {r_2}( v + {r_2}( 2 + {r_2} ) ) ) {r_4}
+\nonumber\\
&& {d_1}( v( {r_1} + 3{r_2} )+ {r_2}( {r_1} - {{r_1}}^2 -
5{r_1}{r_2} + {r_2}( 3 + 2{r_2} ))) {{r_4}}^3 + 2{d_5}{r_1}{r_2}(
2u{r_1} +\nonumber\\
&& ( -1 + {{r_1}}^2 + {{r_3}}^2 ) {r_4} ) ( 2x + v{r_4} +
{{r_2}}^2{r_4} + {r_2}( r_4 - 2 y) )+{d_2}( 4x^2{r_1}( {r_1} - {r_2}
)+\nonumber\\
&&  2x( {{r_1}}^3 + {{r_1}}^4 -{{r_2}}^2( v-1 + {r_2} ) +
{{r_1}}^2( -1 + 2u + v -3( -1 + {r_2} ) {r_2} )+ \nonumber\\
&&  {r_1}{r_2}( -2u + {r_2} - 2{{r_2}}^2 ))
+ v( {{r_1}}^2( 1 + {r_2} )+ 2{r_1}( u + {r_2} )+\nonumber\\
&& {r_2}( -2u + {r_2} - {{r_2}}^2 ) ) {r_4} + {r_2}{r_4}( -2u( {r_2}
+ {r_1}( -1 + {r_4} ) )+ {r_4}( {{r_1}}^2 - 2{{r_1}}^3 +\nonumber\\
&& 2{r_1}{r_2} - 5{{r_1}}^2{r_2} - {{r_2}}^2 -{r_1}{{r_2}}^2 + {r_4} ))))\\
A^{3'}_6 &=& \frac{-( \kappa {{m_t}}^{\frac{7}{2}}}{4{L_2}
{r_1}{r_2}{\sqrt{{r_4}}}}( -2x{d_1}( {r_1} - {r_2} )+ {d_2}(
2{{r_1}}^3 + {{r_1}}^2( {r_2}-1)- 2{r_1}( -u - x + {r_2} +
{{r_2}}^2 )\nonumber\\
&& - {r_2}( r_2 + 2 u - v + x + z ) ) + {d_1}( {r_1} + 3{r_2} )
{{r_4}}^2 + 2{d_5}{r_1}{r_2}( 2u{r_1} + ( {{r_1}}^2-1  +
{{r_3}}^2 ){r_4} )))\\
A^{3'}_{24} &=& \frac{2\kappa
{d_3}{{m_t}}^{\frac{7}{2}}{\sqrt{{r_4}}}}{{L_2}}\\
A^{3'}_{26} &=& -\frac{\kappa {d_5}{{m_t}}^{\frac{7}{2}}}{{L_2}
{\sqrt{{r_4}}}}( 2( x + ( v + x ) {r_1} + v{r_2} )
+{{r_2}}^2{r_4} + {r_2}( -2( u + x )+ {r_4} - 2{{r_4}}^2 ))\\
A^{3'}_{27} &=& - \frac{\kappa {{m_t}}^{\frac{7}{2}}}{{L_2}
{\sqrt{{r_4}}}}( -( {d_3}( -2v + 2x - {r_2} + {r_1}( -1 + {r_4} ) )
{r_4} )+ {d_4}( 2( u + x )- 2x{r_1} +\nonumber\\
&& {r_4}( {{r_3}}^2 + {r_4} - {r_1}{r_4} ))) \\
A^{3'}_{28} &=& \frac{2\kappa
{d_3}{{m_t}}^{\frac{7}{2}}{\sqrt{{r_4}}} } {{L_2}}( x+y )\\
A^{3'}_{30} &=& \frac{\kappa
{d_5}{{m_t}}^{\frac{7}{2}}{\sqrt{{r_4}}}}{{L_2}}( 2( u + x )+ ( 1 +
2{r_1} + {r_2} ) {r_4} )
\end{eqnarray}
and
\begin{eqnarray}
A^2_2&=&-\frac{A^1_2}{L_1}L_2-\frac{2{L_2}{{m_t}}^{\frac{5}{2}}}{{r_1}{r_2}{\sqrt{{r_4}}}}(
-2x{d_1}( -2{r_2} + {r_4} ) +2{d_5}{r_1}{r_2}( 2u{r_1} + ( -1 +
{{r_1}}^2 + {{r_3}}^2 ) {r_4} )+\nonumber\\
&& {d_1}{r_4}( 2{r_2}{r_4} + {{r_4}}^2 )+{d_2}( -2u{r_2} + u{r_4} -
{{r_2}}^2{r_4} + {r_1}{r_4}( -3{r_2} +{r_4} )))\\
A^2_3&=&-\frac{A^1_3}{L_1}L_2+\frac{2{L_2}{{m_t}}^{\frac{5}{2}}}{{r_1}
{r_2}{{r_4}}^{\frac{3}{2}}}( 2x{d_2}( {r_1} - {r_4} ) ( -2{r_2} +
{r_4} )+{d_2}{r_4}( v( 2{r_2} - {r_4} )+\nonumber\\
&& ( {r_1} - {r_2} ) {r_2}( -2{r_1} - 4{r_2} + {r_4} ))+ {r_2}( -(
{d_1}{r_4} ( 2( {r_1} - {r_2} ) {r_2} +( {r_1} + 3{r_2} ) {r_4} ) )
+ \nonumber\\
&& 4{d_5}{r_1}{r_2}( -2u{r_1} -( -1 + {{r_1}}^2 + {{r_3}}^2 ) {r_4} ) ))\\
A^2_6&=&-\frac{A^1_6}{L_1}L_2-\frac{2{d_2}{L_2}{{m_t}}^{\frac{5}{2}}}
{{r_1}{r_2}{\sqrt{{r_4}}}}( -2{r_2} + {r_4} )\\
A^2_{24}&=&-\frac{A^1_{24}}{L_1}L_2\\
A^2_{26}&=&-\frac{A^1_{26}}{L_1}L_2+\frac{8{d_5}{L_2}{{m_t}}^{\frac{5}{2}}{r_2}}{{\sqrt{{r_4}}}}\\
A^2_{27}&=&-\frac{A^1_{27}}{L_1}L_2+\frac{8{L_2}{{m_t}}^{\frac{5}{2}}
}{{\sqrt{{r_4}}}}( {d_3}{r_4} +{d_4}{r_4} )\\
A^2_{28}&=&-\frac{A^1_{28}}{L_1}L_2+16{d_3}{L_2}{{m_t}}^{\frac{5}{2}}{\sqrt{{r_4}}}\\
A^2_{30}&=&-\frac{A^1_{30}}{L_1}L_2+8{d_5}{L_2}{{m_t}}^{\frac{5}{2}}{\sqrt{{r_4}}}\\
A^{4'}_2 &=& \frac{A^{3'}_2}{L_1}L_2+\frac{\kappa
{{m_t}}^{\frac{7}{2}}{\sqrt{{r_4}}}}{2{L_1}{r_1}{r_2}}( {d_1}{r_4}(
2( {r_1} - {r_2} ) {r_2} + ( {r_1} + 3{r_2} ) {r_4} )+ {d_2}{r_4}(x+y))\\
A^{4'}_3 &=& \frac{A^{3'}_3}{L_1}L_2-\frac{\kappa
{{m_t}}^{\frac{7}{2}}}{2 {L_1}{r_1}{r_2}{{r_4}}^{\frac{3}{2}}}(
2{d_5}{r_1}{r_2} ( 2x + {r_2}{r_4} ) ( 2u{r_1} + ( -1 + {{r_1}}^2 +
{{r_3}}^2 ) {r_4} ) + \nonumber\\
&& {d_1}{r_4}( {r_2}{r_4}( 2{r_2}{r_4} + {{r_4}}^2 )+ 2x(
2{r_2}{r_4} + {r_1}( 2{r_2} + {r_4} )))+{d_2}{r_4}( 2x{{r_1}}^2 +
2x{r_1}{r_2} + \nonumber\\
&& {r_1}{r_4}( v + 3{{r_2}}^2 - {r_2}{r_4} )+ {r_2}( {{r_2}}^2{r_4}
-4y {r_2}+ {r_4}( 2u + v + 2x + 2{{r_4}}^2 )) ) )\\
A^{4'}_6 &=& \frac{A^{3'}_6}{L_1}L_2-\frac{\kappa
{d_2}{{m_t}}^{\frac{7}{2}}{{r_4}}^{\frac{3}{2}}}{2{L_1}{r_1}{r_2}}\\
A^{4'}_{24} &=& \frac{A^{3'}_{24}}{L_1}L_2\\
A^{4'}_{26} &=& \frac{A^{3'}_{26}}{L_1}L_2+\frac{2\kappa
{d_5}{{m_t}}^{\frac{7}{2}} }{{L_1}{\sqrt{{r_4}}}}( 2x
+ {r_2}{r_4} )\\
A^{4'}_{27} &=& \frac{A^{3'}_{27}}{L_1}L_2+\frac{2\kappa
{{m_t}}^{\frac{7}{2}}}{{L_1}{\sqrt{{r_4}}}}(
{d_3}{{r_4}}^2 + {d_4}( u+x+y ))\\
A^{4'}_{30} &=& \frac{A^{3'}_{30}}{L_1}L_2-\frac{2\kappa
{d_5}{{m_t}}^{\frac{7}{2}}{{r_4}}^{\frac{3}{2}}}{{L_1}}.
\end{eqnarray}

The remaining coefficients for the case of $[^3P_1]_1$ state:
\begin{eqnarray}
A^1_1 &=& \frac{-2{L_1}{{m_t}}^{\frac{5}{2}}}{{r_1}{r_2}
{\sqrt{{r_4}}}}({d_2}(1 + {r_2} ) ( u + {r_2} + {r_1}( -1 + {r_4} ))
{r_4} - {d_1}{{r_4}}^2(u + ( -1 + {r_1} ) {r_4} )+\nonumber\\
&& 2{r_1}{r_2}( -( {d_4} ( -x + 2u( v + x ) + ( u - v - 3x ) {r_2} +
{{r_1}}^3{r_2} - 3{{r_2}}^3 +{{r_1}}^2( 2( v + x )+  \nonumber\\
&&{r_2} + 4{{r_2}}^2 )+ {{r_2}}^2( 2 u + w - 2 ) + {r_1}( -v - x +
{r_2} ( u-2 r_2 - v + w + 3 z - 2) + \nonumber\\
&& {{r_3}}^2 )))) + {d_3}( x + {{r_1}}^2{r_2} + {r_1}( -v - x +
({r_2} -1) {r_2} ) +{r_2}(r_2 + u - v + 2 z )){r_4} ))\\
A^1_4 &=&
\frac{2{L_1}{{m_t}}^{\frac{5}{2}}}{{r_1}{r_2}{\sqrt{{r_4}}}}( {d_2}
( -x + {r_1}( r_2 + v + x + z )- {r_2}( r_2 - v + 3 z)) {r_4}
+x( 2{d_5}{r_1}{r_2}( 2u +\nonumber\\
&&  ( {r_1} -1 ) ( 1 + 2{r_1} + 3{r_2} ))+ {d_1}{{r_4}}^2 )+ {r_2} (
{d_1}{{r_4}}^3 +2{d_5}{r_1} ( u{r_2}( 1 + 2{r_1} + 3{r_2} )+ \nonumber\\
&& v( 2u + {r_1}(2{r_1} -1)- {r_2} + 3{r_1}{r_2} + {{r_2}}^2 )+
{r_2}( ({r_1} -1 ) ( 2 + {r_1} + 3{r_2} )+ {{r_3}}^2 ) {r_4} ))) \\
A^1_5 &=&\frac{2{L_1}{{m_t}}^{\frac{5}{2}}
{\sqrt{{r_4}}}}{{r_1}{r_2}} ( {d_2} - {d_1}{r_1} +( {d_1} + {d_2} ) {r_2} )\\
A^1_7 &=& \frac{2{L_1}{{m_t}}^{\frac{5}{2}}}{{r_1}{r_2}
{{r_4}}^{\frac{3}{2}}}( 4u{d_5}{r_1}( {r_1} - {r_2} ) {r_2} + {r_4}(
2{d_5}{r_1}( 1 + {r_1} )( -3 + 2{r_1} - {r_2} ) {r_2} +\nonumber\\
&& {d_2}( 2{{r_1}}^2 - {r_1}( -3 + {r_2} )+
{r_2}( 3 + {r_2} ))+ {d_1}( {r_1} - {r_2} ) {r_4} )) \\
A^1_8 &=& \frac{4{L_1}{{m_t}}^{\frac{5}{2}}}
{{r_1}{r_2}{\sqrt{{r_4}}}}( {r_1}{r_2}(
{d_3}( 1 + {r_1} ) {r_4} + {d_4}( 1 + 2u +
{r_1}{r_2} + 2{{r_3}}^2 + {r_4} ))-{d_1}( 1 +{r_2} ) {r_4}) \\
A^1_9 &=& \frac{-2{L_1}{{m_t}}^{\frac{5}{2}}}{{r_1}{r_2}
{{r_4}}^{\frac{3}{2}}}( {d_2} ( {r_1}( 2 + {r_1} )+ {r_2} ) {r_4} +
2{d_5}{r_1}{r_2}( 2u{r_1} + ( -1 + {{r_1}}^2 +
{{r_3}}^2 ) {r_4} ))\\
A^1_{10} &=& \frac{-2{L_1}{{m_t}}^{\frac{5}{2}}}{{r_1}{r_2}
{{r_4}}^{\frac{3}{2}}}( -{d_2} ( {r_1}( -3 + {r_2} )- 3{r_2}( 1 +
{r_2} )) {r_4}+ {d_1}{{r_4}}^3 + 2{r_1}{r_2}( 4u{d_5}{r_1}+\nonumber\\
&& 2{d_5}( -1 + {{r_1}}^2 + {{r_3}}^2 ) {r_4} +
{r_4}( {d_4}( r_2 + v + x + 3 z)-{d_3}{r_2}{r_4} ))) \\
A^1_{11} &=& \frac{4{d_5}{L_1}{{m_t}}^{\frac{5}{2}}}{{\sqrt{{r_4}}}}
( r_2 + v + x + 3 z ) \\
A^1_{12} &=&
\frac{2{d_2}{L_1}{{m_t}}^{\frac{5}{2}}{r_2}}{{r_1}{\sqrt{{r_4}}}}\\
A^1_{13} &=& \frac{{L_1}{{m_t}}^{\frac{5}{2}}}{{r_1}{r_2}
{{r_4}}^{\frac{3}{2}}}( 8u{d_5}{{r_1}}^2{r_2} +{r_4}( {d_2}( -(
{r_1}( -3 + {r_2} ))+3{r_2}( 1 + {r_2} ))+\nonumber\\
&& 4{d_5}{r_1}{r_2}( -1 + {{r_1}}^2 + {{r_3}}^2 )+ {d_1}(
{r_1} - {r_2} ) {r_4} ))\\
A^1_{14} &=&\frac{4{L_1}{{m_t}}^{\frac{5}{2}}} {{\sqrt{{r_4}}}}(
2u{d_4} +( {d_3}( -{r_1} + {r_2} )+
{d_4}( -1 + 2{r_1} + {r_2} )) {r_4} )\\
A^1_{15} &=& \frac{4{L_1}{{m_t}}^{\frac{5}{2}}}{{r_1}
{r_2}{\sqrt{{r_4}}}}( -2u{d_5}{r_1}{r_2} -( {d_2}( {r_1} -
{r_2} )+ {d_5}{r_1}{r_2}( -1 + 2{r_1} + {r_2} )) {r_4} ) \\
A^1_{16} &=& \frac{{L_1}{{m_t}}^{\frac{5}{2}}
{\sqrt{{r_4}}}}{{r_1}{r_2}}( {d_2} - {d_1}{r_1} + 2{d_2}{r_1}
+{d_1}{r_2} -{d_2}{r_2} )\\
A^1_{17} &=& -\frac{{L_1}{{m_t}}^{\frac{5}{2}}}
{{r_1}{r_2}{{r_4}}^{\frac{3}{2}}}( {d_2}( -{{r_1}}^3{r_2}+
{{r_1}}^2( 4( v + x )+ {r_2} )+ {r_2}( v + 4x + 3{{r_2}}^2 - {r_2}(
-2 + \nonumber\\
&& 2u + 3v + {{r_3}}^2 ))+ {r_1}( v + 4x - {r_2}
( u - v + w + x + 3 z - 2 ) )) {r_4} + \nonumber\\
&& {d_1}( {{r_1}}^2{r_2} - {r_2}( v + {r_2} )+ {r_1}( r_2 - 2 v - x
+ 3 z)) {{r_4}}^2 + 2{d_5}{r_1}{r_2}( 2( v + 2x)+\nonumber\\
&& {r_2}( 2 + {r_1} + 3{r_2} )) ( 2u{r_1} + ( -1 + {{r_1}}^2 +
{{r_3}}^2 ) {r_4} ))\\
A^1_{18} &=&
\frac{{L_1}{{m_t}}^{\frac{5}{2}}}{{r_1}{r_2}{{r_4}}^{\frac{3}{2}}}(
-2{d_5}{r_1}{r_2} ( -2u - ( -1 + 2{r_1} + {r_2} ) {r_4} ) ( 2u{r_1}
+ ( -1 + {{r_1}}^2 + {{r_3}}^2 ) {r_4} )+\nonumber\\
&& {d_1}{{r_4}}^2( u( {r_1} - {r_2} )- 2x( 1 + {r_2} )+ ({r_1} -1 -
2{r_2} ) {{r_4}}^2 ) + {d_2}{r_4}( u( 4{{r_1}}^2 -{r_1}( {r_2}-3 )
\nonumber\\
&&  -( {r_2}-3 ) {r_2} )+ {r_4}( ( {r_1} - {r_2} ) ( 1 + {r_2} +
2{{r_3}}^2 ) + {r_1}( 1 +2{r_1} - {r_2} ) {r_4} + 2{{r_4}}^2 )))\\
A^1_{19} &=&\frac{{L_1}{{m_t}}^{\frac{5}{2}}}{
{r_1}{r_2}{{r_4}}^{\frac{3}{2}}}( {d_2} ( x{r_1}( -3 + {r_2} ) -
2{r_1}{{r_2}}^3 - {{r_1}}^2( r_2 + v + x + 3 z) +{r_2}( -3x + \nonumber\\
&& {r_2}( r_2 + 3 v + z ))) {r_4} - {d_1}{{r_4}}^3( x + {r_2}{r_4}
)- 2{d_5}{r_1}{r_2}( 2x + {r_2}{r_4} ) ( 2u{r_1}
+\nonumber\\
&& ( -1 +{{r_1}}^2 + {{r_3}}^2 ) {r_4} ))\\
A^1_{25} &=& \frac{-2{L_1}{{m_t}}^{\frac{5}{2}}}{{r_1}{r_2}
{\sqrt{{r_4}}}}( {d_2}( 1 + {r_2} ) {r_4} -{d_1}{{r_4}}^2 +
2{r_1}{r_2} ( {d_5}( 1 + {r_1} ) ( 1 + {r_2} )-\nonumber\\
&& {d_4}( r_2 - v - x + 3 z )+ {d_3}{r_2}{r_4})) \\
A^1_{29} &=& -4{L_1}{{m_t}}^{\frac{5}{2}}{\sqrt{{r_4}}}(
{d_4} + {d_3}{r_1} - {d_3}{r_2} + {d_4}{r_2} )\\
A^{3'}_1 &=& \frac{-\kappa
{{m_t}}^{\frac{7}{2}}}{2{L_2}{r_1}{r_2}{\sqrt{{r_4}}}}
(2{d_4}{r_1}{r_2}( -2x( x + ( u + x ) {r_1} )+ x( 2 - 2v + 2{r_1} -
{{r_3}}^2 ) {r_4} + \nonumber\\
&&( v - x ) ( 1 + {r_1} ) {{r_4}}^2 - v{{r_4}}^3 + {{r_2}}^2{r_4}( u
+ {r_4} + {r_1}{r_4} )+ {r_2}( u + {r_4} + {r_1}{r_4}
)\cdot\nonumber\\
&& ( -2( u + x )+ {r_4} - 2{{r_4}}^2 ))+ {r_4}( 2x{d_1}( -u +
{{r_4}}^2 )+ {d_1}{{r_4}}^2( u + 2v + {r_1} + \nonumber\\
&& {r_2} + {r_1}{r_2} + {{r_2}}^2 + {{r_4}}^2 )+ {d_2} ( 2u^2 + 2ux
+ u{{r_4}}^2 - {{r_3}}^2{{r_4}}^2 - {r_2}( x+y)+\nonumber\\
&& {r_1}( 1 + {r_4} ) ( x+y ))- 2{d_3}{r_1}{r_2}( -(
x ( 2( u + x )+ {r_1} ))-( u + x ) {{r_2}}^2 +\nonumber\\
&& ( v - x ) {{r_4}}^2 +
{r_2}( x + ( u + x ){r_1} +( 1 + {r_1} ) {{r_4}}^2 )))) \\
A^{3'}_4 &=& \frac{\kappa {{m_t}}^{\frac{7}{2}}}{2{L_2}
{r_1}{r_2}{\sqrt{{r_4}}}}( 2{d_5}{r_1}{r_2} ( 2x( x + ( u + x )
{r_1} )+ x( -2 + 2v - 2{r_1} + {{r_3}}^2 ) {r_4} -\nonumber\\
&& ( v - x ) ( 1 + {r_1} ) {{r_4}}^2 + v{{r_4}}^3 - {{r_2}}^2{r_4}(
u + {r_4} + {r_1}{r_4} )+ {r_2}( u + {r_4} + {r_1}{r_4}
)\cdot\nonumber\\
&& ( 2( u + x )+ {r_4}( -1 + 2{r_4} )))+ {r_4}( {d_1}( 2x^2 +
x{{r_4}}^2 + ( {r_1} - {r_2} ) {r_2}{{r_4}}^2 )+ \nonumber\\
&& {d_2}( -( x( 2( u + x )+ {r_1} ))+ x{r_1}{r_4} + ( 2v - x )
{{r_4}}^2 - {{r_2}}^2( 2( u + x )+ {{r_4}}^2 )+\nonumber\\
&& {r_2}( x + 2( u + x ) {r_1} + ( 1 + 2{r_1} ) {{r_4}}^2 )))) \\
A^{3'}_5 &=& \frac{\kappa {{m_t}}^{\frac{7}{2}}{\sqrt{{r_4}}}}
{2{L_2}{r_1}{r_2}} ( {d_2}( 2(
u + x )+ {r_1}( 2{r_1}-1 )+ {r_2} + 3{r_1}{r_2} + {{r_2}}^2 )+ {d_1}( 2x + {{r_4}}^2 )) \\
A^{3'}_7 &=& \frac{\kappa
{{m_t}}^{\frac{7}{2}}}{2{L_2}{r_1}{r_2}{\sqrt{{r_4}}}} (
-({d_1}{r_4} ( 2x + {{r_4}}^2 ))+ {d_2}{r_4}( 2( u + x )+ {r_2}
+{r_1}( -1 + {r_4} )+ {{r_4}}^2 )\nonumber\\
&& - 2{d_5}{r_1}{r_2} ( 2x({r_1} -1 )+ 2u({r_1}-1 - {r_2} )+
({{r_1}}^2-1 ) {r_4} + ( {r_1} -1) {{r_4}}^2 )) \\
A^{3'}_8 &=& \frac{\kappa {{m_t}}^{\frac{7}{2}}}{2{L_2}{r_1}
{r_2}{\sqrt{{r_4}}}}( 2{d_3}{r_1}{r_2} ( 2x + {{r_1}}^2 + {r_1}(
{r_2}-1 )+ {r_2} + 2{{r_2}}^2 ) {r_4} - 2{d_1}{r_4}( 2x + {{r_4}}^2
)\nonumber\\
&& - 2{d_4}{r_1}{r_2}( -2x{r_1} + 2( u + x + u{r_2} )+
{r_4}( {{r_3}}^2 + {r_4} - {r_1}{r_4} ))) \\
A^{3'}_9 &=& \frac{\kappa {{m_t}}^{\frac{7}{2}}}{2{L_2}
{r_1}{r_2}{{r_4}}^{\frac{3}{2}}}( 2{d_5}{r_1}{r_2}( 1 + {r_2} ) (
2u{r_1} + ({{r_1}}^2-1 + {{r_3}}^2 ) {r_4} )-{d_2}{r_4}( -(
{{r_1}}^2( 1 + {r_2} ))+\nonumber\\
&& ( 1 + {r_2} - {{r_3}}^2 )
{r_4} + {r_1}( 2x + {r_2} + {{r_2}}^2 + {{r_4}}^2 )))\\
A^{3'}_{10} &=& \frac{\kappa {{m_t}}^{\frac{7}{2}}}{2{L_2}{r_1}
{r_2}{\sqrt{{r_4}}}}( 4x{d_4}{r_1}{r_2} + 4x{d_4}{{r_1}}^2{r_2} -
4u{d_4}{r_1}{{r_2}}^2 -
2x{d_1}{r_4} + 4x{d_3}{r_1}{r_2}{r_4} +\nonumber\\
&&  4v{d_4}{r_1}{r_2}{r_4} + 2{d_4}{r_1}{{r_2}}^2{r_4} -
2{d_3}{{r_1}}^2{{r_2}}^2{r_4} + 2{d_3}{r_1}{{r_2}}^3{r_4} +
2{d_4}{r_1}{{r_2}}^3{r_4} -\nonumber\\
&& 4{d_4}{r_1}{{r_2}}^2{{r_4}}^2 + {d_1}{{r_4}}^3 +
2{d_5}{r_1}{r_2}( 2u{r_1} + ( -1 + {{r_1}}^2 + {{r_3}}^2 ) {r_4}
)-\nonumber\\
&& {d_2}{r_4}( 2( u +x )- {r_2}( 1 + 2{r_2} )+ {{r_4}}^2 + {r_1}( 1 + 2{r_2} + {r_4} ))) \\
A^{3'}_{11} &=& -\frac{\kappa {{m_t}}^{\frac{7}{2}}}{{L_2}{r_1}
{\sqrt{{r_4}}}}( {d_2}( {r_1} - {r_2} ) {r_4} +{d_5}{r_1}( 2x( 1 +
{r_1} )-
2u{r_2} + 2v{r_4} + \nonumber\\
&& {r_2}( 1 + {r_2} - 2{r_4} ) {r_4} ))\\
A^{3'}_{12} &=& \frac{\kappa {{m_t}}^{\frac{7}{2}}}{2{L_2}{r_1}{r_2}
{{r_4}}^{\frac{3}{2}}}( {d_2} ( {{r_1}}^2{r_2} - 2{r_1}( z+r_2 )-
{r_2}( 2v + {r_2}( 2 + {r_2} )) ) {r_4}
+\nonumber\\
&& 2{d_5}{r_1}{{r_2}}^2( 2u{r_1} + ( -1 +
{{r_1}}^2 + {{r_3}}^2 ) {r_4} ))\\
A^{3'}_{13} &=& \frac{-( \kappa
{{m_t}}^{\frac{7}{2}}}{4{L_2}{r_1}{r_2}{\sqrt{{r_4}}}}( {d_2}( -2( u
+ x )- {r_1}( 1 + 2{r_1} )+ {r_2} - 5{r_1}{r_2} + {{r_2}}^2 ) {r_4}
+\nonumber\\
&& 2{d_5}{r_1}{r_2}( 2u{r_1} + ( -1 + {{r_1}}^2 + {{r_3}}^2 ) {r_4}
) +{d_1}{r_4}( 2x +{{r_4}}^2 )))\\
A^{3'}_{14} &=& -\frac{\kappa {{m_t}}^{\frac{7}{2}} {\sqrt{{r_4}}}}
{{L_2}{r_1}{r_2}}( {r_1}{r_2}({d_3}{{r_4}}^2 +{d_4}( 2( u + x )+
2{{r_1}}^2 +5{r_1}{r_2} + 3{{r_2}}^2 +{r_4} ))-{d_1}{{r_4}}^2) \\
A^{3'}_{15} &=& \frac{\kappa
{d_5}{{m_t}}^{\frac{7}{2}} {\sqrt{{r_4}}} }{{L_2}}( 2( u + x )+ ( 1 + 2{r_1} + 3{r_2} ) {r_4} )\\
A^{3'}_{16} &=& \frac{\kappa
{{m_t}}^{\frac{7}{2}}}{4{L_2}{r_1}{r_2}{\sqrt{{r_4}}}}( {d_2} ( 2( u
+ x )+ 2{{r_1}}^2 + {r_1}( -1 + {r_2} )+ {r_2} + 3{{r_2}}^2 ) {r_4}
+ \nonumber\\
&& 2{d_5}{r_1}{r_2} ( 2u{r_1} + ( -1 + {{r_1}}^2 +
{{r_3}}^2 ) {r_4} )+{d_1}{r_4}( 2x + {{r_4}}^2 ))\\
A^{3'}_{17} &=& \frac{\kappa {{m_t}}^{\frac{7}{2}}}{4{L_2}{r_1}
{r_2}{{r_4}}^{\frac{3}{2}}}( 2x{d_1}( v + {r_1} + {{r_2}}^2 )
{{r_4}}^2 + {d_1}( v + {r_2} - {r_1}{r_2} + 2{{r_2}}^2 ) {{r_4}}^4
+\nonumber\\
&& 2{d_5}{r_1}{r_2}( 2u{r_1} + ( -1 + {{r_1}}^2 + {{r_3}}^2 ) {r_4}
) ( 4x + {r_4}( v + {r_2} + {{r_2}}^2 - 2{r_2}{r_4} )) +
\nonumber\\
&& {d_2}{r_4}( v( -2u + 3{r_1}( 1 + {r_2} )-{r_2}( 3 + {r_2} ))
{r_4} - {r_2}( 2u + 2{{r_1}}^3 + {{r_1}}^2( {r_2}-1)+\nonumber\\
&& {r_2} - {r_1}( 1 + {r_2}( 4 + {r_2} )+ 2{{r_3}}^2 )+  {r_2}(
-{r_2} + 2( u + {{r_3}}^2 )) ) {r_4} +\nonumber\\
&& 2x( 2{r_1}( {r_1}- {r_2} )+ ( -2 + v +{{r_3}}^2 ) {r_4} )))\\
A^{3'}_{18} &=& \frac{-\kappa {{m_t}}^{\frac{7}{2}}}{4
{L_2}{r_1}{r_2}{\sqrt{{r_4}}}}( 2{d_5}{r_1}{r_2}( u + 2x + {r_1}( 2
+ {r_1} )+ 2{r_2} + 4{r_1}{r_2} + 3{{r_2}}^2 ) ( 2u{r_1}
+\nonumber\\
&& ( -1 + {{r_1}}^2 + {{r_3}}^2 ) {r_4} )+ {d_2}{r_4}( -2u^2 - u (
2{{r_1}}^2 + {r_1}( -3 + {r_2} )+ 3{r_2}( 1 + {r_2} ))-
\nonumber\\
&& 2x( u + {r_2} + {r_1}( {r_4}-1))- ( ( {r_1}-2 ) ( r_1 -1)
{r_1} + 2{r_2} + 4{r_1}{r_2} +3( 1 + {r_1} ) {{r_2}}^2 ) {r_4} \nonumber\\
&&  + {{r_3}}^2{{r_4}}^2 )+ {d_1}{r_4}( 4x^2 + ( u + {r_1} +
{{r_1}}^2 - {r_2} + 3{r_1}{r_2}){{r_4}}^2 + 2x( u + 2{{r_4}}^2 )))\\
A^{3'}_{19} &=&\frac{-( \kappa
{{m_t}}^{\frac{7}{2}}}{4{L_2}{r_1}{r_2}{\sqrt{{r_4}}}}( {d_2}( 2x( u
+ x )+ x{r_2} + 4{{r_1}}^3{r_2} - ( 4u + 2v + x ) {{r_2}}^2 -
3{{r_2}}^3 - 3{{r_2}}^4 +\nonumber\\
&& {{r_1}}^2( -2v + ( -3 + {r_2} ) {r_2} ) - {r_1}( x +
{r_2}( 6 r_2 - 4 u - 2 v - 9 x + 6 z ) )) {r_4} +{d_1}( 2x + \nonumber\\
&&  {{r_1}}^2 - {{r_2}}^2 ) {r_4} ( x + {r_2}{r_4} )+
2{d_5}{r_1}{r_2}( x + 2{r_2}{r_4} ) ( 2u{r_1} + ( {{r_1}}^2-1  +
{{r_3}}^2 ){r_4} )))\\
A^{3'}_{25} &=& \frac{-( \kappa
{{m_t}}^{\frac{7}{2}}}{2{L_2}{r_1}{r_2}{\sqrt{{r_4}}}}(
-4u{d_4}{r_1}{{r_2}}^2 + 4v{d_4}{r_1}{r_2}{r_4} +
2{d_4}{r_1}{{r_2}}^2{r_4} - 2{d_3}{{r_1}}^2{{r_2}}^2{r_4}
+\nonumber\\
&& 2{d_3}{r_1}{{r_2}}^3{r_4} + 2{d_4}{r_1}{{r_2}}^3{r_4} -
4{d_4}{r_1}{{r_2}}^2{{r_4}}^2 + {d_1}{{r_4}}^3 +{d_2}{r_4}( 2( u + x
)+ {r_2} +\nonumber\\
&& {r_1}( -1 + {r_4} )+ {{r_4}}^2 ) + 2{d_5}( -1 + {r_1} )
{r_1}{r_2} ( 2( u + x )+ {r_4}( 1 + {r_1} + {r_4} )) +\nonumber\\
&& 2x( -{d_1}{r_4}+2{r_1}{r_2}( {d_4} + {d_4}{r_1} - {d_4}{r_2} + {d_3}{r_4} )))) \\
A^{3'}_{29} &=& -\frac{\kappa {{m_t}}^{\frac{7}{2}}{\sqrt{{r_4}}}
}{{L_2}{r_1}{r_2}}({r_1}{r_2}( {d_3}{{r_4}}^2 + {d_4}( 2( u + x ) +
2{{r_1}}^2 + 3{r_1}{r_2} + {{r_2}}^2 + {r_4} ) -{d_1}{{r_4}}^2) )
\end{eqnarray}
and
\begin{eqnarray}
A^2_{1}&=&-\frac{A^1_{1}}{L_1}L_2-
\frac{4{L_2}{{m_t}}^{\frac{5}{2}}}{{r_1}{r_2}{\sqrt{{r_4}}}}(
2{r_1}{r_2} ( {d_3}( x{r_1} + x{r_2} - {{r_1}}^2{r_2} + {{r_2}}^3 )
+ {d_4}( - {{r_1}}^2{r_2}+\nonumber\\
&&  {r_1}( z + r_2^2 ) + {r_2}(3 z-u - 2 v ))) +
{d_2}( u + {{r_1}}^2 + {{r_2}}^2 ){r_4} + {d_1}{{r_4}}^3 )\\
A^2_{4}&=&-\frac{A^1_{4}}{L_1}L_2+\frac{4{L_2}
{{m_t}}^{\frac{5}{2}}}{{r_1}{r_2}{\sqrt{{r_4}}}}( {d_2} ( -x + (
{r_1} - {r_2} ) {r_2} ) {r_4} +\nonumber\\
&& 2{d_5}{r_1}{r_2}( -( v + 2x ) {r_4}- 2{{r_2}}^2{r_4} + {r_1}(
x + 2{r_2}{r_4} ) - {r_2}( -u+ x + {{r_4}}^2 ))) \\
A^2_{5}&=&-\frac{A^1_{5}}{L_1}L_2+ \frac{4{d_2}
{L_2}{{m_t}}^{\frac{5}{2}}{\sqrt{{r_4}}}}{{r_1}{r_2}}\\
A^2_{7}&=&-\frac{A^1_{7}}{L_1}L_2-
\frac{4{L_2}{{m_t}}^{\frac{5}{2}}}{{r_1}{r_2}{\sqrt{{r_4}}}}(
-3{d_2}{r_4} + 2{d_5}{r_1}{r_2}{r_4})\\
A^2_{8}&=&-\frac{A^1_{8}}{L_1}L_2+ \frac{8{L_2}{{m_t}}^{\frac{5}{2}}
}{{r_1}{r_2}{\sqrt{{r_4}}}}( -(
{d_1}{r_4} ) +{d_3}{r_1}{r_2}{r_4} + {d_4}{r_1}{r_2}{r_4} )\\
A^2_9&=&-\frac{A^1_9}{L_1}L_2-\frac{4{d_2}{L_2}{{m_t}}^{\frac{5}{2}}
}{{r_1}{r_2}{\sqrt{{r_4}}}}( {r_1} + {r_4} )\\
A^2_{10}&=&-\frac{A^1_{10}}{L_1}L_2-\frac{4{L_2}{{m_t}}^{\frac{5}{2}}
}{{r_1}{r_2}{\sqrt{{r_4}}}}(
2{d_4}{r_1}{{r_2}}^2 + 3{d_2}{r_4} ) \\
A^2_{11}&=&-\frac{A^1_{11}}{L_1}L_2+\frac{8{d_5}
{L_2}{{m_t}}^{\frac{5}{2}}{r_2}}{{\sqrt{{r_4}}}} \\
A^2_{12}&=&-\frac{A^1_{12}}{L_1}L_2\\
A^2_{13}&=&-\frac{A^1_{13}}{L_1}L_2+\frac{6{d_2}{L_2}{{m_t}}^{\frac{5}{2}}{\sqrt{{r_4}}}}{{r_1}{r_2}}\\
A^2_{14}&=&-\frac{A^1_{14}}{L_1}L_2-8{d_4}{L_2}{{m_t}}^{\frac{5}{2}}{\sqrt{{r_4}}} \\
A^2_{15}&=&-\frac{A^1_{15}}{L_1}L_2+8{d_5}{L_2}{{m_t}}^{\frac{5}{2}}{\sqrt{{r_4}}} \\
A^2_{16}&=&-\frac{A^1_{16}}{L_1}L_2+\frac{2{d_2}{L_2}{{m_t}}^{\frac{5}{2}}{\sqrt{{r_4}}}}{{r_1}{r_2}}\\
A^2_{17}&=&-\frac{A^1_{17}}{L_1}L_2-\frac{2{L_2}{{m_t}}^{\frac{5}{2}}}{{r_1}
{r_2}{{r_4}}^{\frac{3}{2}}}( {d_2}{r_4} ( 2{{r_1}}^2{r_2} + ( v + 4x
+ 3{{r_2}}^2 ) {r_4} - {r_1}{r_2}( 2{r_2} + {r_4} ) ) +
\nonumber\\
&& {r_2}( {d_1}( {r_1} - {r_2} ) {{r_4}}^2 + 4{d_5}{r_1}{r_2}(
2u{r_1} + ( -1+ {{r_1}}^2 + {{r_3}}^2 ) {r_4} )))\\
A^2_{18}&=&-\frac{A^1_{18}}{L_1}L_2+\frac{2{L_2}{{m_t}}^{\frac{5}{2}}}{{r_1}{r_2}{\sqrt{{r_4}}}}
(-2{d_5}{r_1}{r_2} ( 2u{r_1} + ( -1 + {{r_1}}^2 + {{r_3}}^2 ) {r_4}
)-{d_1}{r_4}( 2x + {{r_4}}^2 )\nonumber\\
&& +{d_2}{r_4}( 3u - {{r_2}}^2 +
2{{r_4}}^2 + {r_1}( {r_2} + {r_4} ) ))\\
A^2_{19}&=&-\frac{A^1_{19}}{L_1}L_2-\frac{2{d_2}{L_2}{{m_t}}^{\frac{5}{2}}{\sqrt{{r_4}}}}{{r_1}{r_2}}
( 3x + ( {r_1} - {r_2} ) {r_2} )\\
A^2_{25}&=&-\frac{A^1_{25}}{L_1}L_2-\frac{4{L_2}{{m_t}}^{\frac{5}{2}}
}{{r_1}{r_2}{\sqrt{{r_4}}}}(
{d_2}{r_4} +2{r_1}{r_2}( -( {d_4}{r_2} )+ {d_5}{r_4} )) \\
A^2_{29}&=&-\frac{A^1_{29}}{L_1}L_2-8{d_4}{L_2}{{m_t}}^{\frac{5}{2}}{\sqrt{{r_4}}} \\
A^{4'}_{1} &=&\frac{A^{3'}_{1}}{L_1}L_2+\frac{\kappa
{{m_t}}^{\frac{7}{2}}}{{L_1}{r_1}{r_2}{\sqrt{{r_4}}}} (
2{d_4}{r_1}{r_2}( -2x^2 - x{r_4}( -2{r_1} + 2{r_2} + {r_4} )
+{r_4}( - u{r_2}+v{r_4}\nonumber\\
&& - {r_2}{{r_4}}^2 ) )+ {r_4}(
{d_1}{{r_4}}^3 + {d_2}( {r_1} - {r_2} ) ( x+y)+
2{d_3}{r_1}{r_2}( x{r_1} - {r_2}( x + {{r_4}}^2 )) ) )\\
A^{4'}_{4} &=&\frac{A^{3'}_{4}}{L_1}L_2+\frac{\kappa
{{m_t}}^{\frac{7}{2}}}{{L_1}{r_1}{r_2}{\sqrt{{r_4}}}} ( -
{d_2}{r_4}( -x {r_1}+ {r_2}( x + {{r_4}}^2 ) )  +\nonumber\\
&& 2{d_5}{r_1}{r_2}( -2x^2 + x( 2{r_1} - 2{r_2} - {r_4} ) {r_4} +
{r_4}( -u {r_2}+ v{r_4} - {r_2}{{r_4}}^2 )) )\\
A^{4'}_{5} &=&\frac{A^{3'}_{5}}{L_1}L_2-\frac{\kappa
{d_2}{{m_t}}^{\frac{7}{2}} {\sqrt{{r_4}}}}
{{L_1}{r_1}{r_2}}( -{r_1} + {r_2} )\\
A^{4'}_{7} &=&\frac{A^{3'}_{7}}{L_1}L_2-\frac{\kappa
{{m_t}}^{\frac{7}{2}} }{{L_1} {r_1}{r_2}{\sqrt{{r_4}}}}( {d_2}(
{r_2}-{r_1} ) {r_4} +
2{d_5}{r_1}{r_2}( 2( u + x )+ {{r_4}}^2 ))\\
A^{4'}_{8} &=&\frac{A^{3'}_{8}}{L_1}L_2+\frac{2\kappa
{{m_t}}^{\frac{7}{2}}}{{L_1}{\sqrt{{r_4}}}}( {d_3}(
{r_1} - {r_2} ) {r_4} +{d_4}( u + x + y ) )\\
A^{4'}_9 &=& \frac{A^{3'}_9}{L_1}L_2-\frac{\kappa
{{m_t}}^{\frac{7}{2}}}{{L_1}{r_1}{r_2} {{r_4}}^{\frac{3}{2}}}(
{d_2}{r_4} ( {{r_1}}^2 - {r_1}{r_2} - {r_2}{r_4}
)+\nonumber\\
&& 2{d_5}{r_1}{r_2}( 2u{r_1} + ( -1 + {{r_1}}^2 +
{{r_3}}^2 ) {r_4} ))\\
A^{4'}_{10} &=&\frac{A^{3'}_{10}}{L_1}L_2-\frac{\kappa
{{m_t}}^{\frac{7}{2}}( {d_2}( -{r_1} + {r_2} ) {r_4} +
2{d_4}{r_1}{r_2}( 2x + {r_2}{r_4} )) }{{L_1}{r_1}{r_2}
{\sqrt{{r_4}}}}\\
A^{4'}_{11} &=&\frac{A^{3'}_{11}}{L_1}L_2+\frac{2\kappa
{d_5}{{m_t}}^{\frac{7}{2}} } {{L_1}{\sqrt{{r_4}}}}(
2x + {r_2}{r_4} )\\
A^{4'}_{12} &=& \frac{A^{3'}_{12}}{L_1}L_2+\frac{2\kappa {d_2}
{{m_t}}^{\frac{7}{2}}{\sqrt{{r_4}}}}{{L_1}{r_1}}\\
A^{4'}_{13} &=& \frac{A^{3'}_{13}}{L_1}L_2-\frac{\kappa
{d_2}{{m_t}}^{\frac{7}{2}}{\sqrt{{r_4}}}}
{2{L_1}{r_1}{r_2}}( {r_1} - {r_2} )\\
A^{4'}_{14} &=&\frac{A^{3'}_{14}}{L_1}L_2+\frac{2\kappa
{d_4}{{m_t}}^{\frac{7}{2}}{{r_4}}^{\frac{3}{2}}}{{L_1}}\\
A^{4'}_{15} &=&\frac{A^{3'}_{15}}{L_1}L_2-\frac{2\kappa
{d_5}{{m_t}}^{\frac{7}{2}}{{r_4}}^{\frac{3}{2}}}{{L_1}}\\
A^{4'}_{16} &=& \frac{A^{3'}_{16}}{L_1}L_2+\frac{\kappa
{d_2}{{m_t}}^{\frac{7}{2}}
{\sqrt{{r_4}}}}{2{L_1}{r_1}{r_2}}( {r_1} - {r_2} )\\
A^{4'}_{17} &=& \frac{A^{3'}_{17}}{L_1}L_2-\frac{\kappa
{{m_t}}^{\frac{7}{2}}}{2 {L_1}{r_1}{r_2}{{r_4}}^{\frac{3}{2}}}(
2{d_5}{r_1}{r_2} ( 4x + {r_2}{r_4} ) ( 2u{r_1} + ( -1 + {{r_1}}^2 +
{{r_3}}^2 ) {r_4} ) + \nonumber\\
&& {d_1}{{r_4}}^2( 2x{r_1} + {r_2}{{r_4}}^2 )+ {d_2}{r_4}( 4x{r_1}(
{r_1} - {r_2} )+ {r_1}{r_4}( 3v + {{r_2}}^2 - {r_2}{r_4} )
-\nonumber\\
&& {r_2}{r_4}( 4 u + 2 v + x - 2 y + z)))\\
A^{4'}_{18} &=& \frac{A^{3'}_{18}}{L_1}L_2+\frac{\kappa
{{m_t}}^{\frac{7}{2}}{\sqrt{{r_4}}}}{2{L_1}{r_1}{r_2}} ( {d_1}(
{r_1} - {r_2} ) {{r_4}}^2 +4{d_5}{r_1}{r_2}( 2u{r_1} + ( -1 +
{{r_1}}^2 + {{r_3}}^2 ) {r_4} )+ \nonumber\\
&& {d_2}( -2x{r_2} + 2{{r_1}}^2{r_4} - 3{r_2}( u + {{r_4}}^2 )+
{r_1}( 3u + 2x - 2{r_2}{r_4} + {{r_4}}^2)))\\
A^{4'}_{19} &=& \frac{A^{3'}_{19}}{L_1}L_2-\frac{\kappa
{d_2}{{m_t}}^{\frac{7}{2}}{\sqrt{{r_4}}}
}{2{L_1}{r_1}{r_2}}( x{r_1} - x{r_2} + 3{r_2}{{r_4}}^2 )\\
A^{4'}_{25} &=&\frac{A^{3'}_{25}}{L_1}L_2-\frac{\kappa
{{m_t}}^{\frac{7}{2}}}{{L_1}{r_1} {r_2}{\sqrt{{r_4}}}}( {d_2}( {r_1}
- {r_2} ) {r_4} +2{r_1}{r_2}({d_5}( u+x+y )-{d_4}( 2x + {r_2}{r_4}
)))\\
A^{4'}_{29} &=&\frac{A^{3'}_{29}} {L_1}L_2+\frac{2\kappa
{d_4}{{m_t}}^{\frac{7}{2}} {{r_4}}^{\frac{3}{2}}}{{L_1}}.
\end{eqnarray}
Here for convenience an overall factor ${\cal C}_s$ has been
contracted out from these coefficients. Then the square of the
amplitude $|M_i|^2$ can be conveniently obtained with the help of
Eqs.(\ref{amat},\ref{relat1},\ref{relat2}) and Eqs.(\ref{3pja},
\ref{3pjb},\ref{3pjc}).

\end{document}